\documentclass[a4paper,11pt]{article}

\usepackage{color}
\usepackage{ulem}
\usepackage{enumerate}
\usepackage{amsmath}
\usepackage{graphicx}
\usepackage{subfigure}
\usepackage{amssymb}
\usepackage{braket}
\usepackage{enumitem}

\def \be {\begin{equation}}
\def \ee {\end{equation}}
\def \ba {\begin{array}}
\def \ea {\end{array}}
\def \bea {\begin{eqnarray}}
\def \eea {\end{eqnarray}}

\def \ble {\begin{widetext}\begin{equation}}
\def \ele {\end{equation}\end{widetext}}
\def \blea {\begin{widetext}\begin{eqnarray}}
\def \elea {\end{eqnarray}\end{widetext}}

\def \nn {\nonumber}

\def \blea {\begin{widetext}\begin{eqnarray}}
\def \elea {\end{eqnarray}\end{widetext}}
\def \mO {\mathcal{O}}

\usepackage{amsmath,amssymb,mathtools}
\usepackage{graphicx}
\usepackage[colorlinks=true,linkcolor=red,citecolor=blue,urlcolor=blue,bookmarks]{hyperref}

\def \be {\begin{equation}}
\def \ee {\end{equation}}
\def \ba {\begin{array}}
\def \ea {\end{array}}
\def \bea {\begin{eqnarray}}
\def \eea {\end{eqnarray}}
\def \nn {\nonumber}

\def \p {\partial}

\def \arccosh {\mathop{\rm arccosh}}

\def \Tr {{\mathop\textrm{Tr}}}

\def \and {{~\textrm{and}~}}

\hypersetup{linktoc=page}

\begin{document}

\title{Entanglement measures for causally connected subregions and holography}


\author{XiangKun Gong\footnote{gxk964@hust.edu.cn},~  Wu-zhong Guo\footnote{wuzhong@hust.edu.cn},~  Jin Xu\footnote{xujin1@hust.edu.cn}}


\date{}

\maketitle
\vspace{-10mm}
\begin{center}
\it
\it School of Physics, Huazhong University of Science and Technology,\\
 Wuhan, Hubei 430074, China\\\vspace{1mm}
\vspace{10mm}
\end{center}

\begin{abstract}
In this paper, we investigate entanglement for causally connected subregions $A$ and $B$ in quantum field theory and holography. Recent developments have established that a transition operator  $T_{AB}$ can be well-defined for such subregions, which is generally non-Hermitian. By employing the Schwinger-Keldysh formalism and the real-time replica method, we show how to construct $T_{AB}$ and compute associated entanglement measures. In certain configurations, this leads to a notion of timelike entanglement entropy, for which we provide explicit quantum field theory computations and propose a holographic dual via analytic continuation from the Euclidean setup. Both analytical and numerical results are compared and found consistent.

If entanglement between causally connected subregions is to be meaningful, it should also be able to define other entanglement measures. Motivated by the spacelike case, we propose a timelike extension of the entanglement wedge cross section, though we do not expect it to carry the same physical interpretation. In AdS$_3$/CFT$_2$, we compute explicit examples and find that the timelike entanglement wedge cross section is generally positive. Furthermore, we show that the reflected entropy for timelike intervals---obtained via analytic continuation of twist correlators---coincides with twice the timelike entanglement wedge cross section at leading order in $G$, supporting a holographic duality in the timelike case. We also discuss the extension of other entanglement measures, such as logarithmic negativity, to timelike separated regions using replica methods. We highlight conceptual challenges in defining reflected entropy via canonical purification for non-Hermitian operators.
\end{abstract}

\maketitle

\newpage
\tableofcontents
\newpage

\section{Introduction}

Quantum entanglement plays a central role in understanding the holographic nature of gravity. Within the AdS/CFT correspondence, the Ryu–Takayanagi (RT) formula \cite{Ryu:2006bv} and its covariant generalization, the\\ Hubeny–Rangamani–Takayanagi (HRT) formula \cite{Hubeny:2007xt}, establishes a profound connection between the entanglement entropy (EE) in the boundary theory and the extremal surfaces in the bulk. These foundational results provide a powerful framework for probing the microscopic degrees of freedom of quantum gravity via dual field theories.

Extensive studies \cite{VanRaamsdonk:2010pw}-\cite{Dong:2016eik} have significantly advanced our understanding of the holographic structure of gravitational systems. When quantum corrections are taken into account, the framework is generalized to the quantum extremal surface (QES) prescription \cite{Engelhardt:2014gca}, which plays a crucial role in addressing the black hole information paradox \cite{Penington:2019npb}-\cite{Almheiri:2020cfm}. For a more comprehensive overview of recent progress in this field, one may refer to the review essay \cite{Takayanagi:2025ula}.

On the other hand, EE is only one among various useful quantities that characterize the entanglement between subregions. In the case of mixed states, more refined entanglement measures are required to properly quantify quantum correlations beyond EE. From the holographic perspective, RT surfaces typically extend to the AdS boundary and hence yield divergent quantities. As such, they should be regarded as specific bulk geometric constructs that may correspond to well-defined boundary quantities under duality. Motivated by this, various bulk geometric quantities have been proposed as duals to other entanglement measures in the boundary theory.

One prominent example is the entanglement wedge cross section (EWCS), which is defined within the entanglement wedge bounded by the RT surfaces. EWCS is conjectured to be dual to the entanglement of purification \cite{Takayanagi:2017knl,Nguyen:2017yqw} or the reflected entropy \cite{Dutta:2019gen} in the boundary theory. Other entanglement measures that are widely studied in quantum field theories and may potentially admit holographic duals include entanglement negativity \cite{Calabrese:2012ew,Calabrese:2012nk}, Rényi entropy \cite{Dong:2016fnf}, and the entanglement spectrum \cite{Li2008,Calabrese2008}, among others.

EE and other entanglement measures are typically defined on a given Cauchy surface for a subsystem $A$. The reduced density matrix is obtained by tracing out the complement $\bar A$: $\rho_A:=\Tr_{\bar A} \rho$, where $\rho$ is the state of the system. In the framework of QFTs, the Rényi entropy $S_n(\rho_A)=\frac{\log \Tr\rho_A^n}{1-n}$ can be computed using the replica trick, and the EE is obtained as the von Neumann entropy of $\rho_A$ at the limit $S_A(\rho_A)=\lim_{n\to 1}S_n(\rho_A)$. From the perspective of algebraic QFTs, the reduced density matrix $\rho_A$ encodes information only within the causal domain of dependence of the region $A$. Therefore, EE and other entanglement measures primarily capture quantum correlations between spacelike-separated regions. We will refer to this EE as spacelike EE for comparison in what follows.

Time does not appear to play a central role in the conventional understanding of EE and the significance of the RT formula. EE becomes relevant for probing the thermalization dynamics of a system only when time evolution is explicitly considered, such as in quantum quenches \cite{Calabrese:2016xau}. However, in QFTs, time and space are intrinsically intertwined. This motivates the investigation of how temporal aspects can be incorporated into the concept of entanglement.

In \cite{Doi:2022iyj}, the authors introduce the timelike EE in Lorentzian spacetime, which may serve as an example of pseudo entropy \cite{Nakata:2020luh}. In this setup, the entanglement is extended to subsystems of temporal duration. In $(1+1)$-dimensional QFTs, this can be viewed as interchanging the roles of time and space coordinates. From the perspective of evaluating timelike EE, the key object is the correlation function of twist operators at timelike separations.  Moreover, it is expected that timelike EE can also be computed holographically via the RT formula \cite{Doi:2022iyj,Li:2022tsv,Doi:2023zaf,Heller:2024whi,Guo:2025pru,Nunez:2025ppd,Heller:2025kvp}. Exploring the holographic dual of timelike EE reveals novel features in the behavior of RT surfaces. Interestingly, timelike EE respects causality constraints and satisfies an exact sum rule that relates to spacelike EE \cite{Guo:2024lrr,Guo:2025pru}. 

Recently, the paper \cite{Milekhin:2025ycm} introduced a new framework for understanding entanglement in time, which can be applied to arbitrary quantum systems. The central idea is to interpret the entanglement in time as quantum correlations between two subregions, $A$ and $B$, which are separated by time. This perspective allows one to define a transition operator $T_{AB}$, which is generally non-Hermitian. The operator $T_{AB}$ can be viewed as a natural generalization of the reduced density matrix $\rho_{AB}$ when $A$ and $B$ are separated in space.

The construction of $T_{AB}$ is applicable both to quantum systems with finite-dimensional Hilbert spaces and to quantum field theories. In certain cases, it reduces to the notion of timelike EE introduced in \cite{Doi:2022iyj}. We will briefly review this concept in Section~\ref{Section_entanglement_in_time}. This paper continues the investigation of this framework from the perspective of QFTs and holography.

In our view, there are several motivations to consider the concept of entanglement in time:
\begin{itemize}
    \item  Entanglement in time is a fundamental concept that can be applied to arbitrary quantum systems. It generalizes the notion of the density matrix to a more general object often referred to as the ``spacetime density matrix''. In this framework, the transition operator plays a role analogous to that of the density matrix and can be used to define entropy-related quantities, such as the von Neumann entropy. For convenience, we will refer to the entropy associated with the transition operator as the EE for the transition operator, or simply as timelike EE, in contrast to the conventional spacelike EE. In a certain sense, the EE for the transition operator can be regarded as an example of the pseudo entropy introduced in \cite{Nakata:2020luh}.
    \item The transition operator $T_{AB}$ can be constructed from the specification of subregions $A$ and $B$, together with the quantum state of the system. As a result, entropy-related quantities derived from $T_{AB}$ are, in principle, computable. Specifically, in QFTs, the conventional spacelike EE can be computed via state preparation using the Euclidean path integral and the replica method \cite{Calabrese:2004eu}. For the timelike case, the transition operator $T_{AB}$ can be prepared using the Schwinger–Keldysh formalism \cite{Schwinger:1960qe,Feynman:1963fq,Keldysh:1964ud}, and the corresponding entropy-related quantities can be evaluated using the real-time replica method proposed in \cite{Dong:2016hjy,Colin-Ellerin:2020mva,Colin-Ellerin:2021jev}. In this work, we focus on the evaluation of such quantities in QFTs. Furthermore, in certain cases, the computation can be reformulated in terms of correlation functions involving twist operators with timelike separation. This implies that timelike EE is intimately connected to Lorentzian correlation functions, which have not been extensively explored in previous studies.

    \item The transition operator $T_{AB}$ encodes dynamical information and reflects the causal structure of the theory. As mentioned previously, the evaluation of entropy-related quantities associated with $T_{AB}$ requires knowledge of Lorentzian correlators. These correlators exhibit a richer structure than their Euclidean counterparts, as they depend sensitively on the spacetime locations and temporal ordering of the operators. In \cite{Guo:2024lrr,Guo:2025pru}, a relation between timelike and spacelike EE was derived for certain states by exploiting causality constraints. It is expected that this relation may hold more generally \cite{Guo:2025mwp}, suggesting a unified framework for timelike and spacelike entanglement. Notably, when $A$ and $B$ are taken to be Cauchy surfaces at different times, $T_{AB}$ captures the full dynamical evolution of the system. In particular, if $A$ and $B$ correspond to the in-state ($t \to -\infty$) and out-state ($t \to +\infty$) Cauchy surfaces, $T_{AB}$ is expected to be closely related to the scattering $S$-matrix.

    \item It is particularly intriguing to explore the holographic dual of timelike EE and related entanglement measures. Currently, there is no clear prescription for determining the corresponding bulk RT surface. Various proposals have been put forward, including the idea that the RT surface may extend into a complexified bulk geometry \cite{Heller:2024whi}. In this paper, we use a procedure to compute the holographic timelike EE by analytically continuing from its Euclidean counterpart \cite{Guo:2025pru}. Both numerical and analytical methods will be employed in this analysis. Furthermore, we extend several well-established concepts from the spacelike case to the timelike setting, with a particular focus on the entanglement wedge and its cross section. As previously mentioned, these concepts play a crucial role in understanding subregion/subregion duality and the holographic duals of reflected entropy or entanglement of purification. We will present supporting evidence that analogous dualities also exist in the timelike case. In summary, timelike EE provides a promising avenue for extending real-time holography \cite{Skenderis:2008dh,Skenderis:2008dg} and deepening our understanding of the dynamical aspects of holographic duality.
    \item Lastly,  the timelike EE can measured in the labtory for the spin chain system or quantum circuits. Some discussions can be found in \cite{Bou-Comas:2024pxf,Milekhin:2025ycm}.  
\end{itemize}

In this paper, we focus primarily on the EE associated with the transition operator $T_{AB}$ in QFTs. Our starting point is the preparation of $T_{AB}$ using the Schwinger–Keldysh formalism. The real-time replica method, originally introduced in \cite{Dong:2016hjy} to provide a derivation of the HRT formula, can also be used to evaluate spacelike EE. This framework proves particularly well suited to build and analyze the transition operator $T_{AB}$, allowing us to apply the replica method to compute its associated EE. We will examine several simple examples where the EE reduces to the computation of correlation functions of twist operators in Lorentzian spacetime.

We also investigate the holographic dual of timelike EE in several specific examples. By analytically continuing the Euclidean RT surface, we construct and evaluate its Lorentzian counterpart for timelike-separated regions. This method yields results that are in perfect agreement with the CFT calculations in AdS$_3$. We further explore higher-dimensional strip geometries, where both numerical and analytical results are obtained, showing excellent consistency.

With the constructed RT surfaces, we demonstrate that the concept of the entanglement wedge cross section (EWCS) can be generalized to the timelike case. Our definition of the timelike EWCS closely resembles that of its spacelike counterpart, although notable differences emerge. We present the precise definition and evaluate the timelike EWCS in various examples. The physical interpretation of the timelike entanglement wedge and its cross section may differ significantly from the spacelike case. In particular, we do not expect that the subregion/subregion duality or bulk reconstruction theorems \cite{Dong:2016eik} can be directly extended to the timelike context. The meaning of these constructions remains open for further investigation.

If the notion of entanglement in time is considered meaningful, it is natural to ask whether other entanglement measures can also be extended to timelike-separated regions. In this work, we propose methods for defining and computing logarithmic negativity and reflected entropy in QFTs. For logarithmic negativity, we consider the partial transpose of the transition operator $T_{AB}$, which again reduces to computing correlation functions of twist operators in Lorentzian spacetime. For reflected entropy, we apply an analytic continuation of twist operators to obtain the corresponding timelike result. We find that the reflected entropy in the timelike case is given by twice the timelike EWCS, providing further evidence that the holographic dual of reflected entropy can be generalized to timelike configurations.

However, we do not yet provide a definition of timelike reflected entropy via purification. We point out the subtleties and difficulties in extending the spacelike definition to the timelike case, particularly the challenge of constructing a purification for a non-Hermitian transition operator.

Finally, we conclude by outlining several open problems and promising directions for future exploration.

\section{Entanglement in time}\label{Section_entanglement_in_time}
In this section, we will first review different views on entanglement in time. We will mainly focus on QFTs.

\subsection{Views on entanglement in time}

A promising definition of timelike EE was proposed in \cite{Doi:2022iyj}, where the roles of time and space coordinates are exchanged. Under this approach, timelike EE can be understood in a manner analogous to the spacelike case. This definition is particularly effective in $(1+1)$-dimensional QFTs. However, in higher dimensions, where the number of spatial coordinates exceeds that of the temporal ones, extending this definition becomes subtle and problematic.

From the perspective of correlation functions, timelike EE can also be defined and computed via a Wick rotation of correlators involving twist operators. In this formulation, one analytically continues Euclidean correlators to the Lorentzian signature. Several examples of such calculations are presented in \cite{Doi:2023zaf,Guo:2024lrr}. The reconstruction theorem in QFT ensures that correlators in Lorentzian spacetime can be obtained through appropriate analytic continuation procedures; see \cite{Hartman:2015lfa} for more details. Nevertheless, there are some unsatisfactory aspects of this approach. First, it appears difficult to generalize to quantum systems with finite-dimensional Hilbert spaces. Second, the physical interpretation of the Wick rotation remains obscure. In particular, it is still unclear which correlations within the system are captured by the timelike EE.


The recent work \cite{Milekhin:2025ycm} provides a new perspective on entanglement in time, which is applicable to both finite-dimensional quantum systems and QFTs. In certain specific cases, the results for QFTs reduce to those obtained via analytic continuation of the correlators used to define timelike EE.

This idea can be viewed as a natural generalization of the concept of entanglement between two spacelike-separated subsystems. In QFTs, one can choose a Cauchy surface and define the Hilbert space of the theory accordingly. On this surface, two subregions $A$ and $B$ are selected. For a given state  $|\psi\rangle$, the correlation functions between operators $\mathcal{O}_A$ and $\mathcal{O}_B$ localized in $A$ and $B$, respectively, are defined as
\bea\label{correlation}
\langle \mO_A \mO_B \rangle_\psi := \langle \psi| \mO_A \mO_B |\psi\rangle.
\eea
The reduced density matrix $\rho_{AB} := \mathrm{Tr}_{\bar{AB}} \rho$, where $\bar{AB}$ denotes the complement of $AB$, satisfies
\bea
\Tr_{AB}(\rho_{AB} \mO_A \mO_B)= \langle \mO_A \mO_B\rangle_\psi.
\eea
According to the Wightman reconstruction theorem in QFTs $\rho_{AB}$ can be fully determined by the set of correlation functions of all local operators $\mO_A$ and $\mO_B$. In other words, the reduced density matrix $\rho_{AB}$ is completely specified by the correlation data within the subregions $\mO_A$ and $\mO_B$.
According to microcausality $[\mO_A,\mO_B]=0$. However, in general, the correlation function $\langle \mO_A \mO_B\rangle_\psi \ne 0$ does not vanish, reflecting the presence of quantum entanglement between two causally disconnected subsystems. 

\begin{figure}[htbp]
\centering
\begin{minipage}[t]{0.48\textwidth}
\centering
\includegraphics[width=6cm]{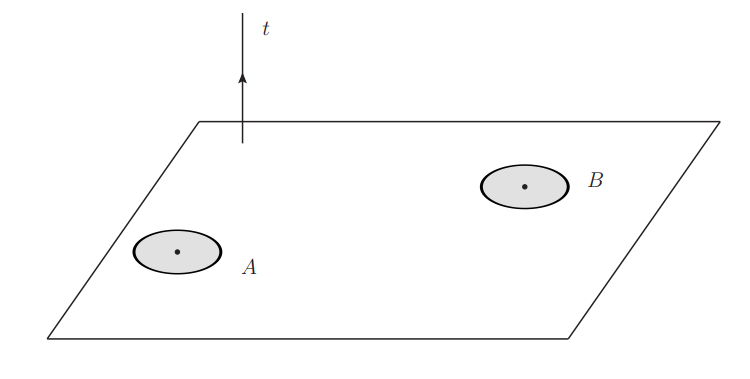}
\end{minipage}
\begin{minipage}[t]{0.48\textwidth}
\centering
\includegraphics[width=6cm]{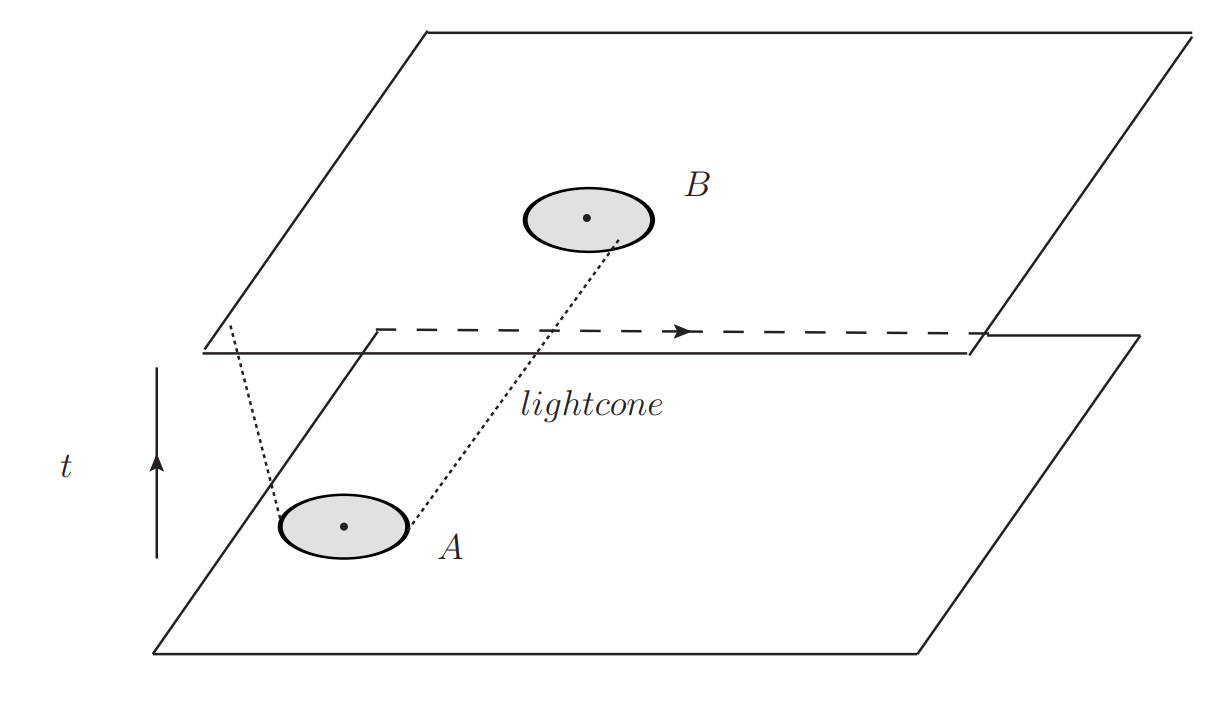}
\end{minipage}
\caption{An illustration of subsystems $A$ and $B$: (left) $A$ and $B$ are causally disconnected; (right) $A$ and $B$ are causally connected.}
\label{subsystemsAandB}
\end{figure}

Entanglement is closely related to correlation functions between subsystems. In QFTs, correlation functions for causally connected subregions are also of fundamental importance. For instance, scattering processes involve correlation functions between timelike-separated operators. The definition of entanglement in time in \cite{Milekhin:2025ycm} is a generalization of the above definition of $\rho_{AB}$ to causally connected subregions $A$ and $B$, see Fig.~\ref{subsystemsAandB} for an illustration.

For causally connected subsystems $A$ and $B$ (Figure on the right of Fig.~\ref{subsystemsAandB} one could also evaluate the correlation functions in the state $|\psi\rangle$~(\ref{correlation}).
 Similar to the spacelike case, we expect the existence of a transition operator $T_{AB}:\mathcal{H}_A\otimes \mathcal{H}_B \to \mathcal{H}_A\otimes \mathcal{H}_B$ such that
\bea\label{transition_definition}
\langle \mO_A\mO_B(t)\rangle_\psi =\Tr(T_{AB}\mO_A\mO_B),
\eea
where $\mO_B(t)=U^\dagger \mO_B U$.
However, since the operators do not commute in general, i.e., $[\mO_A,\mO_B(t)]\ne 0$, the correlator $\langle \mO_A \mO_B(t)\rangle_\psi$ is generally complex-valued. This implies that the transition operator $T_{AB}$ is non-Hermitian.  

\begin{figure}[htbp]
  \centering
  \includegraphics[width=0.8\textwidth]{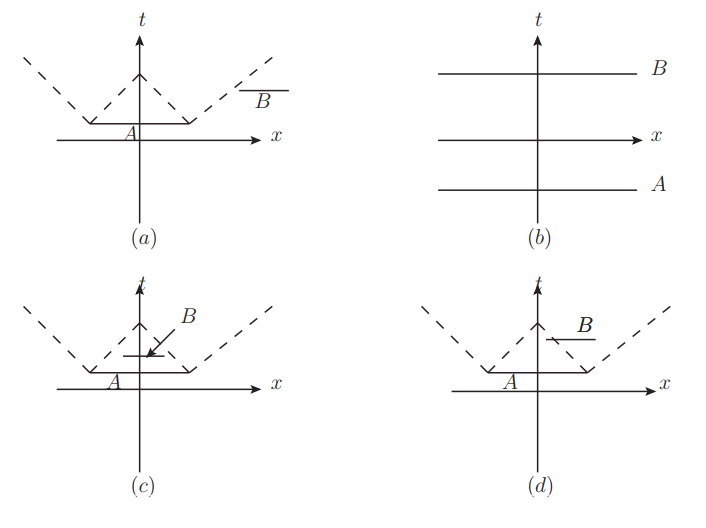}
  \caption{An illustration of the four possible configurations of regions $A$ and $B$: (a) $A$ and $B$ are causally disconnected; (b) $A$ and $B$ are Cauchy surfaces at different times; (c) $A$ lies entirely within the causal domain of $B$; (d) a portion of $B$ lies within the causal domain of $A$.}
  \label{Fourcases_fig}
\end{figure}

The transition operator $T_{AB}$  can be viewed as a generalization of the reduced density matrix to more general configurations of subsystems. The usual reduced density matrix $\rho_{AB}$ emerges as a special case when $A$ and $B$ are causally disconnected.  Depending on the causal relations between subsystems $A$ and $B$ , different scenarios of interest arise:
\begin{enumerate}[label=(\alph*)]
    \item $A$ and $B$ are causally disconnected. In this case, the transition operator reduces to the standard reduced density matrix $\rho_{AB}$.
    \item  $A$ and $B$ represent the entire system on different Cauchy surfaces. In this case, the state at $B$ corresponds to the final state resulting from the time evolution of the initial state at $A$, and thus $T_{AB}$ encodes the dynamical information of the full system. If $A$ and $B$ correspond to the in-state ($t \to -\infty$) and out-state ($t \to +\infty$) Cauchy surfaces in a scattering process, then $T_{AB}$ is expected to be related to the scattering $S$-matrix.
    \item  $A$ and $B$ are causally connected subsystems, with $B$ lying within the causal domain of $A$. This means that the information in $B$ is determined by that in $A$. While this case resembles case 2, the transition operator $T_{AB}$ is expected to encode not only the system’s dynamics but also its non-local properties.
    \item $A$ and $B$ are causally connected subsystems, but $B$ does not lie within the causal domain of $A$. In this case, the information in $B$ cannot be reconstructed solely from $A$. This is the scenario we will primarily focus on in this paper.
\end{enumerate}
One may refer to Fig.~\ref{Fourcases_fig} for an illustration of the four cases.

The above discussion has primarily focused on QFTs. However, the definition of $T_{AB}$ is applicable to systems with finite-dimensional Hilbert spaces. The concept of entanglement in time is both fundamental and meaningful in any quantum mechanical system. Moreover, it may be possible to design experiments to measure entropy-related quantities associated with $T_{AB}$.

\subsection{State preparation via the Schwinger–Keldysh formalism}
We are particularly interested in how to understand and express the transition operator $T_{AB}$ in QFTs.
For $\rho_{AB}$ the standard approach begins with preparing the density matrix $\rho=|\psi\rangle \langle \psi|$ , followed by tracing out the complementary degrees of freedom to obtain $\rho_{AB}$ via Euclidean path integral. However, in our case, the two subsystems $A$ and $B$  are timelike separated, and the Euclidean approach must be extended to incorporate real-time evolution \cite{Milekhin:2025ycm}. In \cite{Dong:2016hjy}, the authors propose using the Schwinger–Keldysh formalism to represent a general time-dependent density matrix $\rho(t)$. Within the Schwinger–Keldysh framework, both the reduced density matrix and the replica method can be generalized, leading to what is known as the real-time replica method. For spacelike-separated subsystems, this method reproduces the same results as the Euclidean replica approach \cite{Colin-Ellerin:2020mva,Colin-Ellerin:2021jev}.

In the following, we would like to show that the transition operator $T_{AB}$ can also be constructed within the Schwinger–Keldysh formalism. Furthermore, we demonstrate how to compute the Rényi entropy associated with $T_{AB}$ using the real-time replica method.

We begin by introducing the path integral representation of the time-evolved density matrix $\rho(t)$. Starting from an initial state $\rho_0$ at time $t_0$, which can be prepared using a Euclidean path integral. The time-evolved density matrix in time $\rho(t)$ is given by 
\bea\label{rho-time-depend}
\rho(t)=U \rho_0 U^\dagger,
\eea
where $U:=e^{iH(t-t_0)}$ is the time evolution operator from $t_0$ to $t$. For simplicity, we focus on the case where the initial state is pure, i.e.
\bea
 \rho_0 = \ket{\Psi_0} \bra{\Psi_0}, \rho(t) = U \ket{\Psi_0} \bra{\Psi_0} U^\dagger.
\eea
For mixed states, one can first purify the state and then proceed analogously.
 
\begin{figure}[htbp]
  \centering
  \includegraphics[width=0.6\textwidth]{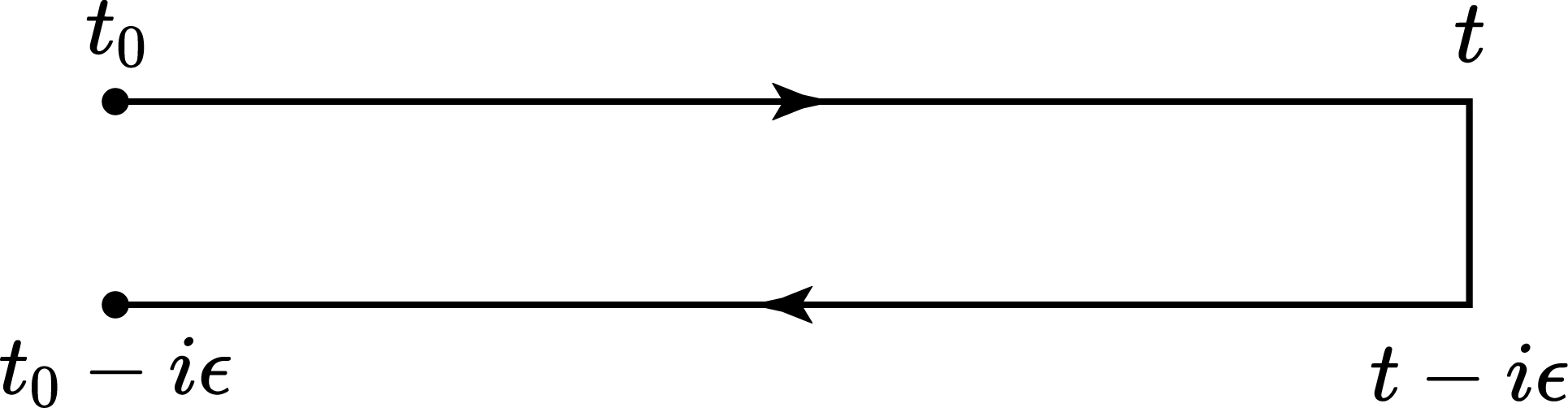}
  \caption{A typical Keldysh contour.}
  \label{Keldysh_contour}
\end{figure}

To represent ket $\ket{\Psi} = U \ket{\Psi_0}$ and its conjugate $\bra{\Psi}=\bra{\Psi_0}U^\dagger$ in the path integral formalism, it is natural to employ a Schwinger–Keldysh contour, as pointed out in \cite{Dong:2016hjy}. This contour evolves the ket state forward in time and the bra state backward, thus fully encoding the real-time dynamics. A typical Keldysh contour is illustrated in Fig.~\ref{Keldysh_contour}.

\begin{figure}[htbp]
  \centering
  \subfigure[]{\includegraphics[scale=0.4]{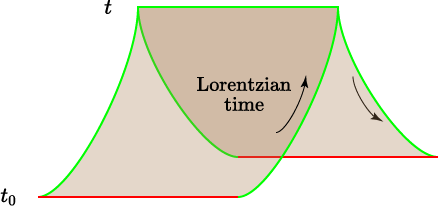}}
  \subfigure[]{\includegraphics[scale=0.4]{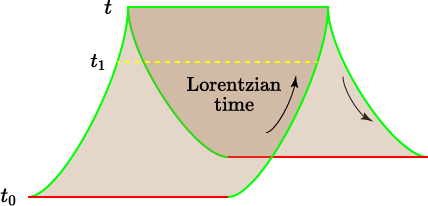}}
  \caption{The Schwinger-Keldysh formalism illustrating the path-integral representation of $\Tr \rho$. (a) Calculate $\Tr \rho(t)$ by~(\ref{Tr-rho-singleU}). (b) Calculate $\Tr \rho(t_1)$ by~(\ref{Tr-rho-doubleU}).}
  \label{fig:both}
\end{figure}

As a result, $\Tr\rho(t)$ can be expressed as
\bea\label{Tr-rho-singleU}
\Tr\rho(t) = \int [D \phi] e^{i S_{t;t_0}(\phi) - i S_{t_0;t}(\phi)},
\eea
where $S_{t;t_0}$ is the action defined as $\int_{t_0}^t dt L[\phi]$.  If we have $t_0<t_1<t$, the $\Tr\rho(t_1)$ could be written as
\bea\label{Tr-rho-doubleU}
\Tr \rho(t_1) = \int [D \phi] e^{i S_{t_1;t_0}(\phi) + i S_{t;t_1}(\phi) - i S_{t_0;t}(\phi)}.
\eea
We could derive it from~(\ref{rho-time-depend}), the time evolution operator $U(t,t_1)$ and its Hermitian conjugate $U^{\dagger}(t,t_1)$ mutually cancel under the trace. It is more convenient to illustrate the path integral representation using diagrams, as shown in Fig.~\ref{fig:both}. In these diagrams, we include the spatial direction of the Keldysh contour to facilitate the definition of the reduced density matrix in the following discussion.

\begin{figure}[htbp]
  \centering
  \includegraphics[width=0.5\textwidth]{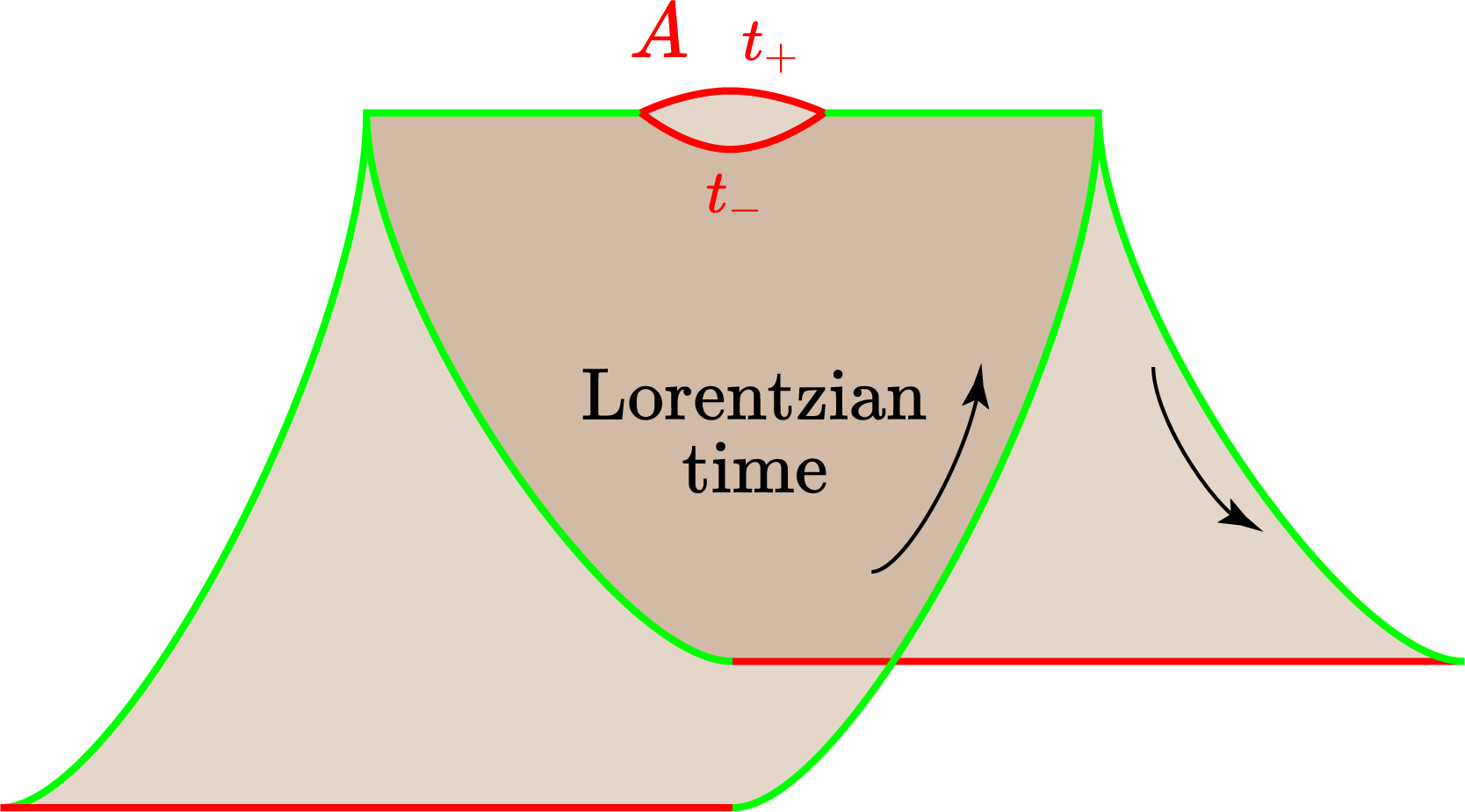}
  \caption{The Schwinger-Keldysh formalism illustrating the path-integral representation of the reduced density matrix $\rho_A(t)$.}
  \label{RhoA-Ltime-path}
\end{figure}

To evaluate the EE for a subsystem $A$ on the Cauchy surface $\Sigma_t$, one needs to construct the reduced density matrix  $\rho_A(t) = \Tr_{\bar{A}} \rho(t)$. The full trace $\Tr \rho(t)$ consists of a forward and a backward evolution, which are glued together along the Cauchy surface $\Sigma_t$. To obtain $\rho_A(t)$, the path integral is cut along the region $A$, and appropriate boundary conditions are imposed at the cut. The reduced density matrix can then be expressed as a path integral:
\begin{align}
\rho_A =& \int [D \Phi_R] [D \Phi_L] e^{i S[\Phi_R] - i S[\Phi_L]} \delta(\Phi_{R,A}(t_-)-\Phi_A(t_-)) \nn \\
&\times \delta(\Phi_{L,A}(t_+)-\Phi_A(t_+)),
\end{align}
where $\Phi_R$ and $\Phi_L$ denote the field configurations on the forward and backward segments of the contour in Fig.~\ref{RhoA-Ltime-path}, respectively. The boundary conditions at subregion $A$ are specified by $\Phi_A(t_-)$ and $\Phi_A(t_+)$, corresponding to the forward and backward branches. An illustration of the reduced density matrix $\rho_A(t)$ in Fig.~\ref{RhoA-Ltime-path}.

\subsection{Representation of \texorpdfstring{$T_{AB}$}{TAB} via Schwinger-Keldysh formalism}

\begin{figure}[htbp]
  \centering
  \includegraphics[width=0.5\textwidth]{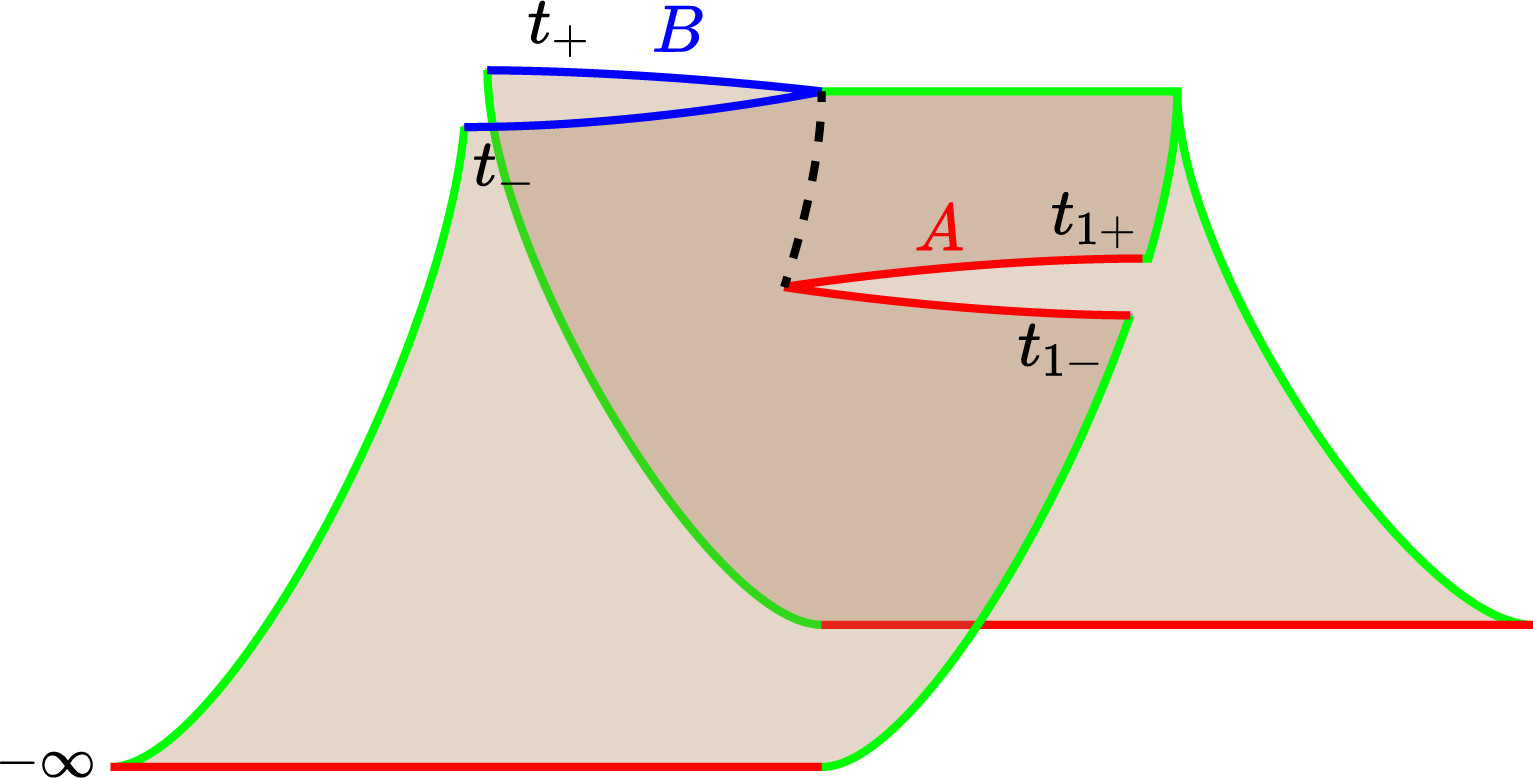}
  \caption{The Schwinger-Keldysh formalism illustrating the path-integral representation of $T_{AB}$. This diagram shows the specific case for a two-point function, where the cuts for subsystems $A$ and $B$ are localized at times $t_1$ and $t$.}
  \label{T_AB-real-time}
\end{figure}

In this section, we demonstrate that the transition operator $T_{AB}$ can also be represented using the Schwinger-Keldysh formalism. We consider subsystems $A$ and $B$ located on Cauchy surfaces at times $t_1$ and $t$, respectively. The initial state $|\psi\rangle$ is prepared at $t\to -\infty$. In the context of this paper, $A$ and $B$ are taken to be semi-infinite intervals, which are appropriate for the computation of two-point correlation functions discussed later. Moreover, such semi-infinite intervals can be obtained as the limit of finite intervals when their lengths are extended to infinity, ensuring that our formalism naturally generalizes from the finite-interval case. In the previous section, we introduced the path integral representation of the reduced density matrix $\rho_A(t)$ as illustrated in Fig.~\ref{RhoA-Ltime-path}. For the construction of $T_{AB}$, the path integral involves two cuts: one at time $t$ on the forward segment, with appropriate boundary conditions imposed on subregion $B$; and another at time $t_1$, also on the forward segment, with boundary conditions specified on subregion $A$. The transition operator $T_{AB}$ is then expressed in terms of the path integral as follows:
\begin{align}
T_{AB} = &\int [D \Phi_R] [D \Phi_L] 
    e^{i S_{t_1;-\infty}[\Phi_R]}
    \delta(\Phi_{R,A}(t_{1-})-\Phi_A(t_{1-})) \nn \\
  &\times \delta(\Phi_{R,A}(t_{1+})-\Phi_A(t_{1+}))
    e^{i S_{t;t_1}[\Phi_R]} 
    \delta(\Phi_{R,B}(t_{-})-\Phi_B(t_{-})) \nn \\
  &\times \delta(\Phi_{L,B}(t_{+})-\Phi_B(t_{+})) 
    e^{-i S_{-\infty;t}[\Phi_L]},
\end{align}
where $\Phi_{A}(t_{1\pm})$ and $\Phi_B(t_{\pm})$ denote the boundary conditions in the cut. We show the presentation of $T_{AB}$ in Fig.~\ref{T_AB-real-time}.

\begin{figure}[htbp]
  \centering
  \subfigure[]{\includegraphics[scale=0.4]{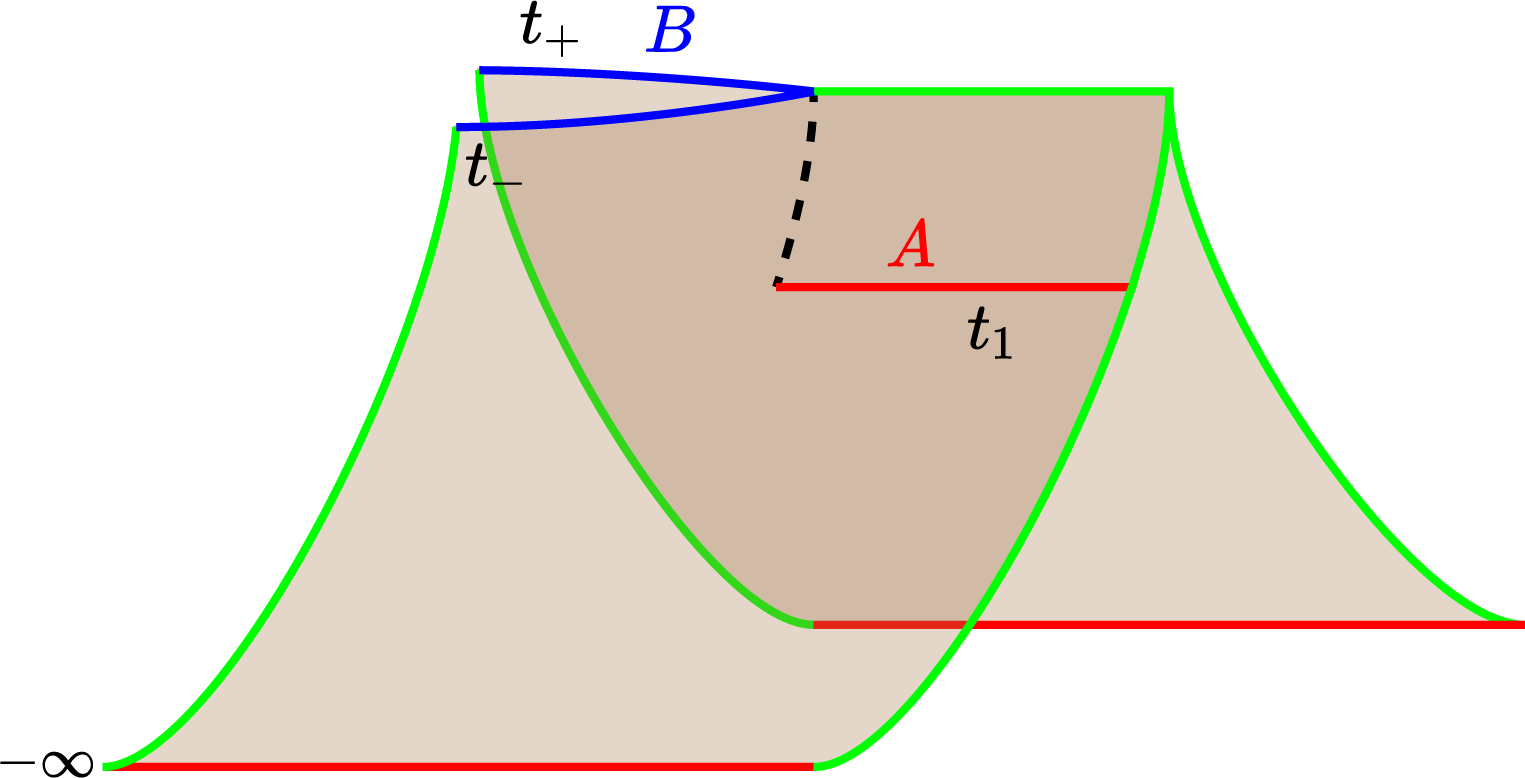}}
  \subfigure[]{\includegraphics[scale=0.4]{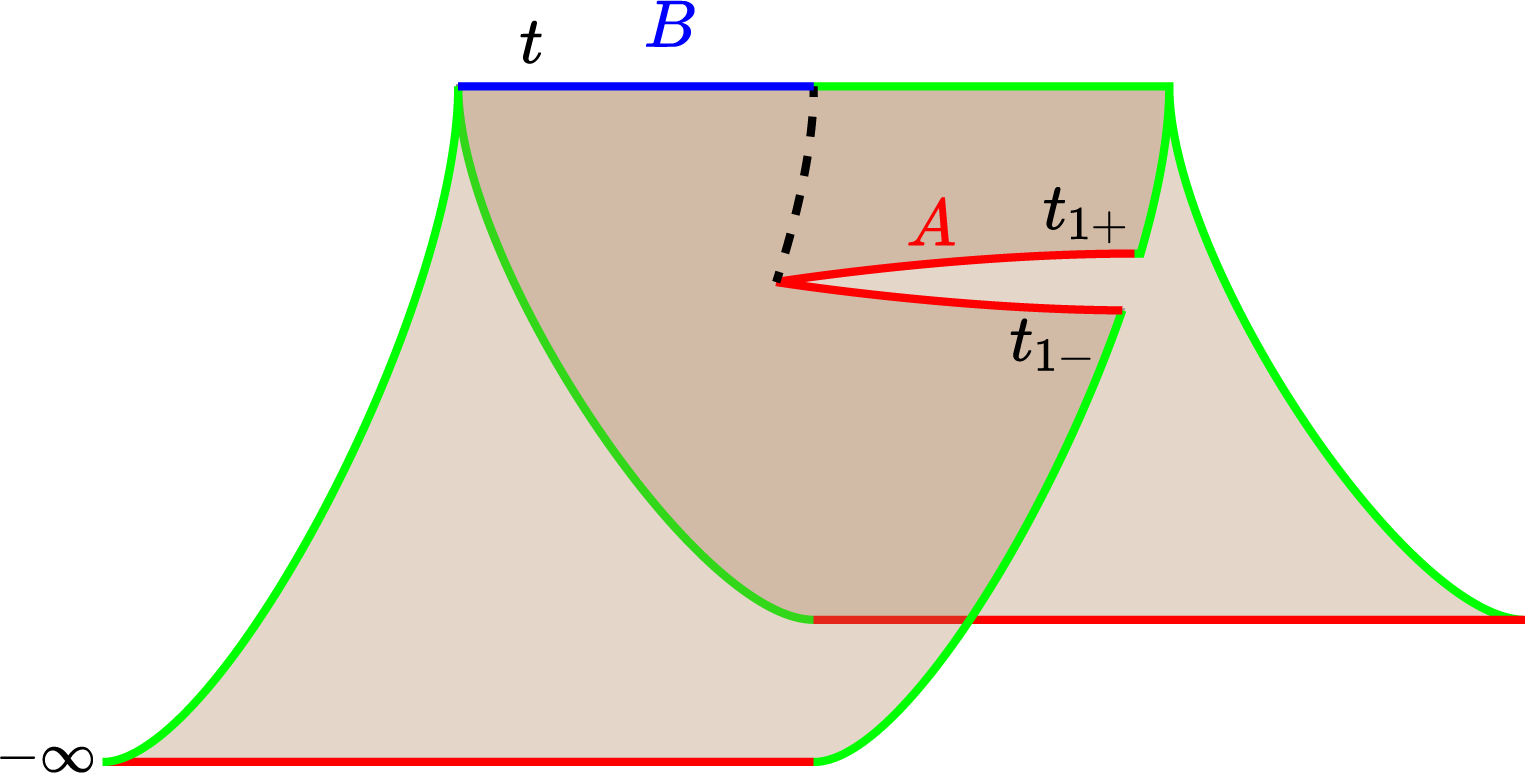}}
  \caption{The partial trace of $T_{AB}$. (a) $\Tr_A T_{AB}$: obtained by gluing the cut along region $A$. (b) $\Tr_B T_{AB}$: obtained by gluing the cut along region $B$. In this case, the resulting Schwinger–Keldysh formalism is equivalent to the geometry without the contribution over region $A$, as can be seen from~(\ref{Tr-rho-doubleU}).}
  \label{partial trace Tab}
\end{figure}

We can verify that the above path integral representation satisfies the basic properties of  $T_{AB}$. According to the definition of the transition operator, inserting local operators $\mO_A$ and $\mO_B$ in subregions $A$ and $B$, respectively, allows us to compute $\Tr(T_{AB} \mO_A\mO_B)$, which reproduces the correlator  $\langle \mO_A \mO_B(t)\rangle_\psi$. This serves as a consistency check for the definition of $T_{AB}$ via~(\ref{transition_definition}). Furthermore,  $T_{AB}$ also satisfies
\bea
&\Tr_AT_{AB}=\rho_B,\\
&\Tr_BT_{AB}=\rho_A.
\eea
The first relation involves taking the partial trace over region $A$. In the path integral representation, this corresponds to gluing along the cut at 
$A$ in Fig.~\ref{partial trace Tab} (a). As a result, the remaining path integral yields the reduced density matrix $\rho_B(t)$. The second relation is analogous: after gluing the cut at $B$, a cut remains at $A$ on the Cauchy surface at time $t_1$.  The contributions from the path integral between $t_1\to t$ on the forward segment and $t\to t_1$ on the backward segment cancel each other. Therefore, the final result $\Tr_B T_{AB}$ is given by the reduced density matrix $\rho_A(t_1)$. We illustrate the partial trace operation diagrammatically in Fig.~\ref{partial trace Tab} (b).

\subsection{Real-time replica and timelike EE}
Recall the definition for pseudo Rényi entropy and EE for $T_{AB}$:
\bea
&&S_n(T_{AB})=\frac{\Tr T_{AB}^n}{1-n},\\
&&S(T_{AB})=\lim_{n\to 1} S_n (T_{AB}).
\eea
Given the transition operator $T_{AB}$ we are now ready to compute the EE for $T_{AB}$, which requires understanding the product of $T_{AB}^n$ via the replica method. 

For our motivation in this paper, we would like to study the cases that would be related to the timelike EE evaluated by Wick rotation.  We will mainly focus on some specific examples in $(1+1)$D QFTs. The generalization to higher dimensions is straightforward. 

\subsubsection{Example I}\label{section_example_I}
\begin{figure}[htbp]
  \centering
  \includegraphics[width=1.0\textwidth]{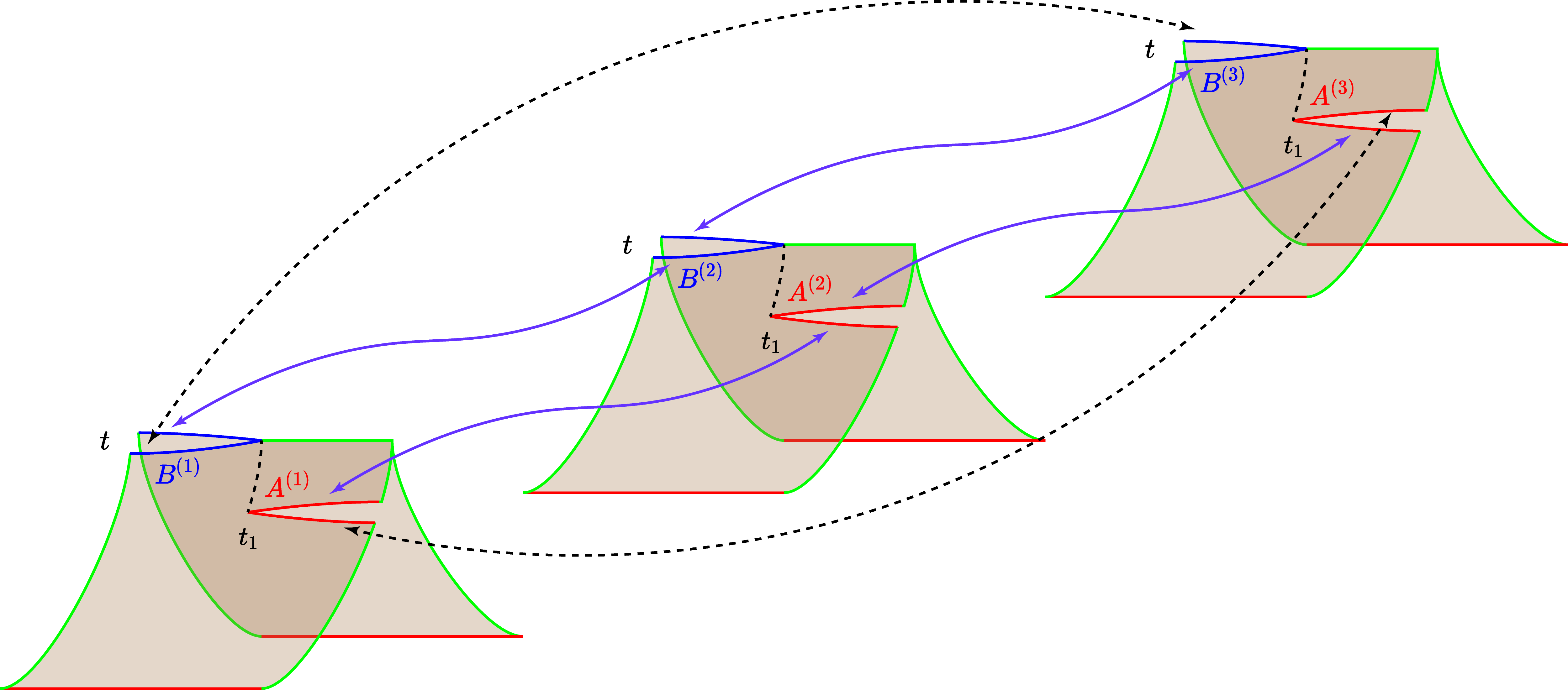}
  \caption{Replica trick with a timelike interval for computing $\Tr (T^3_{AB})$.}
  \label{two-twist-timelike-real-t}
\end{figure}

Let us consider a simple example: $A\in [0,+\infty]$ at time $t_1$ and $B \in [-\infty,0]$ at time $t$. The EE for $T_{AB}$ is related to timelike EE for an interval $[0,t]$ on the time coordinate. To evaluate $\Tr(T_{AB}^n)$ we should make an n-copy of the geometry and fields for $T_{AB}$ with replica index $I=1,2,...,n$. $\Tr(T_{AB}^n)$ involves identification along the cut with $\Phi_{I-}=\Phi_{(I+1)+}$ with $n+1\to 1$. We draw a diagram in Fig.~\ref{two-twist-timelike-real-t} to show the replica process. As a result, the final path integral can be considered as the insertion of local twist operator $\sigma_n(t,0)$ and anti-twist operator $\tilde{\sigma}_n(0,0)$ in the n-copied theory. According to the SK formalism, this can be translated to the two-point correlation function for the twist operator.
Thus, we have
\bea
\Tr(T_{AB}^n)= \langle \sigma_n(0,0)\tilde{\sigma}_n(t,0)\rangle,
\eea
where the right-hand side can be computed via the analytic continuation of the two-point correlator in Euclidean theory, evaluated as follows:
\bea
\langle \sigma_n(0,0)\tilde{\sigma}_n(0,t)\rangle=\langle \sigma_n(\tau_1,0)\tilde{\sigma}_n(\tau_2,0)\rangle|_{\tau_1\to 0 i+ \epsilon_1,\tau_2\to i t +\epsilon_2},
\eea
with $\epsilon_1>\epsilon_2>0$\footnote{The ordering of $\epsilon_i$ determines the correct operator ordering in the correlator.}
In $(1+1)$-dimensional CFTs, the two-point correlation function is universal. Consequently, we obtain the Rényi entropy for $T_{AB}$ as follows:
\bea
S_n(T_{AB})=\frac{c}{6}(1+\frac{1}{n})\log \frac{t}{\delta}+\frac{c}{12}(1+\frac{1}{n})i\pi.
\eea
The EE is given by
\bea\label{EE_oneinterval}
S(T_{AB})=\frac{c}{3}\log\frac{t}{\delta}+\frac{ic\pi}{6}.
\eea







\subsubsection{Example II}\label{section_example_II}

\begin{figure}[htbp]
  \centering
  \includegraphics[width=0.8\textwidth]{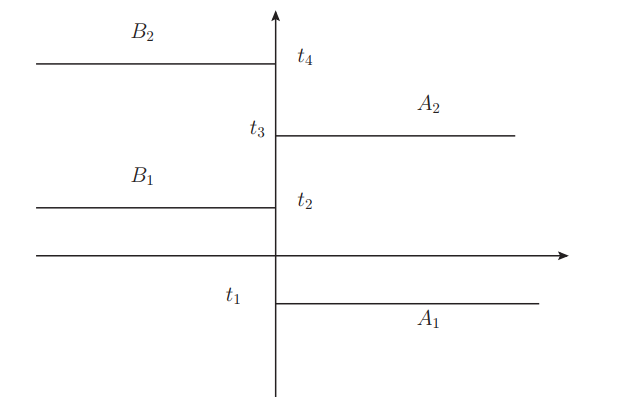}
  \caption{An illustration of the four subregions considered in Section~\ref{section_example_II}.}
  \label{two_interval_write}
\end{figure}

Let us now consider a more involved example. We take four subregions: $A_1\in(0,+\infty)$ at time $t_1$, $A_2\in (0,+\infty)$ at time $t_3$, $B_1:\in (-\infty,0)$ at time $t_2$, and $B_2\in (-\infty,0)$ at time $t_1$. We assume the order of time $t>t_4>t_3>t_2>t_1>-\infty$. For clarity, the configuration of the subregions is illustrated in Fig.~\ref{two_interval_write}.

\begin{figure}[htbp]
  \centering
  \includegraphics[width=0.8\textwidth]{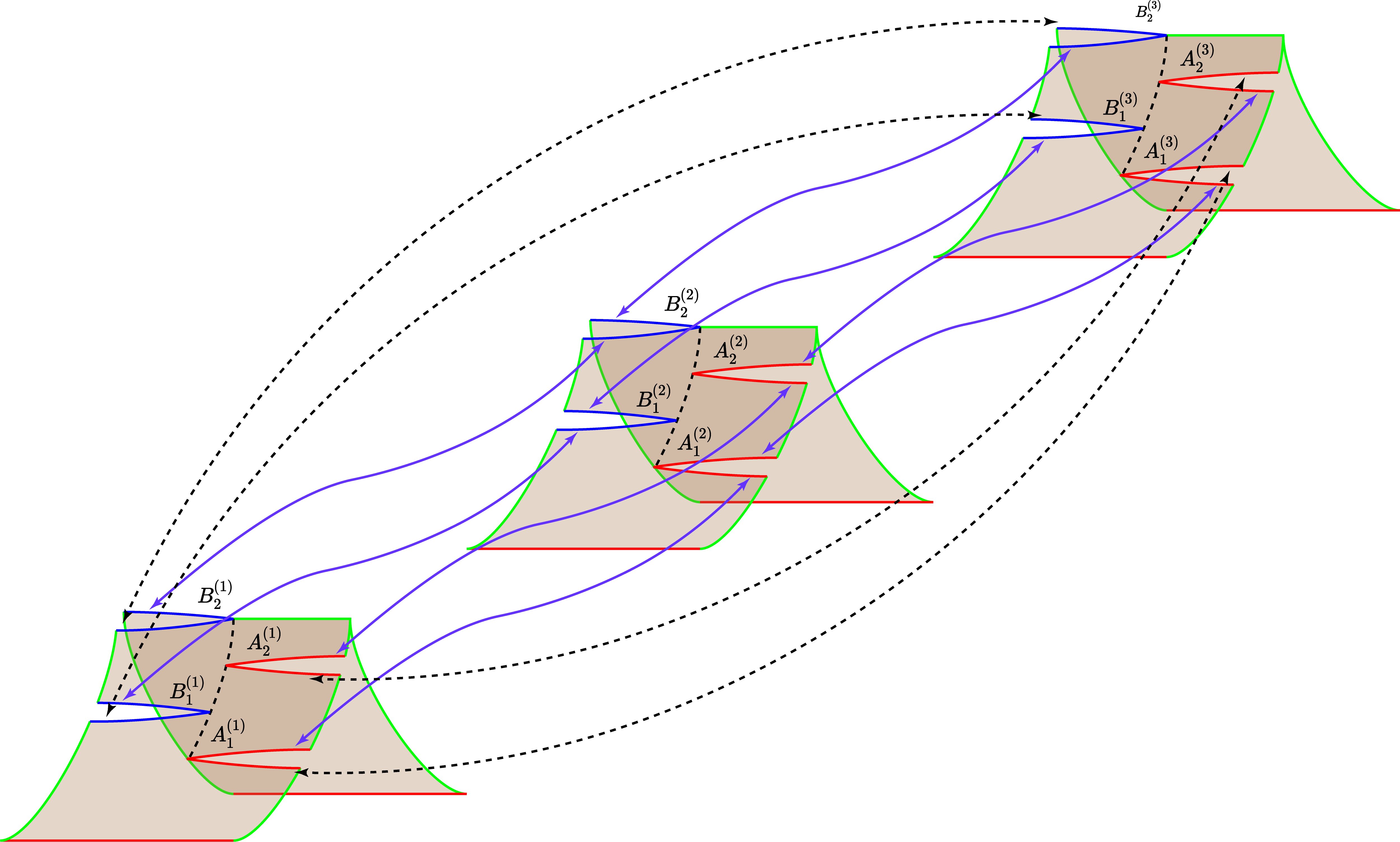}
  \caption{Replica trick with two timelike intervals for computing $\Tr (T^3_{A_1A_2B_1B_2})$.}
  \label{four-twist-real-time}
\end{figure}

We can now define and prepare the transition operator $T_{A_1A_2B_1B_2}$ in the same way as in Example I. A diagram illustrating the path integral representation is shown in Fig.~\ref{four-twist-real-time}. The procedure for evaluating $\Tr(T_{A_1A_2B_1B_2})^n$ is nearly identical to that in Example I, so we will omit the details of the replica method here. After applying the replica trick, the problem reduces to evaluating the four-point, time-ordered correlation function of twist operators:
\bea\label{ExampleII_trTn}
\Tr(T_{A_1A_2B_1B_2})^n=\langle \sigma_n(t_1,0)
\tilde{\sigma}_n(t_2,0)\sigma_n(t_3,0)\tilde{\sigma}_n(t_4,0)\rangle,
\eea
which can be computed via analytical continuation of the four-point function in the Euclidean theory by taking $\tau=it$. We have
\bea\label{ExampleII_fourpoint}
&&\langle \sigma_n(t_1,0)
\tilde{\sigma}_n(t_2,0)\sigma_n(t_3,0)\tilde{\sigma}_n(t_4,0)\rangle\nn \\
&&=\langle \sigma_n(\tau_1,0)
\tilde{\sigma}_n(\tau_2,0)\sigma_n(\tau_3,0)\tilde{\sigma}_n(\tau_4,0)\rangle|_{\tau_i\to i t_i +\epsilon_i},
\eea
with the order $\epsilon_1>\epsilon_2>\epsilon_3>\epsilon_4>0$.
Recall the definition of EE, we have
\bea
S(T_{A_1A_2B_1B_2})=-\partial_n \langle \sigma_n(t_1,0)
\tilde{\sigma}_n(t_2,0)\sigma_n(t_3,0)\tilde{\sigma}_n(t_4,0)\rangle|_{n=1}.
\eea
It is particularly interesting to define mutual information as
\bea\label{mutual_information_definition}
I(A_1B_1;A_2B_2):=S(T_{A_1B_1})+S(T_{A_2B_2})-S(T_{A_1A_2B_1B_2}),
\eea
which captures the correlation between the composite subsystems $A_1B_1$ and $A_2B_2$. 

In (1+1)D CFTs, there are scenarios in which we can explicitly compute $S(T_{A_1A_2B_1B_2})$ and therefore $I(A_1B_1;A_2B_2)$. In the vacuum state, mutual information between two spacelike subregions is known to be conformally invariant. This suggests that $I(A_1B_1;A_2B_2)$ should depend only on the conformal cross-ratio. We provide further details on the evaluation of these quantities in Appendix~\ref{Mutualinformation_section}. 

In this paper, we focus on holographic CFTs. The four-point function in~(\ref{ExampleII_fourpoint}) can be computed using the semi-classical approximation for conformal blocks. Detailed calculations and analytic continuation are presented in Appendix~\ref{Mutualinformation_section}. Our result is consistent with the definition of holographic timelike EE for two time intervals, as evaluated by holography in \cite{Doi:2023zaf}.
There exist two distinct phases of the EE, depending on the relative positions of the two time intervals $[t_1,t_2]$ and $[t_3,t_4]$. For case $\frac{(t_1-t_2)(t_3-t_4)}{(t_1-t_3)(t_2-t_4)}\sim 0$ we have
\bea\label{timelikeEE_two_interval_phaseI}
S(T_{A_1A_2B_1B_2})=\frac{c}{3}\log\frac{t_2-t_1}{\delta}+\frac{c}{3}\log \frac{t_4-t_3}{\delta}+\frac{i\pi c}{3},
\eea
while for $\frac{(t_1-t_2)(t_3-t_4)}{(t_1-t_3)(t_2-t_4)}\sim 1$ we have
\bea\label{timelikeEE_two_interval_phaseII}
S(T_{A_1A_2B_1B_2})=\frac{c}{3}\log\frac{t_3-t_2}{\delta}+\frac{c}{3}\log \frac{t_4-t_1}{\delta}+\frac{i\pi c}{3}.
\eea
Note that the imaginary part remains constant and is twice that of the single-interval case~(\ref{EE_oneinterval}). As a result, the mutual information is real in both phases. This result is consistent with previous studies on timelike EE for two intervals along the time direction.  
In the following sections, we will explore the holographic aspects of the EE for $T_{A_1A_2B_1B_2}$. The phase transition of the EE corresponds to different bulk Ryu–Takayanagi surfaces. 
\section{Ryu-Takayanagi surface for timelike EE}\label{section_RT_surface}

In~\cite{Doi:2022iyj}, the holographic EE for a time interval in AdS$_3$ was computed as
\bea
S_A^{(T)} = \frac{c}{3} \log \frac{t_0}{\delta} + \frac{c}{6}i\pi,
\eea
where $t_0$ denotes the proper length of the timelike interval $A$, and $\delta$ is the UV cutoff.

\begin{figure}[htbp]
  \centering
  \includegraphics[width=0.8\textwidth]{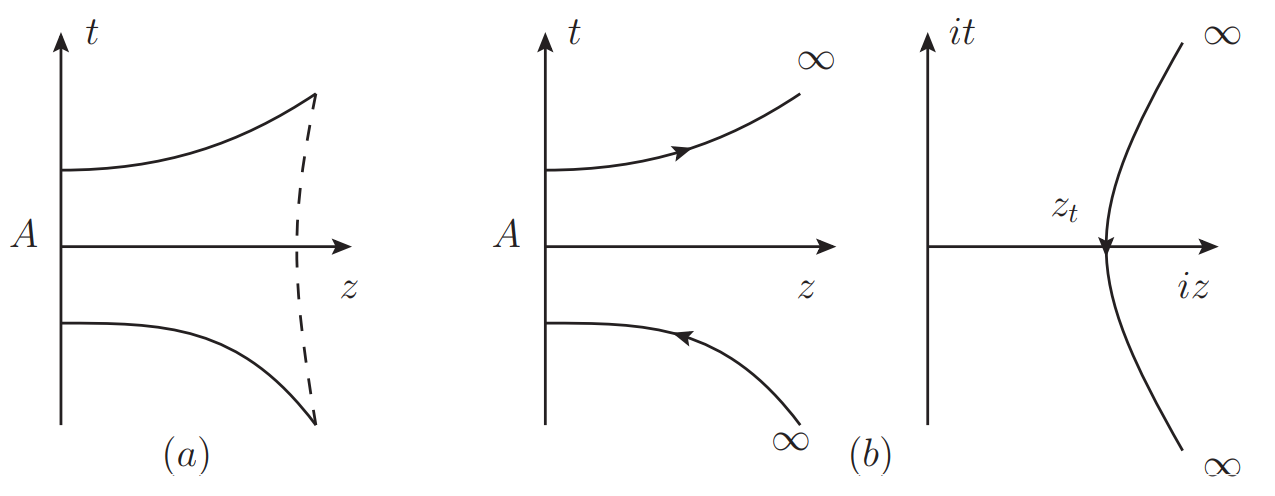}
  \caption{(a) RT surface for a timelike interval $A$ in AdS$_3$ spacetime consists of both spacelike and timelike geodesic segments. (b) Complexified RT surface passing through real and imaginary domains, connected via points at infinity.}
  \label{twoproposals}
\end{figure}

This result contains both a real and an imaginary part. Although the logarithmic real part has a standard geometric interpretation in terms of the RT prescription, the imaginary contribution is more subtle. The imaginary term can be interpreted as arising from the timelike geodesic connecting the two infinities, as illustrated in Fig.~\ref{twoproposals}(a) \cite{Doi:2022iyj}. Another proposal is presented in~\cite{Heller:2024whi}, where the RT surface is extended to a complexified bulk geometry.

In this construction, the embedding functions of the surface are analytically continued in the complex domain. The resulting surfaces can be classified according to whether the continuation modifies their real or imaginary components. Although the RT surface becomes complex-valued, it is argued to acquire physical meaning only when restricted to specific real slices of the complexified geometry, namely the totally real or totally imaginary submanifolds. These slices are connected via analytic continuation, similar to a Wick rotation, and allow the surface to traverse different geometric domains through asymptotic infinity.

This leads to a complexified RT surface that transitions between real and imaginary domains through points at complex infinity. As shown in Fig.~\ref{twoproposals}(b), the surface initially follows the upper branch of a hyperbola in the real Lorentzian geometry. Then it passes through a point at infinity, analytically continues in the imaginary time direction, and reaches a turning point $z_t$ on the imaginary branch. After crossing the complex domain, it re-emerges through another point at infinity into the lower branch of the real geometry. Throughout this path, the surface remains smooth and continuous within the complexified spacetime.



Because the RT surface is symmetric, it has a single turning point $z_t$ on its imaginary branch that splits the surface into two identical halves. Consequently, the total area $\mathcal{A}$ of the RT surface anchored to region $A$ can be written as
\begin{align}\label{A1}
\mathcal{A}=2\int_{\delta}^{z_t}dz\mathcal{L},
\end{align}
where $\mathcal{L}$ is the Lagrangian density along the surface. By the Ryu–Takayanagi prescription, the corresponding timelike EE is  $S_A^{(T)}=\frac{\mathcal{A}}{4G}$, with $G$ the cosmological constant. To proceed, we now derive the explicit form of $\mathcal{L}(z)$ and determine the location of the turning point $z_t$.

\subsection{Procedure to construct the bulk RT surface for timelike subregions}

There remains some ambiguity regarding how to determine the complex RT surface in the complexified geometry. In this section, we present a procedure for constructing the RT surface via analytical continuation from its Euclidean counterpart \cite{Guo:2025pru}.

We consider the case of an AdS$_{d+1}$ black hole, where the bulk metric takes the form of
\begin{align}\label{ads_d+1_m}
ds^2 = \frac{1}{z^2} \left( -f(z)dt^2 + \frac{dz^2}{g(z)} + dx^2 + d\vec{y}^2 \right),
\end{align}
where $d\vec{y}^2 = \sum_{i=1}^{d-2} y_i^2$, and the event horizon is located at $z = z_H$ such that $f(z_H) = 0$. For a neutral (Schwarzschild) black hole, we take $f(z) = g(z) = 1 - \frac{z^d}{z_H^d}$. In the limit $z_H \to \infty$, the metric reduces to the Poincaré AdS case.

We define the boundary subregion $A$ as a timelike strip with $x = 0$ and $t \in [0, t_0]$, spanning all directions in $\vec{y}$. The endpoints of this interval are located at $(t=0, x=0, \vec{y})$ and $(t = t_0, x = 0, \vec{y})$. For simplicity, we will label the two endpoints as $(0, 0)$ and $(t_0, 0)$, suppressing the transverse coordinates.

We analytically continue the spacetime~(\ref{ads_d+1_m}) to Euclidean spacetime via $t \to -i\tau$, which yields the following
\begin{align}\label{ads_d+1_m_e}
ds^2 = \frac{1}{z^2} \left( f(z)d\tau^2 + \frac{dz^2}{g(z)} + dx^2 + d\vec{y}^2 \right).
\end{align}
The endpoints of region $A$ are now located at $(0, 0)$ and $(\tau_0, 0)$. By symmetry, the corresponding RT surface can be parametrized as
\begin{align}
\tau= \tau(z) \ \text{and} \  x = 0.
\end{align}
The induced metric on the RT surface then becomes
\begin{align}
ds^2_{ind} = \frac{1}{z^2} \left( f(z)\tau'(z)^2 + \frac{1}{g(z)}\right) dz^2  + \frac{1}{z^2}d\vec{y}^2 .
\end{align}
Accordingly, the Lagrangian is
\begin{align}\label{L1}
\mathcal{L}&=\frac{1}{z^{d-1}}\sqrt{f(z)\tau'(z)^2 + \frac{1}{g(z)}}.
\end{align}

To proceed, we introduce a conserved quantity $p_{\tau}$ associated with translation invariance in $\tau$, which is defined by
\begin{align}\label{p_tau}
p_{\tau}:&=\frac{\p\mathcal{L}}{\p\tau'}=\frac{f(z)\tau'(z)}{z^{d-1}\sqrt{f(z)\tau'(z)^2+g(z)^{-1}}}.
\end{align}
Solving for $\tau'(z)$ gives
\begin{align}\label{tau'^2}
\tau'(z)^2&=\frac{p_{\tau}^2}{f(z)g(z)[f(z)z^{2(1-d)}-p_{\tau}^2]}.
\end{align}
Substituting this into~(\ref{L1}), the Lagrangian becomes
\begin{align}\label{L2}
\mathcal{L}=\frac{1}{z^{2(d-1)}}\sqrt{\frac{f(z)}{g(z)[f(z)z^{2(1-d)}-p_{\tau}^2]}}.
\end{align}

At the turning point $z = z_{\tau}$ of the RT surface, we have $\tau'(z_{\tau})\to \infty$. From~(\ref{tau'^2}), this divergence implies
\begin{align}\label{p_z}
p_{\tau}^2&=f(z_\tau)z_{\tau}^{2(1-d)}.
\end{align}
This condition allows us to eliminate $p_\tau$ in favor of $z_\tau$, allowing the Lagrangian to be regarded as a function of $z$ and $z_\tau$, that is, $\mathcal{L} = \mathcal{L}(z, z_\tau)$.

To determine $z_\tau$, we apply the boundary conditions $\tau(z=0) = 0$ and $\tau(z = z_\tau) = \frac{\tau_0}{2}$. This yields the integral constraint
\begin{align}\label{boundary}
\int_0^{z_\tau}\tau'(z)dz=\frac{\tau_0}{2}.
\end{align}
Since $\tau'(z)$ can be expressed in terms of $p_\tau$, and hence in terms of $z_\tau$, the above relation allows one to solve for $z_\tau$ as a function of the Euclidean boundary separation $\tau_0$, that is, $z_\tau = z_\tau(\tau_0)$.

To obtain the physical surface in Lorentzian signature, we perform an analytic continuation back from Euclidean spacetime. An important subtlety arises here: both $z_\tau$ and $p_\tau$, initially real in the Euclidean setting, become complex after continuation. In Lorentzian spacetime, multiple extremal surfaces may emerge depending on the choice of analytic branch. Instead of attempting to specify these surfaces directly, we adopt a constructive procedure: Starting from the Euclidean solution, we perform a Wick rotation by deforming $\tau \to i t$ in the complex plane. This method provides a canonical analytic continuation of $z_\tau$ and $p_\tau$, thereby selecting a physically significant branch for the extremal surface.

Under this Wick rotation, the conserved quantity and the turning point transform as
\begin{align}
p_t := -i p_\tau\big|_{\tau_0 \to i t_0}, \quad
z_t := z_\tau\big|_{\tau_0 \to i t_0}.
\end{align}

Once $z_t$ is determined, the area of the RT surface is given by
\begin{align}
\mathcal{A}(0,0; t_0, 0) = 2 \int_C dz\, \mathcal{L}(z, z_t),
\end{align}
where $C$ denotes an integration contour in the complex $z$ plane that connects the UV cutoff $z = \delta$ to the complex turning point $z = z_t$. The integrand $\mathcal{L}(z, z_t)$ is generally complex and features branch points due to the square-root structure. To ensure a well-defined result, the contour $C$ must be chosen such that it avoids all branch cuts in the complex plane.

\subsection{Numerical Method}

Analytic solutions for extremal surfaces are rarely available in higher-dimensional AdS black hole spacetimes. Although exact results can be obtained in special cases such as AdS$_3$ or AdS$_{d+1}$ in the Poincaré patch, generic black hole backgrounds typically necessitate perturbative or numerical techniques. In this section, we implement a numerical strategy based on the shooting method to determine the turning point $z_t$ and evaluate the associated geometric quantities.

The key idea is to remake the original boundary value problem for the embedding function $t(z)$ into an initial value problem, which is more amenable to numerical integration. The required inputs include the interval length $t_0$, the spacetime dimension $d$, the horizon location $z_H$, and the background functions $f(z)$ and $g(z)$. According to the boundary condition~(\ref{boundary}), the extremal surface satisfies $t(0) = 0$ and $t(z_t) = \frac{t_0}{2}$.

To proceed, we use the conservation relation~(\ref{p_z}) to express the conserved quantity $p_t$ in terms of the unknown turning point $z_t$. Substituting this into the expression for $t'(z)$ yields
\begin{align}
t'(z)^2 = \frac{p_t^2}{f(z) g(z) \left[ f(z) z^{2(1-d)} + p_t^2 \right]}.
\end{align}
With an initial guess $z_t^{(0)}$, the corresponding $p_t$ is fixed, making $t'(z)$ a fully determined function. We then numerically integrate $t(z)$ from $z = 0$ with $t(0) = 0$ up to $z = z_t$, resulting in a numerical estimate $\bar{t}(z_t)$. The deviation from the required boundary value is quantified by $\epsilon \equiv \bar{t}(z_t) - \frac{t_0}{2}$.

The turning point is then refined iteratively using root-finding algorithms such as the Newton--Raphson or secant method, until the error satisfies $|\epsilon| < \text{tol}$, where $\text{tol}$ is a prescribed numerical tolerance. The convergent value of $z_t$ is taken as the physical turning point of the extremal surface.

Once $z_t$ is fixed, physical observables, such as the area function $\mathcal{A}$ can be evaluated. By varying the interval size $t_0$, one can extract the dependence of $\mathcal{A}(t_0)$ and other geometric quantities, enabling comparisons with analytical or perturbative results.

In the following sections, we apply this procedure to explicit examples and verify the consistency of our numerical results against known analytical limits.

\subsection{Pure AdS}
In the following examples, we consider Schwarzschild black holes with metric functions satisfying $f(z) = g(z) = 1 - \frac{z^d}{z_H^d}$.

In this subsection, we focus specifically on the Poincaré case taking the limit $z_H \to \infty$, under which the metric simplifies to $f(z) = g(z) = 1$.

As discussed previously, the RT surface area $\mathcal{A}$ for AdS$_{d+1}$ in the Poincaré patch can be analytically calculated for arbitrary $d \geq 2$. Therefore, these cases serve as benchmarks to verify the accuracy of our numerical implementation.

\subsubsection{The AdS\texorpdfstring{$_3$}{3} case}\label{section_AdS3_RT}
We begin with the Poincaré AdS$_3$ ($d=2$) metric $ds^2 = \frac{1}{z^2}\left(-dt^2 + dz^2 + dx^2\right)$. Under $t \to -i\tau$, the Euclidean metric becomes $ds^2 = \frac{1}{z^2}\left(d\tau^2 + dz^2 + dx^2\right)$.

For a timelike interval between $(0,0)$ and $(\tau_0, 0)$, the extremal surface can be parameterized as $\tau = \tau(z)$ with $x = 0$. The corresponding Lagrangian reads $\mathcal{L} = \frac{1}{z}\sqrt{\tau'(z)^2 + 1}$.

The conserved quantity associated with $\tau$ is $p_\tau = \frac{\partial\mathcal{L}}{\partial\tau'} = \frac{\tau'}{z\sqrt{\tau'^2 + 1}}$, which can be inverted to produce
\begin{equation}
\tau'(z)^2 = \frac{p_\tau^2 z^2}{1 - p_\tau^2 z^2}.
\end{equation}
Since the turning point satisfies $\tau'(z_\tau) \to \infty$, the denominator must vanish, resulting in
\begin{align}
z_\tau = \frac{1}{p_\tau}.
\end{align}

Imposing the boundary condition $\int_0^{z_\tau} \tau'(z) dz=\frac{\tau_0}{2}$, and evaluating the integral analytically, we obtain
\begin{equation}
z_\tau = \frac{\tau_0}{2}.
\end{equation}
Applying the Wick rotation $\tau_0 \to i t_0$ then gives the Lorentzian turning point
\begin{equation}
z_{t} =z_{\tau}\big|_{\tau_0\to it_0}= \frac{it_0}{2}.
\end{equation}
Interestingly, the turning point becomes complex under this continuation. The area of the RT surface becomes
\begin{align}\label{ads3_p}
\mathcal{A}(0,0;t_0,0)&=2\int_{\delta}^{z_t}dz\mathcal{L}=2\int_{\delta}^{\frac{it_0}{2}}\frac{dz}{z\sqrt{1 + (\frac{2z}{t_0})^2}}\nn\\
&=2\log(\frac{t_0}{\delta})+i\pi.
\end{align}
This result consists of a real and an imaginary component, respectively, corresponding to the contributions from the segments of the RT surface in the real and imaginary regions of the complexified geometry, as illustrated in Fig.~\ref{twoproposals}(b).

\begin{figure}[h]
\centering
\subfigure[]{\includegraphics[scale=0.21]{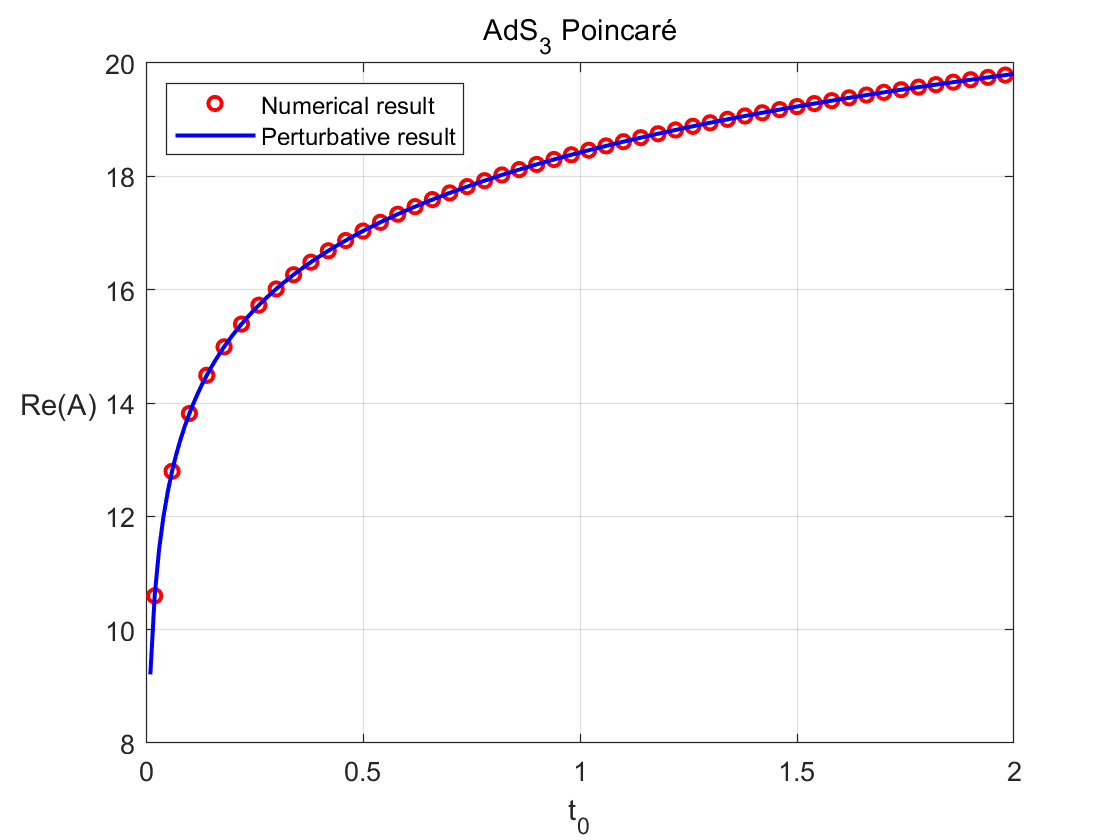}}
\subfigure[]{\includegraphics[scale=0.21]{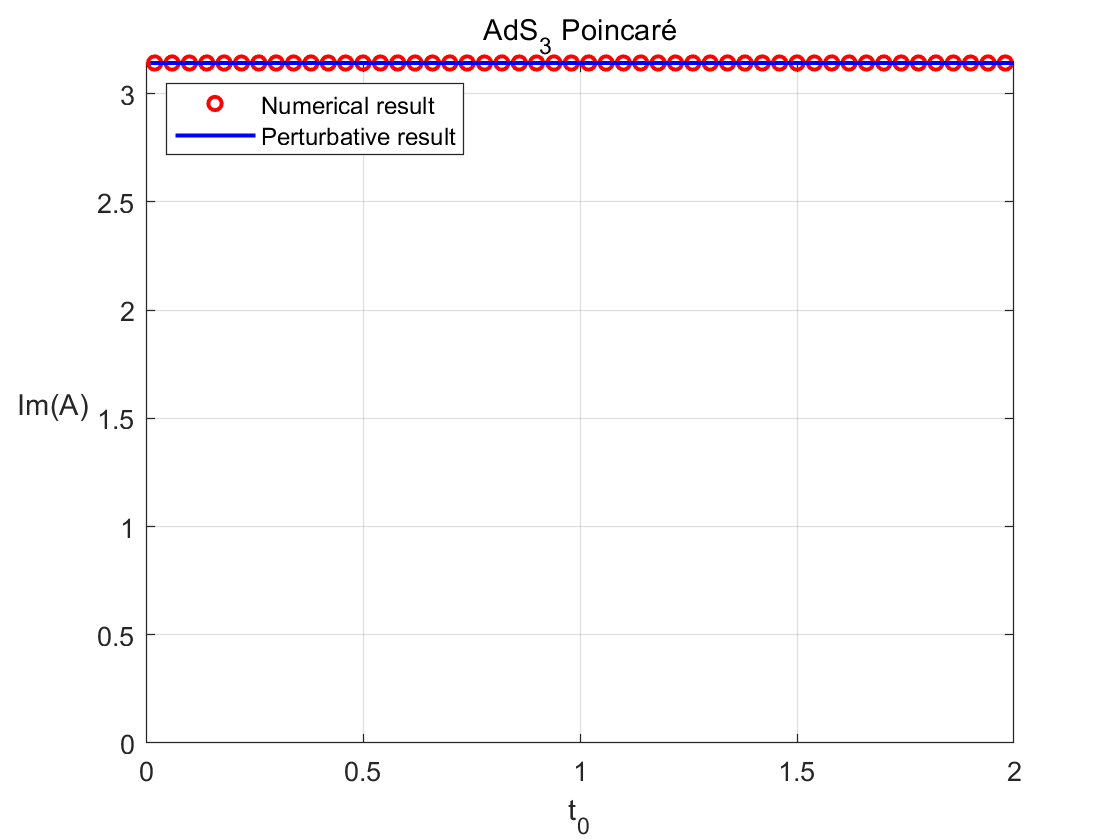}}
\caption{Area $\mathcal{A}$ of the RT surface as a function of interval length $t_0$ in the timelike AdS$_3$ Poincaré case. The blue curve represents the analytical result~(\ref{ads3_p}), while the red circles are obtained numerically. Panel (a) shows the real component, and panel (b) displays the imaginary part. The interval $t_0$ is sampled over $(0,2)$ with step size 0.01, and we use a UV cutoff $\delta = 10^{-4}$.}
\label{fig_t_3_p}
\end{figure}

To verify the numerical implementation, we compare this with this analytical result. The numerical data are shown in Fig.~\ref{fig_t_3_p}, where red circles denote numerical values, and blue curves represent the analytic expression~(\ref{ads3_p}). As expected, the numerical and analytical results agree precisely, with the real part reproducing $2\log(\frac{t_0}{\delta})$ and the imaginary part producing $i\pi$. This agreement validates the numerical method in the Poincaré AdS$_3$ case.

\begin{figure}[h]
\centering
\subfigure[]{\includegraphics[scale=0.21]{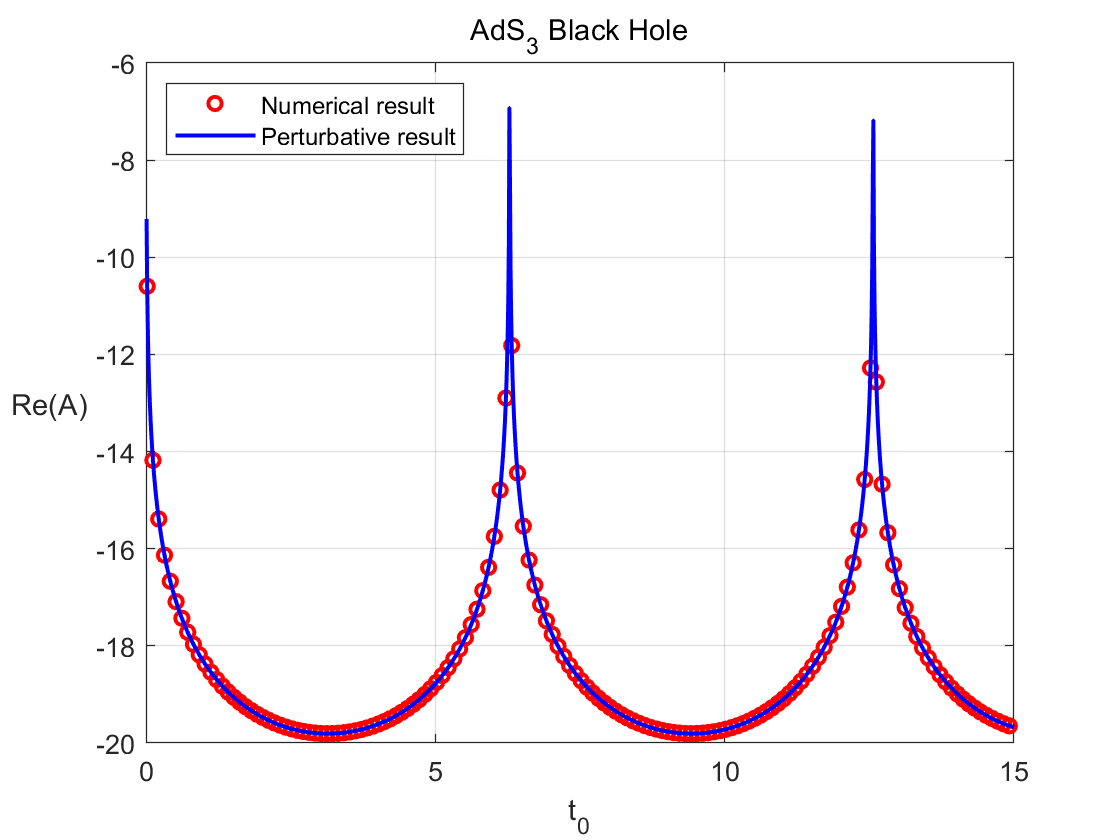}}
\subfigure[]{\includegraphics[scale=0.21]{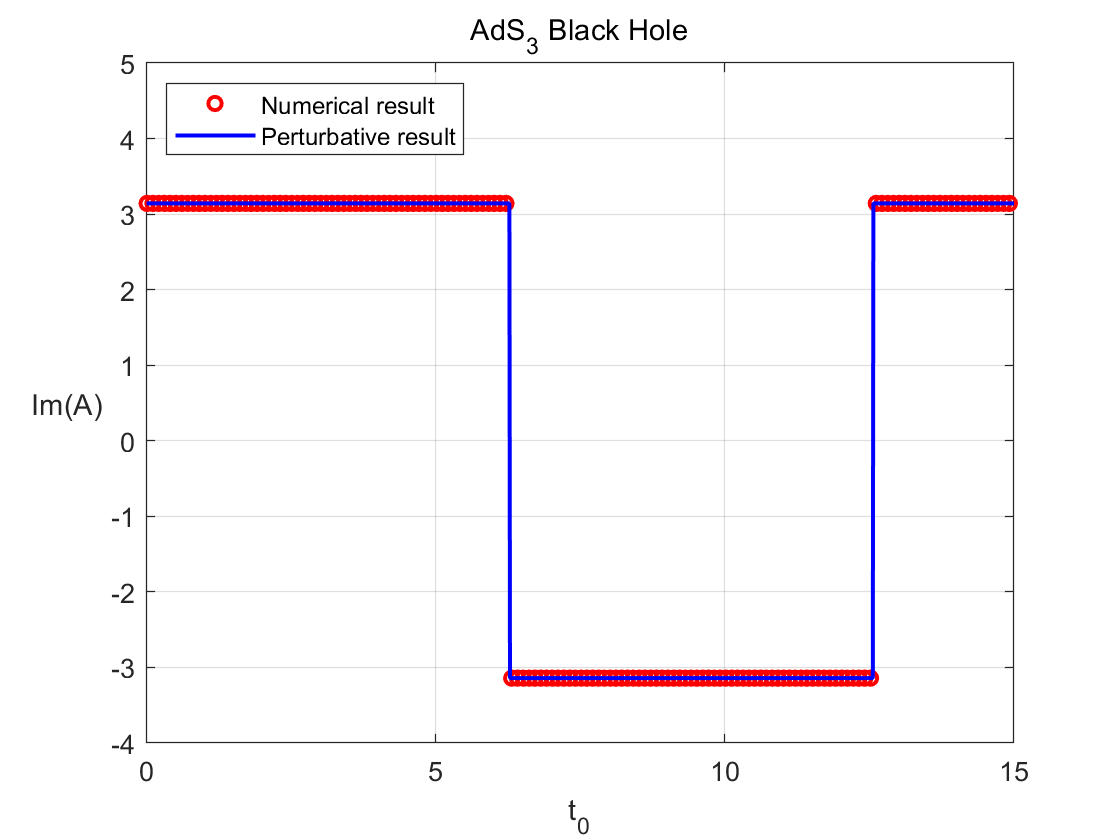}}
\caption{Area $\mathcal{A}$ of the RT surface as a function of interval length $t_0$ in the timelike AdS$_3$ global coordinate. The blue curve shows the analytic result~(\ref{ads3_g}), while the red circles indicate numerical data. Panel (a) shows the real part of $\mathcal{A}$, and panel (b) presents the imaginary part. We set $\delta = 10^{-4}$ and sample $t_0 \in (0,15)$ with step size 0.01.}
\label{fig_t_3_g}
\end{figure}

One can construct the RT surface in the global coordinate with the metric
\bea
ds^2 = -(1+r^2) dt^2 +\frac{1}{1+r^2} dr^2 + r^2 d\theta^2,
\eea
where $0<r<+\infty$, $0\le \theta<2\pi$. CFT lives on the AdS boundary $r=r_\infty$, where $r_\infty$ is the bulk cut-off.   Consider the time interval $[0,t_0]$. By the same procedure we can find the turning point of the RT surface:
\bea
r_t=r_{\tau}\big|_{\tau_0\to it_0}=i \csc(\frac{t_0}{2}),
\eea
which is imaginary. The area of RT surface is given by 
\begin{align}\label{ads3_g}
\mathcal{A}(0,0;t_0,0)&=2\int_{C}\sqrt{\frac{1}{r^2-r_t^2}}=-2\log{(\frac{2\sin(\frac{t_0}{2})}{\delta})}+i\pi,
\end{align}
where the contour $C$ is chosen as $r_\infty \to 0 \to r_t$ and $r_\infty=\frac{1}{\delta}$. As shown in Fig.~\ref{fig_t_3_g}, the numerical results (red circles) perfectly agree with the analytical expression (blue curves). In particular, the surface area exhibits a periodic structure with period $4\pi$ due to the dependence $\sin(\frac{t_0}{2})$. The real part shows a sequence of logarithmic divergences at $t_0 = 2n\pi$ ($n\in\mathbb{Z}$), corresponding to the vanishing of the sine function. Meanwhile, the imaginary part displays step-like jumps of $-2\pi i$ across each period, reflecting the branch cut behavior of the logarithm in the complexified geometry.

Although the global AdS$_3$ time coordinate is compactified with period $2\pi$, the area function exhibits a $4\pi$ periodicity due to its $\sin(\frac{t_0}{2})$ dependence. This arises because $\sin(\frac{t_0}{2})$ is $4\pi$-periodic, and its appearance inside a logarithm leads to branch cut jumps of $2\pi i$ across each $4\pi$ cycle. As a result, the complexified surface area encodes a richer periodic structure than the underlying spacetime geometry.

We can also solve the RT surface $r(t)$. The results are shown in Appendix~\ref{Appendix_global}, where we also show how to find the RT surface for an arbitrary interval in the global coordinate.

\subsubsection{Higher-dimensional AdS\texorpdfstring{$_{d+1}$}{d+1} case}

We now generalize the analysis to higher-dimensional Poincaré AdS$_{d+1}$ geometries. The analytic solution can be obtained for arbitrary $d$, and we numerically validate it in the AdS$_4$ ($d=3$) case as a representative example.

The AdS$_{d+1}$ Poincaré metric takes the form
\begin{equation*}
ds^2 = \frac{1}{z^2}\left(-dt^2 + dz^2 + dx^2 +d\vec{y}^2\right).
\end{equation*}
The corresponding Euclidean metric is given by
\begin{equation*}
ds^2 = \frac{1}{z^2}\left(d\tau^2 + dz^2 + dx^2 +d\vec{y}^2\right).
\end{equation*}

For a timelike interval from $(0,0)$ to $(\tau_0, 0)$, the extremal surface can be parameterized as $\tau = \tau(z)$ with $x = y_i = 0$. The corresponding Lagrangian becomes $\mathcal{L} = \frac{1}{z^{d-1}} \sqrt{1 + \tau'(z)^2}$.

The conserved constant associated with $\tau$ is $p_\tau = \frac{\tau'}{z^{d-1}\sqrt{\tau'^2 + 1}}$, which leads to
\begin{equation}
\tau'(z)^2 = \frac{p_\tau^2 z^{2(d-1)}}{1 - p_\tau^2 z^{2(d-1)}}.
\end{equation}
The turning point $z_\tau$ corresponds to the divergence of $\tau'(z)$, giving
\begin{align}
z_\tau &= \left(\frac{1}{p_\tau}\right)^{\frac{1}{d-1}}.
\end{align}

Imposing the boundary condition $\int_0^{z_\tau} \tau'(z)\, dz = \frac{\tau_0}{2}$, and evaluating the integral analytically, we find
\begin{equation}
z_{\tau} = \frac{\Gamma\left(\frac{1}{2(d-1)}\right)}{2\sqrt{\pi} \Gamma\left(\frac{d}{2(d-1)}\right)}\tau_0.
\end{equation}
After Wick rotation $\tau_0 \to i t_0$, the turning point becomes
\begin{equation}
z_t = \frac{i \Gamma\left(\frac{1}{2(d-1)}\right)}{2\sqrt{\pi} \Gamma\left(\frac{d}{2(d-1)}\right)}t_0,
\end{equation}
which lies in the complex plane. The area of RT surface splits into real and imaginary contributions:
\begin{align}
\mathcal{A}(0,0;t_0,0)&=2\int_{\delta}^{z_t}dz\mathcal{L}\nn\\
&=\frac{2}{(d-2)\delta^{d-2}}+2\kappa_d \frac{(-i)^{d}}{t_0^{d-2}},
\end{align}
with $\kappa_d=\frac{\pi^{\frac{d-1}{2}} 2^{d-2}}{d-2}(\frac{\Gamma(\frac{d}{2(d-1)})}{\Gamma(\frac{1}{2(d-1)})})^{d-1}$.

\begin{figure}
\centering
\subfigure[]{\includegraphics[scale=0.21]{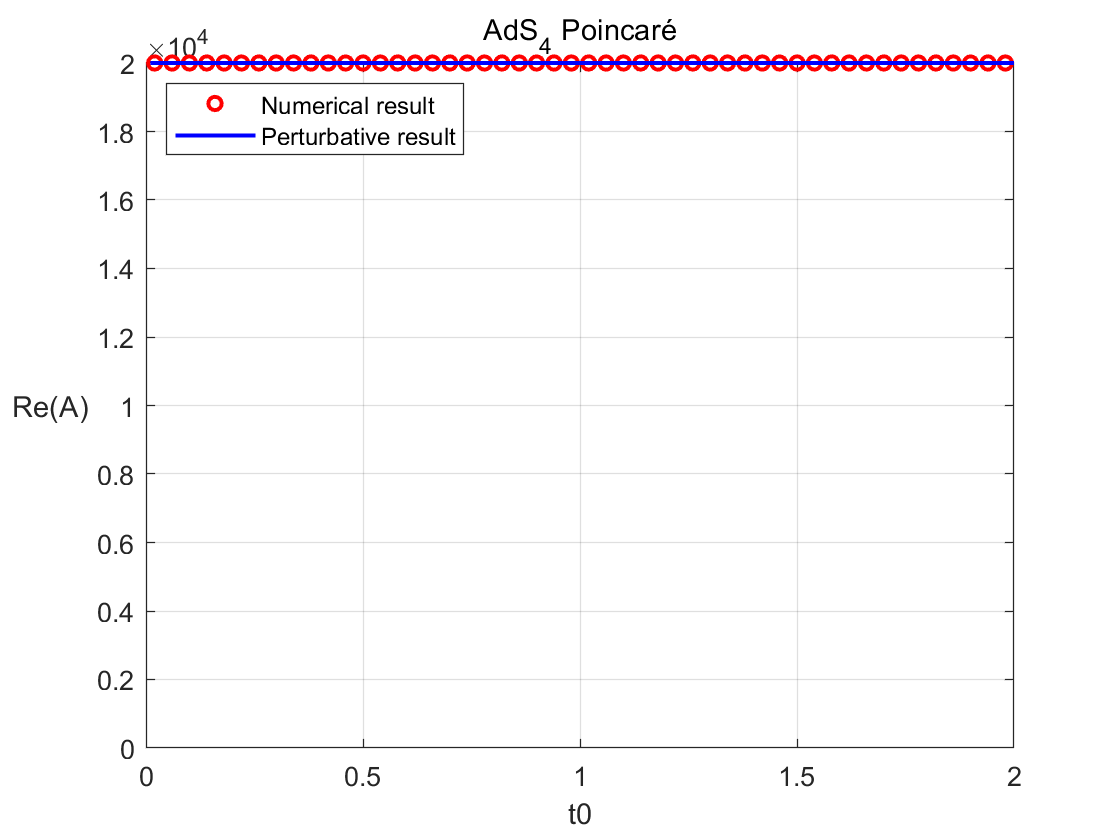}}
\subfigure[]{\includegraphics[scale=0.21]{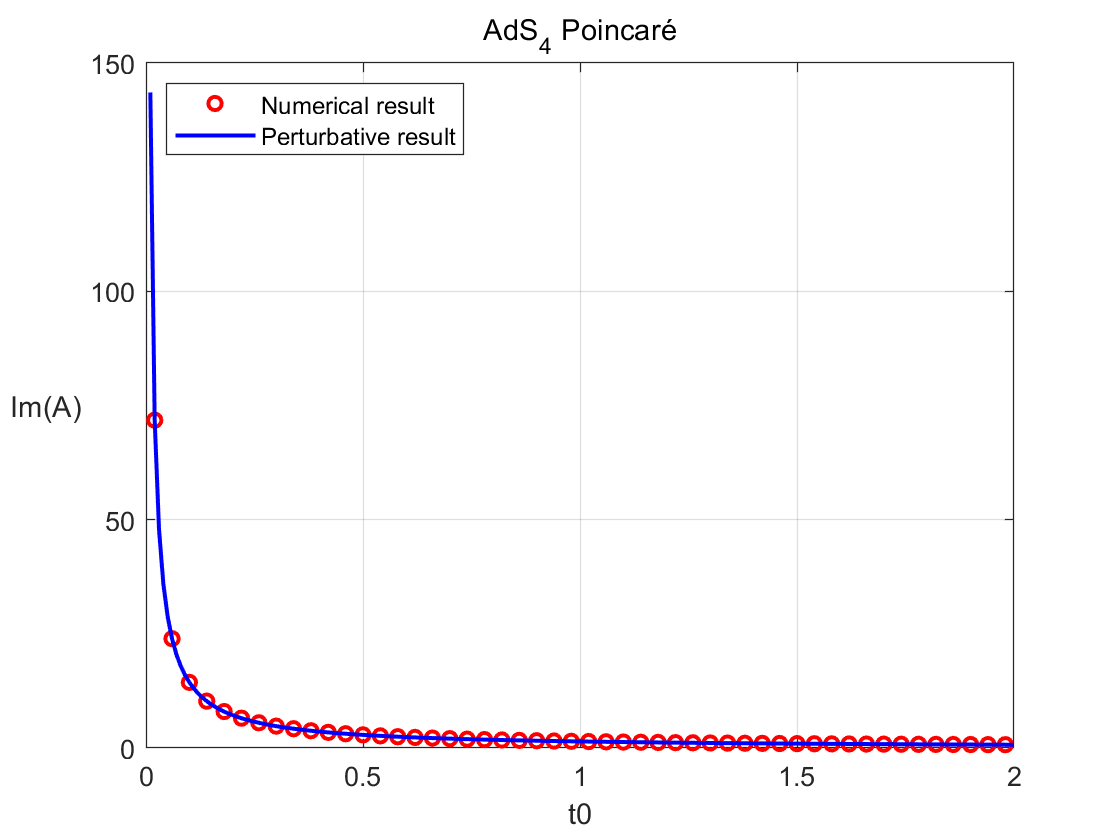}}
\caption{Area $\mathcal{A}$ of the RT surface as a function of interval length $t_0$ in the timelike AdS$_4$ Poincaré case. The blue curve represents the analytical result~(\ref{ads4_p}), while the red circles are obtained numerically. Panel (a) shows the real component, and panel (b) displays the imaginary part. The interval $t_0$ is sampled over $(0,2)$ with step size 0.01, and we use a UV cutoff $\delta = 10^{-4}$.}
\label{fig_t_4_p}
\end{figure}

As a concrete example, we take $d = 3$ (AdS$_4$), for which the area becomes
\begin{equation}\label{ads4_p}
\mathcal{A}_4(0,0;t_0,0)=\frac{2}{\delta}+\frac{i8\pi^3}{t_0\Gamma(\frac{1}{4})^4}.  
\end{equation}
To validate the result, we compare with the numerical data shown in Fig.~\ref{fig_t_4_p}, where the red circles represent the numerical values and the blue curve denotes the analytical expression~(\ref{ads4_p}). As expected, the numerical and analytical results agree precisely: the real part reproduces the UV divergence $\frac{2}{\delta}$, while the imaginary part follows the inverse-$t_0$ behavior with coefficient proportional to $\frac{8\pi^3}{\Gamma(\frac{1}{4})^4}$.

\subsection{Examples of  black hole}
We now turn to the case of AdS black hole geometries. As discussed earlier, we consider metrics of the form $f(z) = g(z) = 1 - \frac{z^d}{z_H^d}$, where $z_H$ denotes the position of the horizon.

In the AdS$_3$ BTZ black hole, the extremal surface equations possess fully analytic solutions. However, for higher-dimensional AdS$_{d+1}$ black holes, the equations become analytically intractable and require a perturbative treatment. As a result, the comparison between analytic and numerical results shows only approximate agreement in higher dimensions.

\subsubsection{BTZ black hole}
We now consider the AdS$_3$ BTZ black hole, whose metric takes the form $ds^2 = \frac{1}{z^2}\left[-f(z)dt^2 + \frac{dz^2}{f(z)} + dx^2\right]$, where $f(z) = 1-\frac{z^2}{z_H^2}$ and $z_H$ denote the horizon radius. After Wick rotates $t \to -i\tau$, the Euclidean metric becomes $ds^2 = \frac{1}{z^2}\left[f(z)d\tau^2 + \frac{dz^2}{f(z)} + dx^2\right]$.

For an interval $(0,0;\tau_0,0)$, the RT surface is parameterized as $\tau = \tau(z)$ with $x = 0$, leading to Lagrangian $\mathcal{L} = \frac{1}{z}\sqrt{f(z)\tau'(z)^2 + \frac{1}{f(z)}}$. So, the conserved constant is $p_\tau = \frac{f(z)\tau'}{z\sqrt{f(z)\tau'^2 + f(z)^{-1}}}$, which leads to
\begin{equation}
\tau'(z)^2 = \frac{p_\tau^2}{f(z)^2[z^{-2}f(z) - p_\tau^2]}.
\end{equation}
The turning point is defined by $\tau'(z_\tau)\to\infty$, which gives
\begin{align}
z_\tau &= \frac{z_H}{\sqrt{1 + (p_\tau z_H)^2}}.
\end{align}

Applying the boundary condition $\int_0^{z_\tau} \tau'(z) dz = \frac{\tau_0}{2}$, one finds the turning point
\begin{equation}
z_{\tau} = z_H \sin\left(\frac{\tau_0}{2z_H}\right).
\end{equation}
After Wick rotation $\tau_0 \to i t_0$, the turning point becomes complex:
\begin{equation}
z_{t} =z_{\tau}\big|_{\tau_0\to it_0}= iz_H \sinh\left(\frac{t_0}{2z_H}\right).
\end{equation}
The area function becomes
\begin{align}\label{ads3_b}
\mathcal{A}(0,0;t_0,0)=2\int_{\delta}^{z_t}dz\mathcal{L}=&2\int_{\delta}^{iz_H\sinh(\frac{t_0}{2z_H})}\frac{dz}{z\sqrt{1 - \frac{z^2}{z_H^2} + p_t^2 z^2}}\nn\\
=&2\ln\left(\frac{2z_H}{\delta}\sinh\left(\frac{t_0}{2z_H}\right)\right) + i\pi.
\end{align}

\begin{figure}
\centering
\subfigure[]{\includegraphics[scale=0.21]{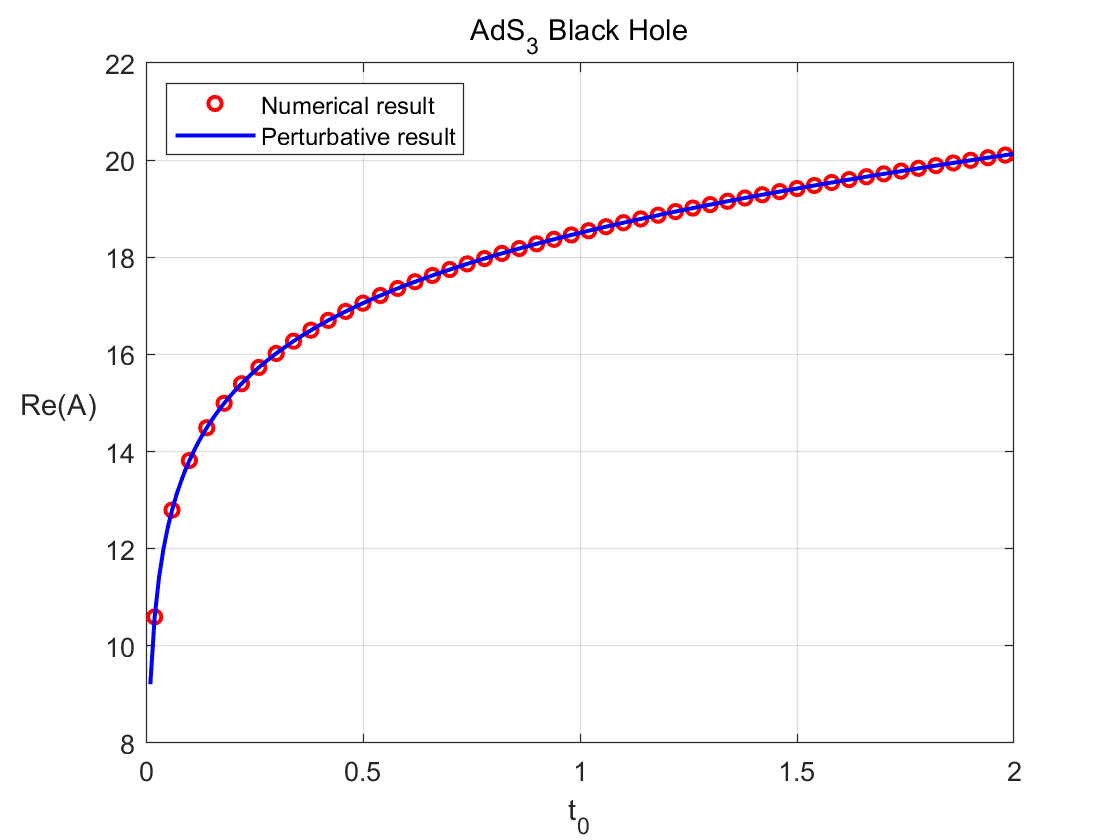}}
\subfigure[]{\includegraphics[scale=0.21]{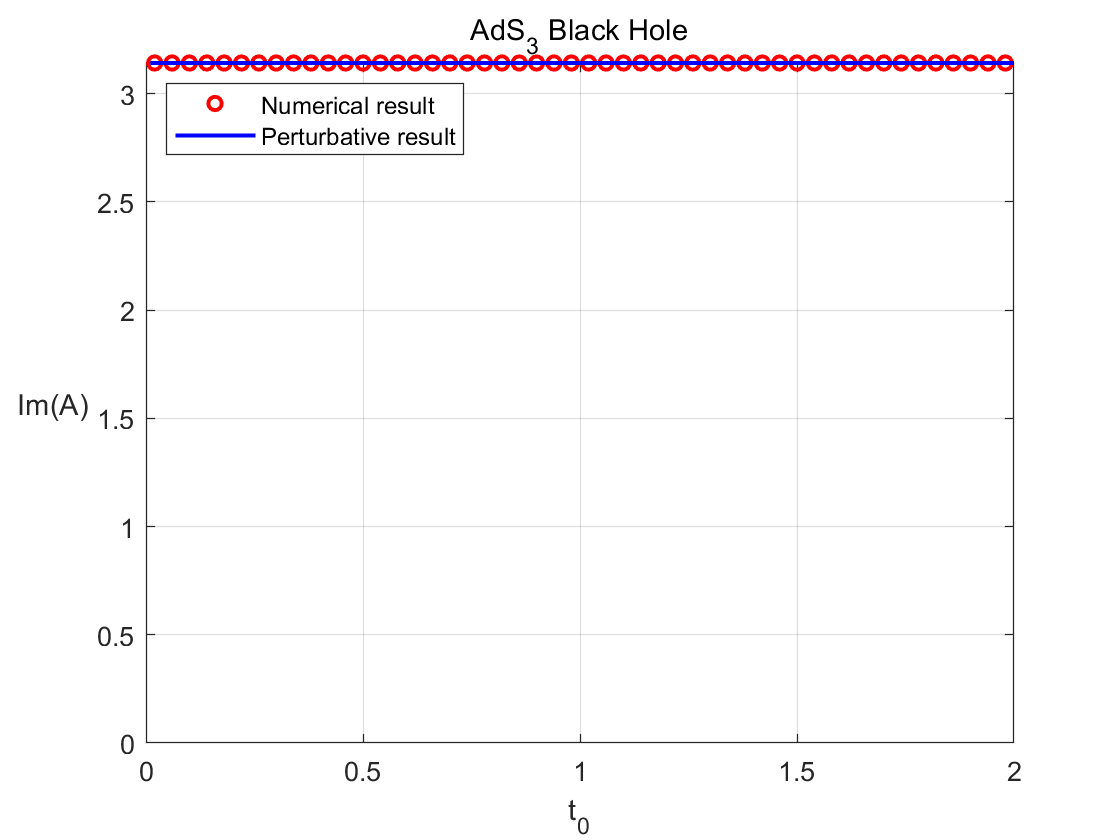}}
\caption{Area $\mathcal{A}$ of the RT surface as a function of interval length $t_0$ for the timelike AdS$_3$ black hole case. The blue curve shows the analytical expression~(\ref{ads3_b}), while red circles denote the numerical results. Panel (a) displays the real component; panel (b) shows the imaginary part. The interval $t_0$ is sampled over $(0,2)$ with spacing 0.01, and we use a UV cutoff $\delta=10^{-4}$.}
\label{fig_t_3_b}
\end{figure}

The numerical results are shown in Fig.~\ref{fig_t_3_b}, where red circles denote numerical values and the blue curve represents the analytical expression~(\ref{ads3_b}). Both real and imaginary components are in strong agreement.

\subsubsection{Higher-dimensional AdS\texorpdfstring{$_{d+1}$}{d+1} black hole}\label{section_higher_d_black_hole}
This section provides a detailed analysis of higher-dimensional AdS$_{d+1}$ black hole geometries. Since fully analytical solutions are generally inaccessible in this case, we adopt a perturbative approach in the small interval limit $t_0 \ll z_h$. Specifically, we expand the RT surface equations perturbatively around the pure AdS background and compute the leading corrections due to the black hole geometry.

The analytical expressions will then be compared with numerical solutions. To ensure the validity of the perturbative expansion, the numerical results presented in this section are restricted to the small interval regime with $t_0 \lesssim z_h = 1$.

We now turn to higher-dimensional AdS$_{d+1}$ black hole geometries. The metric takes the form $ds^2 = \frac{1}{z^2} \left( -f(z)dt^2 + \frac{dz^2}{g(z)} + dx^2 + d\vec{y}^2 \right)$, where $f(z)=g(z)=1-\frac{z^d}{z_H^d}$. For simplicity, we set $z_H = 1$ as follows. The corresponding Euclidean metric is given by $ds^2 = \frac{1}{z^2} \left( f(z)d\tau^2 + \frac{dz^2}{g(z)} + dx^2 + d\vec{y}^2 \right)$.

Talking into account a symmetric timelike interval $(0,0;\tau_0,0)$, we parameterize the extremal surface by $\tau = \tau(z)$ and $x = 0$. The associated Lagrangian becomes $\mathcal{L} = \frac{1}{z^{d-1}}\sqrt{(1-z^d)\tau'(z)^2 + \frac{1}{1-z^d}}$. So, the conserved constant is $p_\tau = \frac{(1-z^d)\tau'}{z^{d-1}\sqrt{(1-z^d)\tau'^2 + \frac{1}{1-z^d}}}$, which yields
\begin{equation}
\tau'(z)^2 = \frac{p_\tau^2}{(1-z^d)^2[(1-z^d)z^{2(1-d)} - p_\tau^2]}.
\end{equation}

The turning point $z_\tau$ is defined by the divergence of $\tau'(z)$, leading to the condition
\begin{align}
p_\tau^2=(1-z_\tau^d)z_\tau^{2(1-d)}.
\end{align}
Substituting this into the area function, we obtain the following
\begin{align}
\mathcal{A}(0,0;\tau_0,0)=2\int_{\delta}^{z_\tau}dz\frac{1}{z^{2(d-1)}}\sqrt{\frac{1}{(1-z^d)z^{2(1-d)}-(1-z_\tau^d)z_\tau^{2(1-d)}}}.
\end{align}
This integral can be perturbatively evaluated in the small $\tau_0$ regime, yielding
\begin{align}\label{perturbative1}
\mathcal{A}(0,0;\tau_0,0) =& \frac{2}{(d-2)\delta^{d-2}}-\frac{2\sqrt{\pi} \Gamma \left(\frac{d}{2 (d-1)}\right)}{z_{\tau_0}^{d-2} (d-2) \Gamma \left(\frac{1}{2 (d-1)}\right)}\nn\\
&+\frac{1}{2} \sqrt{\pi} z_{\tau_0}^2 \left[\frac{(d-3) \Gamma \left(\frac{d}{d-1}\right)}{(d-1) \Gamma \left(\frac{d+1}{2 (d-1)}\right)}+\frac{2 \Gamma \left(\frac{d}{2 (d-1)}\right)}{(d-1) \Gamma \left(\frac{1}{2 (d-1)}\right)}\right].
\end{align}

On the other hand, imposing the boundary condition $\int_0^{z_\tau} \tau'(z) dz = \frac{\tau_0}{2}$ allows us to solve for $z_\tau$ perturbatively:
\begin{align}\label{perturbative2}
z_{\tau_0}=\zeta_0 \tau_0+ \zeta_d \tau_0^{d+1},
\end{align}
where
\begin{align}
&\zeta_0=\frac{\Gamma(\frac{1}{2(d-1)})}{2\sqrt{\pi}\Gamma(\frac{d}{2(d-1)})},\nn\\
&\zeta_d=-\frac{\zeta_0^{d+1} \left(2 d-2+\frac{(d-3) \Gamma \left(\frac{1}{2 (d-1)}\right) \Gamma \left(\frac{d}{d-1}\right)}{\Gamma \left(\frac{1-3 d}{2-2 d}\right) \Gamma \left(\frac{d}{2 (d-1)}\right)}\right)}{4 (d-1)^2}.
\end{align}

Finally, performing the Wick rotation $\tau_0 \to it_0$ gives the analytic expressions in Lorentzian signature, which we now proceed to evaluate for specific examples.

For the AdS$_4$ case ($d=3$), the perturbative expansion yields
\begin{align}
\mathcal{A}(0,0;t_0,0)
&= \frac{2}{\delta}
+ i\,\frac{8\pi^3}{\Gamma\left(\tfrac{1}{4}\right)^4}\,\frac{1}{t_0}.
\end{align}
\begin{figure}[htbp]
  \centering
\subfigure[]{\includegraphics[scale=0.21]{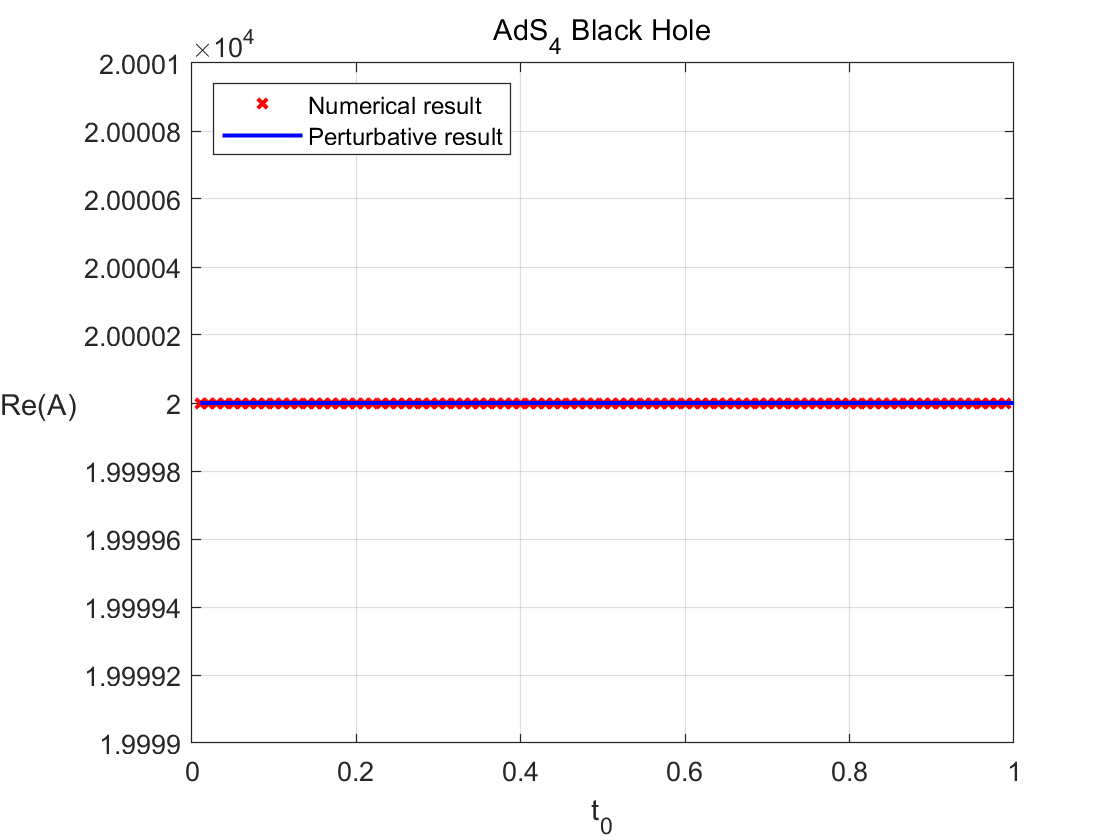}}
\subfigure[]{\includegraphics[scale=0.21]{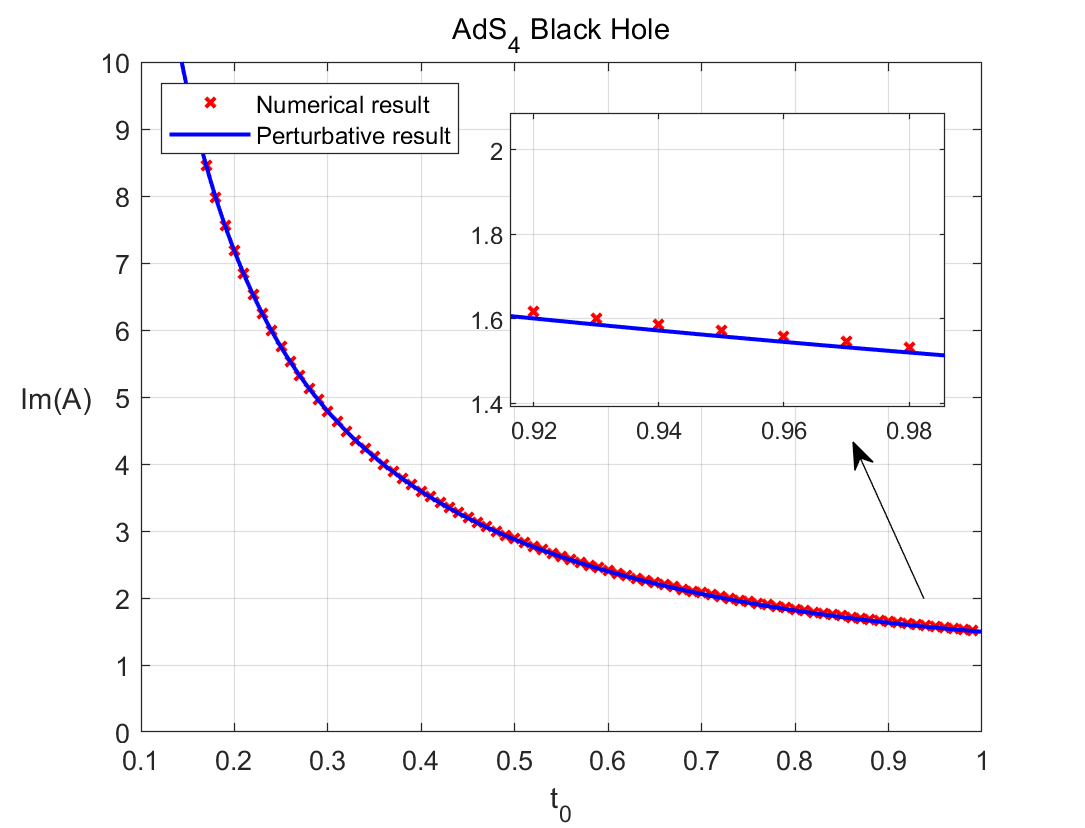}}
  \caption{Area $\mathcal{A}(0,0;t_0,0)$ for the timelike RT surface in the AdS$_4$ black hole background. The red crosses denote numerical results, while the blue curves represent the perturbative analytical result. Panel (a) shows the real part; panel (b) displays the imaginary part. We set $z_H=1$ and $\delta=10^{-4}$, with $t_0$ sampled over $(0,z_H=1)$ in steps of 0.01.}\label{fig_AdS4}
\end{figure}

Fig.~\ref{fig_AdS4} compares this analytical expression with numerical results. We observe excellent agreement in the small-$t_0$ regime, confirming the validity of the perturbative expansion. As $t_0$ increases toward the horizon scale $z_H = 1$, slight deviations begin to appear, although overall behavior remains qualitatively consistent and the approximation is still reasonably accurate.

Obviously, our result matches precisely the asymptotic behavior obtained in Fig.~4 of Ref.~\cite{Heller:2024whi} as $\Delta t \to 0$, corresponding to the regularized area $\mathcal{A}_\text{reg}^{\mathrm{v.c.}}$ of the vacuum-connected solution in their terminology.

According to previous work~\cite{Guo:2024ITE}, in higher-dimensional AdS$_{d+1}$ spacetimes with Poincaré coordinates, the EE (excluding UV-divergent parts) exhibits a dependence on dimensional parity: it becomes purely imaginary for even-dimensional spacetimes AdS$_{\text{even}}$ ($d$ odd), while remaining strictly real for odd-dimensional spacetimes AdS$_{\text{odd}}$ ($d$ even). For higher-dimensional AdS$_{d+1}$ black hole scenarios, the situation becomes more complex. Nevertheless, we still observe a clear distinction between odd and even dimensions. Therefore, in addition to the AdS$_4$ ($d=3$) case, we particularly examine the AdS$_5$ ($d=4$) scenario to highlight this contrast.

For the AdS$_5$ case ($d=4$), using~(\ref{perturbative1}) and~(\ref{perturbative2}), we obtain the following perturbative expression:
\begin{align}\label{ads5_p}
\mathcal{A}(0,0;t_0,0)
= \frac{1}{\delta^2}
+ 0.321\,\frac{1}{t_0^2} - 0.508\,t_0^2 + 0.505\,t_0^6.
\end{align}
Here, we omit the explicit expressions for the $\Gamma$-function and present the coefficients numerically for clarity.

\begin{figure}[htbp]
  \centering
  \subfigure[]{\includegraphics[scale=0.21]{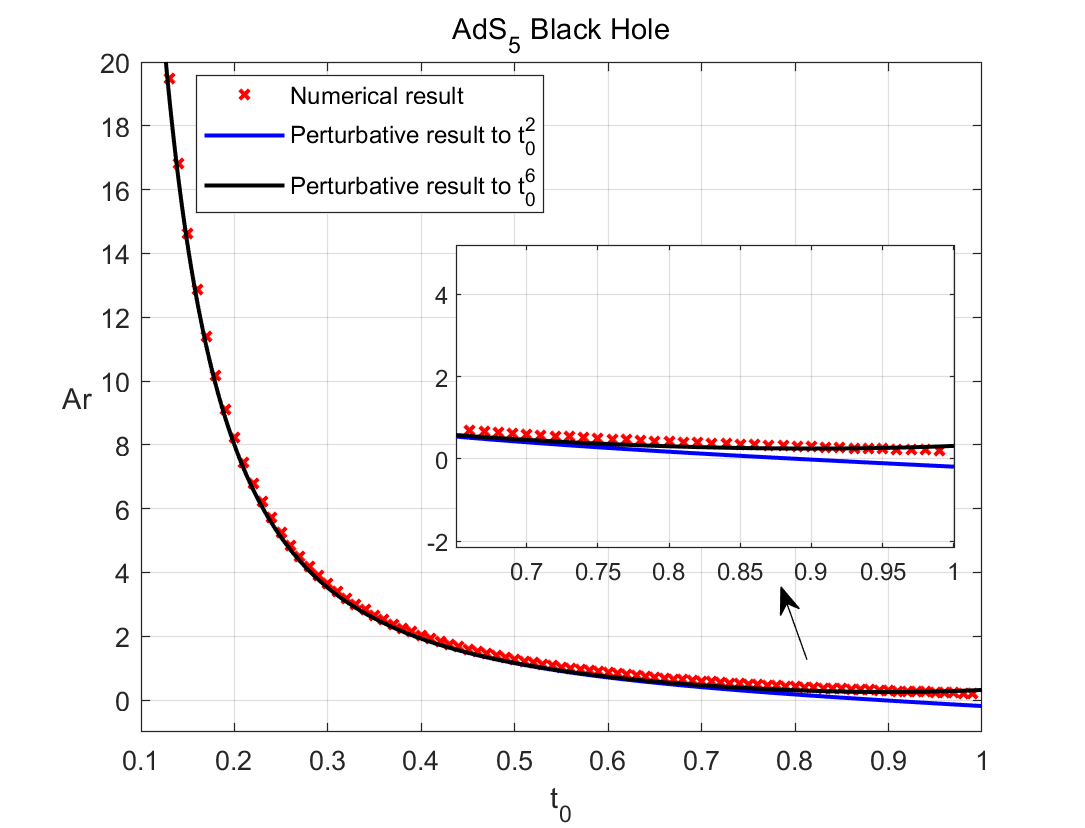}}
\subfigure[]{\includegraphics[scale=0.21]{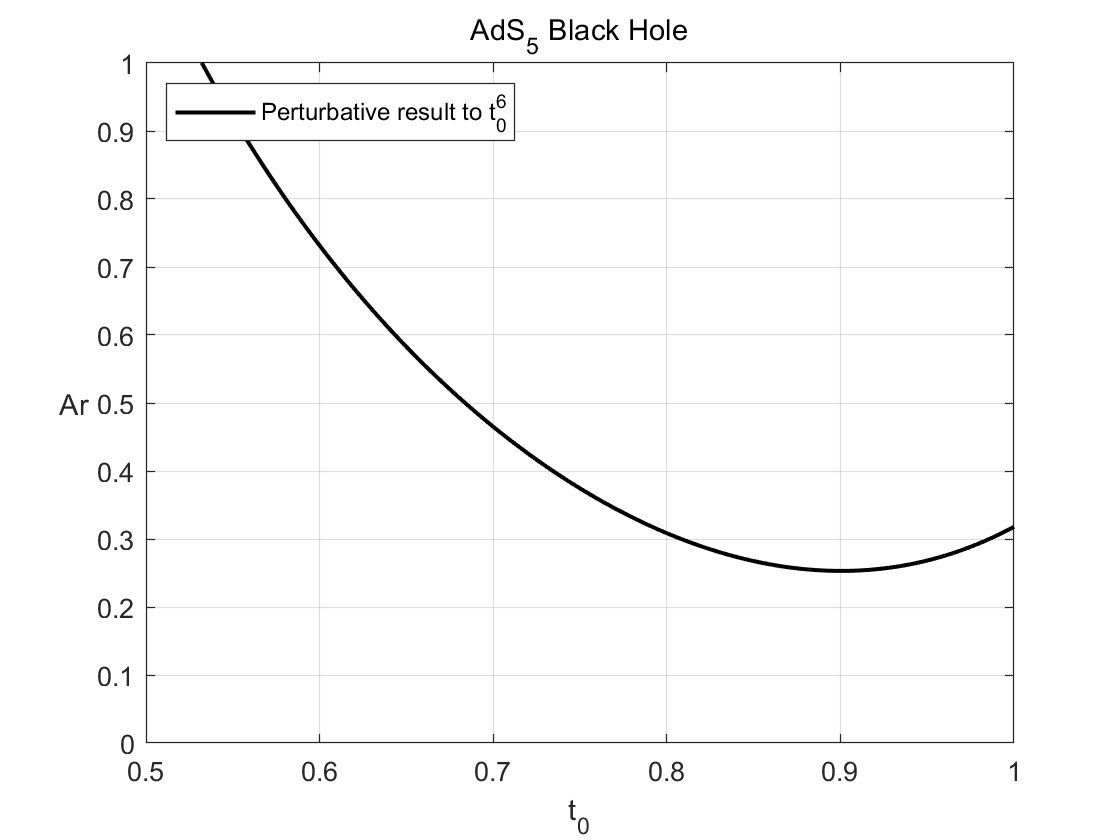}}
  \caption{Regularized area $\mathcal{A}_r = \mathcal{A}(0,0;t_0,0) - \frac{1}{\delta^{2}}$ as a function of interval length $t_0$ in the AdS$_5$ black hole case. 
  Panel (a) compares the full numerical results (red circles) with the perturbative expressions truncated at order $t_0^2$ (blue curve) and $t_0^6$ (black curve). 
  Panel (b) shows the analytic perturbative expansion up to $t_0^6$ only, with the vertical axis rescaled to highlight fine features such as local extrema.
  We set $z_H = R = 1$, sample $t_0$ over $(0,z_H=1)$ with step size $0.01$, and use a UV cutoff $\delta = 10^{-4}$.}
  \label{fig_AdS5}
\end{figure}

 In contrast to the AdS$_4$ case, where the RT surface area develops an imaginary part, the area in AdS$_5$ remains strictly real at all computed orders, consistent with the parity behavior observed in the vacuum case. In Fig.~\ref{fig_AdS5}(a), we compare the numerical results with the perturbative predictions. As in the AdS$_4$ case, the perturbative expansion agrees well with the numerics in the small $t_0 \ll z_H$ regime. Moreover, we observe that including higher order terms in $t_0$ significantly improves agreement across a wider range of $t_0$, with the $t_0^6$ truncation (black curve) closely following the numerical data up to $t_0 \sim z_H$. Interestingly, the perturbative expansion up to order $t_0^6$ shows a local extremum in Fig.~\ref{fig_AdS5}(b), which is absent in the numerical data. We interpret this feature as a spurious artifact of the truncated expansion, and we expect it to disappear once higher-order terms are included, restoring the smooth monotonic behavior seen numerically.

\section{Timelike entanglement wedge and its cross section}\label{Section_EWCS}

For the spacelike case, various entanglement measures have been introduced to better capture quantum correlations between different regions, such as logarithmic negativity and reflected entropy. Some of these quantities are also expected to admit gravitational duals. In particular, both the reflected entropy and entanglement of purification are conjectured to be dual to the so-called entanglement wedge cross section (EWCS).

Given that these entanglement measures can be extended to a timelike setting, it becomes natural to ask whether their gravitational duals also admit similar extensions. In this paper, we propose a definition of EWCS for the timelike case. Although the construction differs from that in the spacelike case in several important aspects, it nevertheless leads to a well-defined quantity. Furthermore, we provide evidence that the timelike reflected entropy is dual to the timelike EWCS.

In Section~\ref{section_EM}, we demonstrate that other entanglement measures can also be extended to the context of time. In QFTs, these quantities can be computed via analytic continuation of the twist operator correlation functions.

\subsection{Entanglement wedge and its cross section for spacelike subregions}
The entanglement wedge is the bulk region that is dual to the reduced density matrix $\rho_A$. For two subregions $A$ and $B$ located on a common Cauchy surface $\Sigma_t$, the corresponding RT surface can exhibit two distinct phases: a connected phase and a disconnected phase. The definition of EWCS is based on the connected phase of the entanglement wedge and is believed to capture the quantum correlations between $A$ and $B$.
In this paper, we focus mainly on the AdS$_3$/CFT$_2$ correspondence.
\begin{figure}[htbp]
  \centering
  \includegraphics[width=0.7\textwidth]{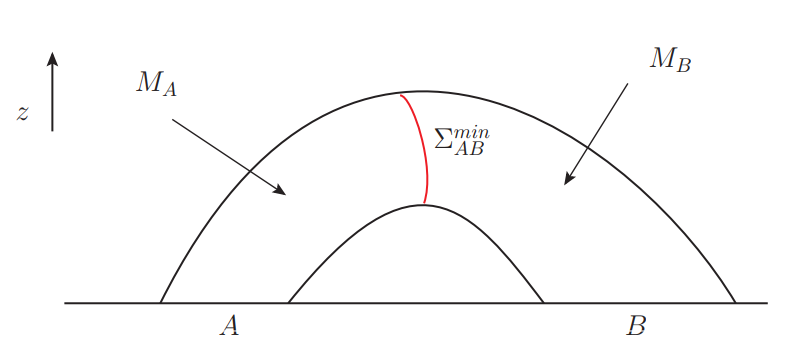}
  \caption{Illustration of the entanglement wedge and its cross section for two subregions $A$ and $B$. The RT surface is in the connected phase, and $\Sigma_{AB}^{\text{min}}$ (red line) denotes the EWCS.
}\label{fig_EW_spacelike}
\end{figure}

For the spacelike case, consider two disjoint subsystems $A$ and $B$ on the boundary of AdS$_3$. Let $\Gamma_{AB}$ denote the RT surface associated with the union $A \cup B$. The entanglement wedge $M_{AB}$ is then defined as the bulk region enclosed by the boundary subregions $A$, $B$, and the corresponding RT surface $\Gamma_{AB}$:
\bea
\partial M_{AB} = A \cup B \cup \Gamma_{AB}.
\eea
Within $M_{AB}$, consider the minimal surface $\Sigma_{AB}$ that divides $M_{AB}$ into two subregions, $M_A$ and $M_B$, such that $A \subset M_A, B \subset M_B$. This surface $\Sigma_{AB}$ is called the EWCS (EWCS). The setup is illustrated in Fig.~\ref{fig_EW_spacelike}, which helps visualize the definitions and geometric structure.

The EWCS $\Sigma_{AB}^{\text{min}}$ is defined as the minimal area surface among all possible surfaces $\Sigma_{AB}$ within the entanglement wedge that divides $M_{AB}$ into two regions $M_A$ and $M_B$, which contain $A$ and $B$, respectively. 
Operationally, this reduces to finding the minimal surface within the entanglement wedge that partitions it appropriately. The EWCS is then defined by
\bea
E_W(A:B) = \frac{\text{Area}(\Sigma_{AB}^{\text{min}})}{4G}.
\eea
This can be seen as an extension of the RT formula to bulk subregions.

\subsection{Timelike entanglement wedge cross section} \label{section_TEW}

In this section, our aim is to extend the definition of the EWCS from the spacelike case to the timelike case. The overall structure of the definition remains largely similar to that of the spacelike scenario. However, new features arise in the timelike setup, which we will elaborate on below. Before presenting the formal definition and construction, we first highlight some key similarities and differences between the spacelike and timelike cases.

First, we need to generalize the concept of the entanglement wedge to the timelike setting. The construction of the entanglement wedge is fundamentally based on the RT surfaces anchored to the boundary subregions. In the spacelike case, the entanglement wedge resides entirely within the Lorentzian geometry. However, for the timelike case, the RT surface is expected to extend into the complexified geometry \cite{Heller:2024whi}, see also Section~\ref{section_RT_surface}. Consequently, the timelike entanglement wedge may also contain components that go beyond the usual Lorentzian geometry. This observation plays an important role in our definition and derivation of its cross section later.

\begin{figure}[htbp]
  \centering
  \includegraphics[width=0.7\textwidth]{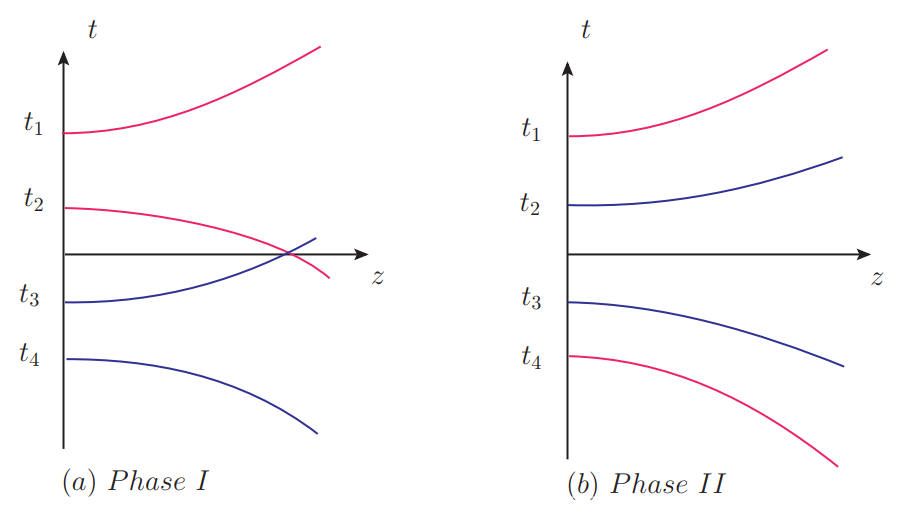}
  \caption{Illustration of two different phases of the bulk RT surfaces, with only the Lorentzian parts shown. (a) corresponds to Phase I, matching the CFT result in~(\ref{timelikeEE_two_interval_phaseI}); (b) corresponds to Phase II, matching~(\ref{timelikeEE_two_interval_phaseII}).}
  \label{twophases}
\end{figure}

As shown in Section~\ref{section_example_II}, the EE for the transition operator $T_{A_1A_2B_1B_2}$ can be computed by following the analytic function of the four-point function of twist operators. In this setup, the pairs $A_1B_1$ and $A_2B_2$ correspond to the time intervals $T_1 = [t_1, t_2]$ and $T_2 = [t_3, t_4]$, respectively. There exist two distinct phases for the EE, as demonstrated by our field-theoretic calculations in Section~\ref{section_example_II}. Then it is straightforward to identify the corresponding holographic RT surfaces for each phase, illustrated in Fig.~\ref{twophases}.

In phase I, the RT surfaces consist simply of the union of the RT surfaces for $A_1B_1$ and $A_2B_2$. This configuration is analogous to the spacelike case where two subregions are sufficiently far apart, in which the entanglement wedge becomes disconnected. However, the timelike configuration exhibits qualitatively different behavior: the RT surfaces corresponding to timelike-separated regions necessarily intersect or overlap in the bulk, as illustrated in Fig.~\ref{twophases}. This geometric overlap arises even when the intervals $T_1$ and $T_2$ are separated in time and are located far from each other. In such cases, the mutual information vanishes, indicating the absence of quantum correlations between the two subregions. However, since the regions remain causally connected, the bulk intersection of the RT surfaces may be related to the underlying causal structure. The entanglement wedge in this phase becomes homotopic to two disconnected disks (or circles), and there is no non-trivial minimal cross section. Consequently, we expect the timelike EWCS to vanish, consistent with the expectation that distant intervals do not share significant quantum entanglement.

In phase II, the entanglement wedge $M_{T_1T_2}$ is defined as the bulk region enclosed by the boundary intervals $T_1$, $T_2$, and the associated RT surface $\Gamma_{T_1T_2}$. In the timelike case, the EE is more appropriately understood as being defined for the subsystems $A_1B_1$ and $A_2B_2$, as discussed in Section~\ref{section_example_II}. Compared with the definition of the entanglement wedge for the spacelike case, it seems that we should consider the timelike entanglement wedge as the bulk region enclosed by RT surfaces and $A_1B_1$ and $A_2B_2$. However, since the subsystems $A_1B_1$ and $A_2B_2$ are associated with the boundary time intervals $T_1$ and $T_2$, we denote the entanglement wedge $M_{T_1T_2}$ with the boundary condition
\bea
\partial M_{T_1T_2} = T_1 \cup T_2 \cup \Gamma_{T_1T_2}.
\eea
To fully account for the RT surface in this phase, its extension must be included in the complexified geometry. In this phase, the mutual information is non-vanishing, indicating the presence of correlations between $A_1B_1$ and $A_2B_2$. In the following, we will examine concrete examples to demonstrate that it is possible to define a cross section of the timelike entanglement wedge, which captures the quantum correlations between the subsystems and may serve as the bulk dual of the reflected entropy.

Note that the definition of the timelike entanglement wedge should primarily be viewed as a proposal to construct certain bulk geometric quantities. It remains unclear whether the physical interpretation of timelike entanglement wedge fully parallels that of the spacelike case. In the spacelike setting, the entanglement wedge plays a crucial role in understanding the bulk reconstruction theorem, which underlies the subregion/subregion duality in AdS/CFT \cite{Dong:2016eik}. However, we do not expect the timelike entanglement wedge to fulfill the same function. The bulk reconstruction theorem relies on the equivalence between bulk and boundary relative entropies. Yet, in the timelike context, the relevant object is the transition operator $T_{AB}$, which is generally non-Hermitian. Consequently, there is no well-defined notion of relative entropy for such operators, and the standard arguments for reconstruction no longer apply. We therefore expect the timelike case to be fundamentally different from its spacelike counterpart. However, it is worthwhile to explore the potential physical interpretation and implications of the timelike entanglement wedge. We will offer further comments on this topic in Section~\ref{section_summary}.

Finally, we define the cross section of the timelike entanglement wedge. Within $M_{T_1T_2}$, consider a surface $\Sigma_{T_1T_2}$ that divides the wedge into two parts. Unlike in the spacelike case, where the minimal surface $\Sigma_{AB}$ (as shown in Fig.~\ref{fig_EW_spacelike}) is always spacelike and therefore has a real area. The surface $\Sigma_{T_1T_2}$ in the timelike setting may be timelike or even lie within the complexified geometry. As a result, the area of $\Sigma_{T_1T_2}$ can become complex-valued. This observation suggests that the notion of minimality may no longer be appropriate in timelike context, and a more general extremality or stationarity condition may be needed.

In this paper, we focus mainly on the AdS$_3$ case. In this setting, the surface $\Sigma_{T_1T_2}$ can be parameterized by two complex parameters $\lambda_1$ and $\lambda_2$, which label the points on the RT surfaces anchored to the boundary intervals. The area of $\Sigma_{T_1T_2}$ then becomes a function of these parameters:
\bea
\mathcal{A}_{T_1T_2}(\lambda_1,\lambda_2) := \text{Area}(\Sigma_{T_1T_2}).
\eea
We propose to define the cross section of the timelike entanglement wedge by imposing the following stationarity condition:
\bea\label{stationary_condition}
\partial_{\lambda_1} \mathcal{A}_{T_1T_2}(\lambda_1,\lambda_2) = \partial_{\lambda_2} \mathcal{A}_{T_1T_2}(\lambda_1,\lambda_2) = 0.
\eea
The surface $\Sigma_{T_1T_2}$ that satisfies this condition is denoted $\Sigma_{T_1T_2}^{s}$. The corresponding cross section of the timelike entanglement wedge is then defined as
\bea
E_W(T_1 : T_2) = \frac{\text{Area}(\Sigma_{T_1T_2}^{s})}{4G}.
\eea

The function $\mathcal{A}_{T_1T_2}(\lambda_1,\lambda_2)$ is, in general, a complex analytic function of the complex parameters $\lambda_1$ and $\lambda_2$. Therefore, the stationarity condition~(\ref{stationary_condition}) does not necessarily correspond to minimizing the modulus $|\mathcal{A}_{T_1T_2}|$. However, in the spacelike case, the area function is always real and positive, and the stationarity condition naturally reduces to the minimization of the area. In this way, the standard definition of the spacelike EWCS emerges as a special case of~(\ref{stationary_condition}).
\footnote{If multiple solutions to the stationarity condition exist, one should select the appropriate surface based on further physical or geometric principles. In the examples considered in this paper, we find only one solution that is physically reasonable and consistent with the expected properties.
}

It is also worth noting that a similar stationarity condition has been used in \cite{Doi:2023zaf} to determine the RT surface for timelike intervals. This suggests that the condition may originate from a more fundamental principle. In the following, we will explicitly illustrate the above construction through concrete examples in AdS$_3$.

\subsubsection{Pure AdS\texorpdfstring{$_3$}{3} Poincare coordinate}
\begin{figure}[htbp]
  \centering
  \includegraphics[width=0.7\textwidth]{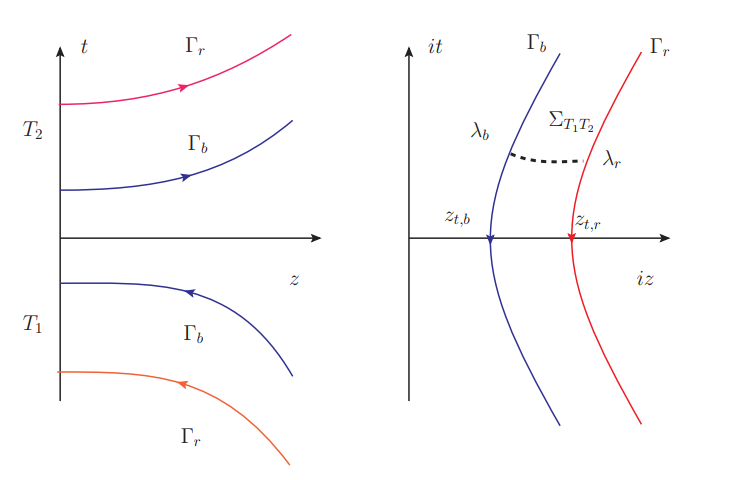}
  \caption{Figure to show the connected phase for two time intervals $T_1$ and $T_2$. $\Gamma_a\cup \Gamma_b$ is the RT surface for the two time intervals. $\lambda_b$ and $\lambda_r$ are the parameters on the RT surface. $\Sigma_{T_1T_2}$ is the extreme surface in the entanglement wedge. $z_{t,b}$ and $z_{t,r}$ denotes the turning points of the RT surface in the complexified geometry.}
  \label{connectedphase}
\end{figure}

Let us first consider the two time intervals in the vacuum state. The bulk metric in the Poincare coordinate is 
\bea
ds^2=\frac{-dt^2+dz^2+dx^2}{z^2}.
\eea
Without loss of generality we will choose symmetric intervals with $-t_1=t_4=\Delta t_1$ and $-t_2=t_3=\Delta t_2$. Just like the spacelike EE, the timelike EE for two intervals also have two phases similar to the spacelike EE regarding the values of $\Delta t_1$ and $\Delta t_2$. We will consider the connected phase with $\frac{\Delta t_1}{\Delta t_2}>3+2\sqrt{2}$. We can write down the equations for the geodesic lines:
\bea
&&\Gamma_r:\ z=\sqrt{t^2-\Delta t_1^2},\nn \\
&&\Gamma_b:\ z=\sqrt{t^2-\Delta t_2^2},
\eea
which are shown in Fig.~\ref{connectedphase}. By using RT formula, the above geodesic lines can give the correct timelike EE for the two intervals. The imaginary parts are given by the geodesic line in the complex geometry. 

In the connected phase, the minimal surface is shown in Fig.~\ref{connectedphase}. There exists a bulk region surrounded by the minimal surface $\Sigma_{AB}$ (blue and red lines in Fig.~\ref{connectedphase}) and $AB$. 
Within the entanglement wedge one could find extreme surfaces $\Sigma_{T_1T_2}$ ending on the RT surface $\Gamma_a\cup \Gamma_b$. The parameters of the ending points are denoted $\lambda_r$ and $\lambda_b$ in Fig.~\ref{connectedphase}. It should be noted that the extreme surface $\Sigma_{AB}$ may also be complex, and thus its area $\mathcal{A}(\lambda_r,\lambda_b):=\text{Area}(\Sigma_{AB})$ may be complex valued. In the spacelike case, an important geometric quantity is the minimal cross section of entanglement wedge, which is supposed to be related to many interesting information quantities in field theory.  In the timelike case among all the extreme surfaces $\Sigma_{AB}$ we would like to choose the stationary one using the condition~(\ref{stationary_condition}).  
We can obtain the cross section of the timelike entanglement wedge as
\bea
E_W(T_1,T_2)=\frac{\text{Area}(\Sigma_{T_1T_2}^{s})}{4G}.
\eea 
For the case we consider here, we can choose the time $t$ as the parameters of $\Gamma_r$ and $\Gamma_b$, that is, $\lambda_r=t_r$ and $\lambda_b=t_b$. $\Sigma_{AB}$ is given by the geodesic line $t=t(z)$ connecting $(t,z)=(t_r,\sqrt{t_r^2-\Delta t_1^2})$ and $(t,z)=(t_b,\sqrt{t_b^2-\Delta t_2^2})$. $\Sigma_{AB}$ satisfies the relation
\bea
\frac{t'(z)^2}{z^2(1-t'(z)^2)}=p^2,
\eea
where $p$ is the constant, which is given by
\bea
p^2=\frac{4 \left(t_r-t_b\right)^2}{-2 \Delta t_1^2 \left(2 t_b \left(t_r-t_b\right)+\Delta t_2^2\right)+4 \Delta t_2^2 t_r \left(t_r-t_b\right)+\Delta t_1^4+\Delta t_2^4}.\nn
\eea
With this equation we can obtain the area function $\mathcal{A}(t_r,t_b)$ of $\Sigma_{AB}$. The expression of $\mathcal{A}(t_r,t_b)$ could be found in the Appendix~\ref{Poincare_crosssection}. We can also show that 
\bea
\p_{t_r}\mathcal{A}(0,0)=\p_{t_b}\mathcal{A}(0,0)=0.
\eea
The saddle point is given by $t_r=t_b=0$, which are the turning points of $\Gamma_r$ and $\Gamma_b$ on the complexified geometry. The geodesic line $\Sigma_{AB}^{\text{s}}$ is given by the geodesic line connecting these two turning points.  It is straightforward to calculate the length of the $\Sigma_{AB}^{\text{s}}$,
\bea\label{TEWCS_vacuum}
E_W(T_1:T_2)=\frac{\mathcal{A}(0,0)}{4G}=\frac{1}{4G}\log\frac{\Delta t_1}{\Delta t_2}.
\eea
\subsubsection{Global coordinate}
In Section~\ref{section_AdS3_RT}, we discuss the RT surface for a single time interval in global coordinates, obtained by an analytic continuation procedure. The details of the RT surface construction are provided in Appendix~\ref{Appendix_global}.

In this section, we consider two time intervals in a constant spatial slice $\theta = \theta_0$. Specifically, we take the intervals to be $T_1 = [t_1, t_2]$ and $T_2 = [t_3, t_4]$, with the relations $-t_1 = t_4 = \Delta t_1$ and $-t_2 = t_3 = \Delta t_2$. The coordinates of the endpoints of these intervals on the AdS$_3$ boundary are
$$(t_1, r_\infty, \theta_0),\quad (t_2, r_\infty, \theta_0),\quad (t_3, r_\infty, \theta_0),\quad (t_4, r_\infty, \theta_0).$$

We are particularly interested in the connected phase of the RT surfaces, where it is possible to construct the complete RT surface configuration and evaluate the area of the extremal surface $\Sigma_{T_1T_2}$. Detailed expressions and calculations are provided in the Appendix~\ref{Appendix_global_EWCS}.

We can determine the stationary point by solving the condition in~(\ref{stationary_condition}), which gives the geodesic $\Sigma_{T_1T_2}^s$. The timelike EWCS in global coordinates is then given by
\bea
E_W(T_1:T_2) = \frac{1}{4G} \log \left( \frac{\tan\left( \frac{\Delta t_1}{2} \right)}{\tan\left( \frac{\Delta t_2}{2} \right)} \right).
\eea


\subsubsection{BTZ black hole}
Now let us consider a single interval in the thermal state, which is dual to the planar BTZ black hole:
\bea
ds^2=\frac{-f(z)dt^2+\frac{dz^2}{f(z)}+dx^2}{z^2},
\eea
with $f(z)=1-\frac{z^2}{z_H^2}$. The black hole horizon is at $z=z_H$, which is related to the inverse temperature $\beta=2\pi z_H$.

\begin{figure}[htbp]
  \centering
  \includegraphics[width=0.4\textwidth]{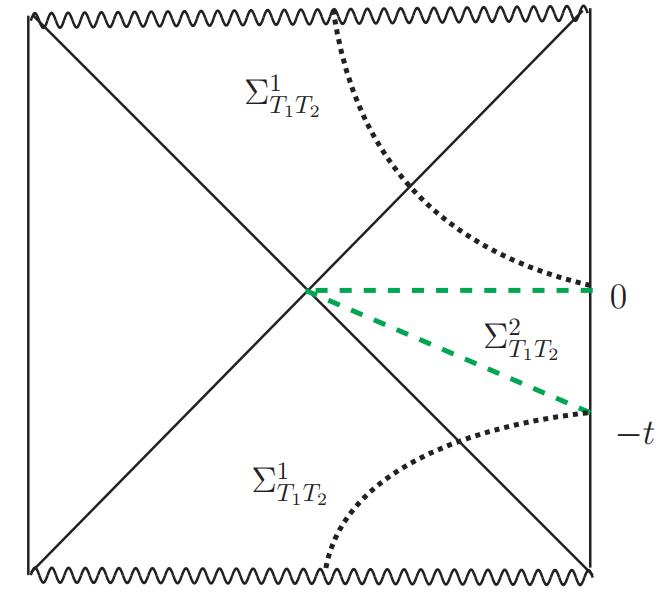}
  \caption{The EWCS for the intervals $T_1 = [-t, 0]$ and $T_2 = (-\infty, -t] \cup [0, +\infty)$. Two types of geodesics are shown: $\Sigma^1_{T_1T_2}$ (black dotted line) and $\Sigma^2_{T_1T_2}$ (green dashed line).}
  \label{blackhole}
\end{figure}

Consider that the timelike interval $T_1=[-t,0]$, $T_2$ is taken to be its complement. According to the spirit of EWCS, we need to separate extreme surface diving in the entanglement wedge into two parts, including $T_1$ and $T_2$ separately.
We can define the extreme surface $\Sigma_{T_1T_2}$, which anchored at the boundary of $T_1$ and $T_2$, satisfying
\bea
\frac{f(z)^3 t'(z)^2}{(1-f(z)^2 t'(z)^2)z^2}=p^2,
\eea
where $p$ is a constant. 
Similar as in the spacelike case there exist two kinds of surface. One is the surface with $p\ne 0$ labeled by $\Sigma_{T_1T_2}^{(1)}$, which starts from the boundary $z=\delta$ and ends on the future and past singularity. The other is the surface labeled by $\Sigma_{T_1T_2}^{(2)}$, which corresponds to $p=0$ or $t$ being constant. See Fig.~\ref{blackhole} for an illustration. For the geodesic line $\Sigma_{T_1T_2}^{(2)}$, the length is given by
\bea
L^{(2)}=2\log\frac{\beta}{\delta \pi}.
\eea
For $\Sigma_{T_1T_2}^{(1)}$ one could vary the ending points on the past and future singularities. One could obtain the  stationary one $\Sigma_{T_1T_2}^{(1),\text{s}}$, which is discussed in \cite{Doi:2023zaf}. The length is
\bea
L^{(1)}=2\log \frac{\beta}{\pi \delta}\sinh\left(\frac{\pi t}{\beta}\right).
\eea
Note that in this example, the area of the two surfaces are real and positive.
It is expected to define the cross section of TEW in this case as
\bea
E_W(T_1:T_2)=\frac{1}{4G}\text{min}\{L^{(1)},L^{(2)} \}.
\eea


Let us consider two intervals in planar BTZ black hole, to simplify the calculation, we also consider the symmetry time intervals $T_1=[t_1,t_2]$ and $T_2=[t_3,t_4]$ with $-t_1 = t_4 = \Delta t_1$ and $-t_2=t_3=\Delta t_2$.

We will consider the connected phase for the RT surfaces.
We can write down the RT surface for two time intervals by using the result in the Appendix~\ref{section_RT_BTZ}:
\bea
&& \Gamma_r:\ z_r = z_H \sqrt{\frac{\cosh(\frac{2t_r}{z_H})-\cosh(\frac{\Delta t_1}{2 z_H})}{1+ \cosh(\frac{2t_r}{z_H})}},\nn \\
&& \Gamma_b:\ z_b = z_H \sqrt{\frac{\cosh(\frac{2t_b}{z_H})-\cosh(\frac{\Delta t_2}{2 z_H})}{1+ \cosh(\frac{2t_b}{z_H})}},
\eea
where $\Gamma_r\cup \Gamma_b$ is the RT surface similar to the Poincare case. We can choose the time $t_r$ and $t_b$ as parameters to evaluate the length of $\Sigma_{T_1T_2}$. The stationary point of the length of $\Sigma_{T_1T_2}$ gives the solution $t_r=t_b=0$. The timelike EWCS is given by
\bea\label{TEWCS_thermal}
E_W(T_1,T_2) =\frac{1}{4G} \log \left( \frac{\tanh(\frac{\Delta t_1}{2z_H})}{\tanh(\frac{\Delta t_2}{2z_H})} \right).
\eea
The calculation details can be found in the Appendix~\ref{Appendix_global_EWCS}


\section{Entanglement measures for timelike subregions}\label{section_EM}

EE is one of the most fundamental measures for characterizing quantum entanglement between different regions. For a pure state, EE is sufficient, as it captures all relevant entanglement properties. However, in the case of a mixed state, EE also includes contributions from thermal entropy, making it less effective in isolating quantum correlations. In such cases, alternative measures, such as logarithmic negativity and reflected entropy, can be more appropriate for characterizing quantum entanglement.

If the notion of entanglement in time is meaningful, it is natural to ask whether these alternative measures can also be extended to the timelike setting. In this section, we show how to generalize such entanglement measures to the timelike case. In general, these measures involve certain operations on the transition operator $T_{AB}$ , which can also be formulated within the Schwinger–Keldysh formalism. Consequently, the problem reduces to computing timelike correlation functions of twist operators.

\subsection{Negativity and real-time replica method}

In spacelike situation, negativity is an effective measure of mixed quantum correlation \cite{Calabrese:2012nk}. Consider the Hilbert space $\mathcal{H}_A \otimes \mathcal{H}_B$ with bases $\ket{e_i^{(A)}}$ and $\ket{e_j^{(B)}}$. The partial transpose of the density matrix $\rho_{AB}$ is defined as
\bea\label{partial_transpose}
\bra{e_i^{(A)} e_j^{(B)}} \rho^{T_B} \ket{e_k^{(A)} e_l^{(B)}} = \bra{e_i^{(A)} e_l^{(B)}} \rho \ket{e_k^{(A)} e_j^{(B)}}.
\eea 
The logarithmic negativity is defined as
\bea
\mathcal{E} \equiv \log || \rho_{AB}^{T_B} || = \log \Tr |\rho_{AB}^{T_B}|,
\eea
where the norm $||M||$ is defined as $||M|| := \sum |\lambda_i|$ and $\lambda_i$ are eigenvalues of operator $M$. $\mathcal{E}$ could be evaluated by
\bea\label{negativity_definition}
\mathcal{E} = \lim_{n_e \to 1} \log \Tr (\rho_{AB}^{T_B})^{n_e}, 
\eea
where $n_e$ is an even integer. In QFTs this expression provides a method to calculate negativity using a replica trick \cite{Calabrese:2012nk}.

In $(1+1)$D QFTs, consider two intervals in the time slice $t=0$ with $A=[x_1,x_2]$ $B=[x_3,x_4]$. logarithmic negativity could be calculated using~(\ref{negativity_definition}). 
$\Tr (\rho_{AB}^{T_B})^{n_e}$ is usually calculated by the path integral,
which can be translated to the correlator of the twist operators:
\bea
\Tr (\rho_{AB}^{T_B})^n = \braket{\sigma_n(0,x_1) \tilde{\sigma}_n(0,x_2) \tilde{\sigma}_n(0,x_3) \sigma_n(0,x_4)}.
\eea
The details can be found in the Appendix~\ref{negativity-replica-trick-section}.

The definition of logarithmic negativity can be generalized to the timelike case. Now we consider the four subregions $A_1,B_1,A_2,B_2$ as shown in Fig.~\ref{two_interval_write}.  $T_{A_1A_2B_1B_2}$ can be seen as operator in the Hilbert space $H_{A_1}\otimes H_{A_2}\otimes H_{B_1} \otimes H_{B_2}$ on the basis $|e^{(A_1)}_{i_1}\rangle$, $|e^{(A_2)}_{i_2}\rangle$ and $|e^{(B_1)}_{j_1}\rangle$ ,$|e^{(B_2)}_{j_2}\rangle$. Thus, we could expand the operator $T_{A_1A_2B_1B_2}$ in these bases. Now we could define the transpose operation similar to~(\ref{partial_transpose}). We can take $A$ as the subregions $A_1B_1$ and $B$ as $A_2B_2$. Then we can define the partial transpose on $B:=A_2B_2$, $T_{A_1A_2B_1B_2}^{T_B}:=T_{A_1A_2B_1B_2}^{T_{A_2B_2}}$.
Therefore, the logarithmic negativity can be defined as
\bea
\mathcal{E}_T=\lim_{n_e\to 1}\log \Tr (T_{A_1A_2B_1B_2}^{T_{A_2B_2}})^{n_e}.
\eea

\begin{figure}[htbp]
  \centering
  \includegraphics[width=0.8\textwidth]{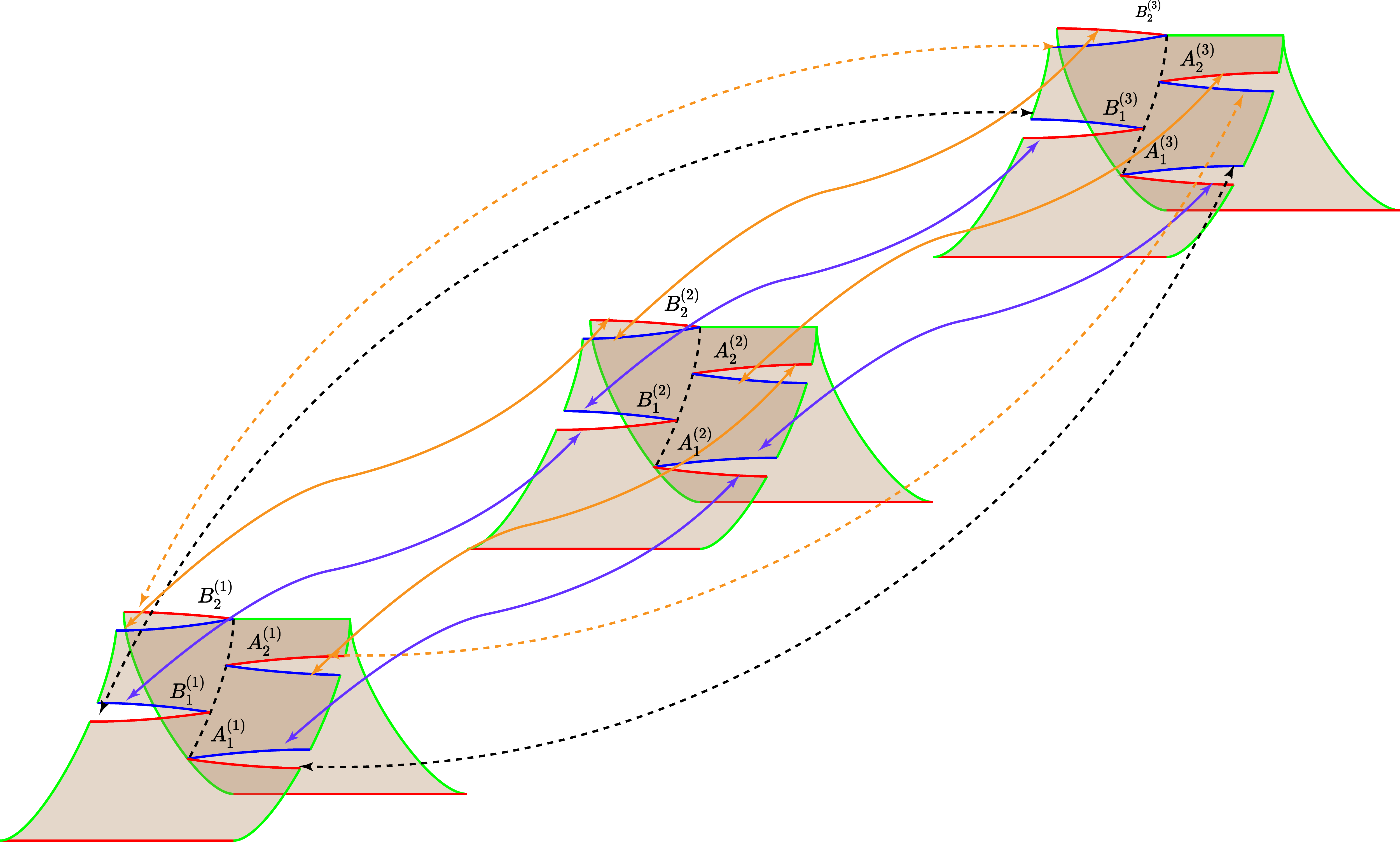}
  \caption{Real-time replica method for computing the logarithmic negativity. The blue lines represent the original upper boundaries, while the red lines indicate the original lower boundaries. The blue boundary of $A^{(i)}$ (or $B^{(i)}$) is sewn to the red boundary of $A^{(i+1)}$ (or $B^{(i+1)}$), respectively. The orange lines denote the sewing of boundary conditions corresponding to the partially transposed intervals.}
  \label{timelike-negativity-replica-trick}
\end{figure}

It is also possible to evaluate $\mathcal{E}_T$ by using SK path integral formalism. In Section~\ref{section_example_II} we represent the transition operator by the path integral shown in Fig.~\ref{timelike-negativity-replica-trick}. The operation of partial transpose in $A_2B_2$ just changes the boundary condition in the subregions $A_2$ and $B_2$ as shown in Fig.~\ref{timelike-negativity-replica-trick}. To translate the evaluation to twist operators, one could introduce the reversed partial transpose $T_{A_1A_2B_1B_2}^{C_{A_2B_2}}=C T_{A_1A_2B_1B_2}^{T_{A_2B_2}} C$, where $C$ is the operation that reverses the order of the indices on the lower and upper cut. It is obvious that $C^2=1$.  Thus we have
\bea
\Tr(T_{A_1A_2B_1B_2}^{C_{A_2B_2}})^n=\Tr(T_{A_1A_2B_1B_2}^{T_{A_2B_2}})^n.
\eea
The left hand side of the above equation can be translated to the path integral with gluing the boundary of the cuts, see the Appendix~\ref{negativity-replica-trick-section}. The method of gluing is different from $\Tr(T_{A_1A_2B_1B_2})^n$. As a result, 
\bea\label{negativity_time_twist}
\Tr(T_{A_1A_2B_1B_2}^{T_{A_2B_2}})^n=\braket{\sigma_n(t_1,0) \tilde{\sigma}_n(t_2,0) \tilde{\sigma}_n(t_3,0) \sigma_n(t_4,0)}.
\eea

Thus, the problem of evaluating logarithm negativity for the timelike case is translated to evaluate the correlation function of twist operators~(\ref{negativity_time_twist}), which can be obtained by analytical continuation of Euclidean correlators. The correlation function can be computed in some models such as free compacitified boson and critical Ising models. It is also interesting to explore the results for holographic CFTs \cite{Kulaxizi:2014nma,Kudler-Flam:2018qjo,Kusuki:2019zsp}. Some studies exist on the above correlators. However, it is pointed out in recent paper \cite{Dong:2024gud} that there may exist some subtle in the discussions. We leave the studies on details of the logarithm negativity in specific models for future work.

A specific case is the adjacent time intervals, that is, to take the limit $t_2\to t_3$. In this limit we would have
\bea
\Tr(T_{A_1A_2B_1B_2}^{T_{A_2B_2}})^n\propto\langle \sigma_n(t_1,0) \tilde{\sigma}^{2}_n(t_2,0) \sigma_n(t_4,0)\rangle,
\eea
where $\tilde{\sigma}^{2}_n$ is the local operator with scaling dimension $h_{\tilde{\sigma}^2_{n_e}}=\bar h_{\tilde{\sigma}^2_{n_e}}=\frac{c}{12}(\frac{n_e}{2}-\frac{2}{n_e})$. The three-point correlation functions are universal. The logarithm negativity is obtained in the limit $t_2\to t_3$:
\bea
\mathcal{E}_T=\frac{c}{4}\log\frac{2(t_2-t_1)(t_4-t_2)}{(t_4-t_1)}+\text{constant}.
\eea
In this case the logarithm negativity is related to the timelike mutual information~(\ref{mutual_information_definition}) 
\bea 
\mathcal{E}_T=\frac{3}{4}\lim_{t_2\to t_3}I(A_1B_1;A_2B_2)+\text{constant},
\eea
where ``constant'' is associated with the coupling constant.

\subsection{Reflected entropy via analytical continuation}

Reflected entropy is another important quantity that captures quantum correlations between subregions $A$ and $B$. It is also expected to be holographically dual to the EWCS. Let us first review the definition of reflected entropy for two spacelike subregions lying on a Cauchy surface at $t=0$.

Consider a density matrix $\rho_{AB}$ defined in a bipartite Hilbert space $\mathcal{H}_A \otimes \mathcal{H}_B$. Let $\{\ket{i}_A\}$ and $\{\ket{i}_B\}$ denote orthonormal bases of $\mathcal{H}_A$ and $\mathcal{H}_B$, respectively. A generic basis state in $\mathcal{H}_A \otimes \mathcal{H}_B$ can be expressed in the Schmidt decomposition form as
\bea
\ket{\phi_a} = \sum_i \sqrt{\lambda_a^i} \ket{i_a}_A \ket{i_a}_B,
\eea
where $\lambda_a^i$ are the Schmidt coefficients associated with the state $\ket{\phi_a}$, satisfying the normalization condition $\sum_i \lambda_a^i = 1$.

The density matrix $\rho_{AB}$ can then be written as
\bea
\rho_{AB} = \sum_{a,i,j} p_a \sqrt{\lambda_a^i \lambda_a^j} \ket{i_a}_A \ket{i_a}_B \bra{j_a}_A \bra{j_a}_B,
\eea
where $p_a$ are the classical probabilities (eigenvalues), satisfying $\sum_a p_a = 1$.

To define the reflected entropy, we consider a purification of $\rho_{AB}$. This can be done by promoting $\bra{j_a}_{A(B)}$ to $\ket{j_a}_{A^*(B^*)}$ in auxiliary Hilbert spaces $\mathcal{H}_{A(B)}^*$. The purified state in $\mathcal{H}_A \otimes \mathcal{H}_B \otimes \mathcal{H}_A^* \otimes \mathcal{H}_B^*$ is then given by
\bea\label{purification_rho_AB}
\ket{\sqrt{\rho_{AB}}} := \sum_{a,i,j} \sqrt{p_a \lambda_a^i \lambda_a^j} \ket{i_a}_A \ket{i_a}_B \ket{j_a}_{A^*} \ket{j_a}_{B^*}.
\eea
This is the so-called canonical purification of $\rho_{AB}$, in the sense that
\bea
\Tr_{A^* B^*} \ket{\sqrt{\rho_{AB}}} \bra{\sqrt{\rho_{AB}}} = \rho_{AB}.
\eea
The reflected entropy $S_R(A:B)$ for $\rho_{AB}$ is defined by
\bea\label{reflected_entropy_definition}
\rho_{AA^*} := \Tr_{BB^*} \ket{\sqrt{\rho_{AB}}} \bra{\sqrt{\rho_{AB}}}, \nn \\
S_R(A:B) := - \Tr_{AA^*} [\rho_{AA^*} \log \rho_{AA^*}].
\eea
In QFTs one could prepare the reduced density matrix $\rho_{AB}$ as well as its purification $|\sqrt{\rho_{AB}}\rangle$ by path integral.  Specifically, for $(1+1)D$ CFTs, the reflected entropy can be computed via the replica trick. Consider two intervals $A=[x_1,x_2]$, $B=[x_3,x_4]$ and $x_1 < x_2 < x_3 < x_4$. As discussed in \cite{Dutta:2019gen}, the reflected entropy is given by
\begin{align}
S_R(A:B) =& \lim_{n \to 1} S_n(AA^*)_{\psi_m}\nn \\
=& \lim_{n \to 1} \frac{1}{1-n} \log \frac{\braket{\sigma_{g_A}(0,x_1) \sigma_{g_A^{-1}}(0,x_2) \sigma_{g_B}(0,x_3) \sigma_{g_B^{-1}}(0,x_4)}}{\left(\braket{\sigma_{g_m}(0,x_1) \sigma_{g_m^{-1}}(0,x_2) \sigma_{g_m}(0,x_3) \sigma_{g_m^{-1}}(0,x_4)}\right)^n},\nn \\
~
\end{align}
where the twist operators $\sigma_{g_A(g_B)}$ implement cyclic permutations in the replica manifold and depend on the replica symmetry group. The conformal dimension of the twist operators is given by 
\bea
h_{g_A^{-1}}=h_{g_B}=\frac{cn(m^2-1)}{24m}.
\eea
One could refer to \cite{Dutta:2019gen} for more details.

In the previous section, we have shown that the EE and logarithmic negativity can be obtained via analytical continuation of twist operator correlation functions into the timelike regime. Therefore, it is natural to expect that the reflected entropy for two time intervals can also be computed using a similar analytic continuation procedure. We will discuss the subtleties of the replica method in the timelike case in the next section.

For the timelike case, we still consider the two time intervals $T_1=[t_1,t_2]$ and $T_2=[t_3,t_4]$ with $t_1<t_2<t_3<t_4$.

To generalize to Lorentzian or complexified coordinates, we promote $x_i$ to complex variables $w_i = x_i + i \tau_i$, with corresponding antiholomorphic coordinates $\bar{w}_i = x_i - i \tau_i$ . The correlation function becomes
\begin{multline}\label{reflected_entropy_twist}
S_R(A:B) = \lim_{n \to 1} \frac{1}{1-n} \log \Bigg[ \\
\frac{\braket{\sigma_{g_A}(w_1,\bar{w}_1) \sigma_{g_A^{-1}}(w_2,\bar{w}_2) \sigma_{g_B}(w_3,\bar{w}_3) \sigma_{g_B^{-1}}(w_4,\bar{w}_4)}}
{\left(\braket{\sigma_{g_m}(w_1,\bar{w}_1) \sigma_{g_m^{-1}}(w_2,\bar{w}_2) \sigma_{g_m}(w_3,\bar{w}_3) \sigma_{g_m^{-1}}(w_4,\bar{w}_4)}\right)^n} \Bigg].
\end{multline}
Expanding the four-point function in the t-channel and keeping only the dominant conformal block, we obtain the reflected entropy
\bea
S_R(A:B) \approx \frac{c}{6}\left( \log \left( \frac{1+\sqrt{\eta}}{1-\sqrt{\eta}} \right) +  \log \left( \frac{1+\sqrt{\bar{\eta}}}{1-\sqrt{\bar{\eta}}} \right) \right),
\eea
where $\eta$ and $\bar \eta$ are cross ratios:
\bea
\eta := \frac{(w_1 - w_2)(w_3 - w_4)}{(w_1 - w_3)(w_2 - w_4)},\ \bar{\eta} := \frac{(\bar{w}_1 - \bar{w}_2)(\bar{w}_3 - \bar{w}_4)}{(\bar{w}_1 - \bar{w}_3)(\bar{w}_2 - \bar{w}_4)}.
\eea

In the timelike setting, let the intervals be $A=[t_1,t_2], B = [t_3,t_4]$ with ordering $t_1<t_2<t_3<t_4$. We would like to analytically continue the  correlation functions to the timelike case. The reflected entropy for the timelike case is
\bea\label{reflected_entropy_definition_analytical}
S_R(T_1:T_2):= S_R(A:B)|_{\tau_i\to it_i+\epsilon_i},
\eea
with $\epsilon_1>\epsilon_2>\epsilon_3>\epsilon_4>0$. 
Then the cross ratios in the Lorentzian spacetime become 
\bea
&&\eta_t  = \frac{(t_1 - t_2)(t_3 - t_4)}{(t_1 - t_3)(t_2 - t_4)}+i\tilde{\epsilon},\nn \\
&&\bar \eta_t  = \frac{(t_1 - t_2)(t_3 - t_4)}{(t_1 - t_3)(t_2 - t_4)}-i\tilde{\epsilon},
\eea
where $\tilde{\epsilon}$ depends $\epsilon_i$ and $t_i$, which is vanishing in the limit $\epsilon_i\to 0$.

Choosing a symmetric configuration such that $-t_1=t_4 = \Delta t_1$ and $-t_2 = t_3 = \Delta t_2$, we find
\bea\label{SRTT_vacuum}
S_R(T_1:T_2) = \frac{c}{3} \log \frac{\Delta t_1}{\Delta t_2}.
\eea

In a thermal state with temperature $\beta^{-1}$, we apply the conformal map $\xi = \exp(-\frac{2\pi w}{\beta}), \bar{\xi}=\exp(-\frac{2\pi \bar{w}}{\beta})$ to evaluate $S_R(A:B)$~(\ref{reflected_entropy_twist}) in the thermal state.  
The cross ratios in the timelike case becomes
\bea
&&\eta_t = \frac{(e^{\frac{2 \pi \Delta t_1}{\beta}}-e^{\frac{2 \pi \Delta t_2}{\beta}})^2}{(e^{\frac{2 \pi (\Delta t_1 + \Delta t_2)}{\beta}}-1)^2}+i\tilde{\epsilon}',\nn \\
&&\bar \eta_t = \frac{(e^{\frac{2 \pi \Delta t_1}{\beta}}-e^{\frac{2 \pi \Delta t_2}{\beta}})^2}{(e^{\frac{2 \pi (\Delta t_1 + \Delta t_2)}{\beta}}-1)^2}-i\tilde{\epsilon}'.
\eea
Thus we obtain $S_R(T_1:T_2)$ in the thermal state:
\bea\label{SRTT_thermal}
S_R(T_1:T_2) = \frac{c}{3} \log \frac{\tanh \frac{\pi \Delta t_1}{\beta}}{\tanh \frac{\pi \Delta t_2}{\beta}}.
\eea

It is remarkable that the analytical continuation result $S_R(T_1:T_2)$ is related to the timelike EWCS defined and computed in Section~\ref{Section_EWCS}. More precisely, by comparing~(\ref{SRTT_vacuum}) with~(\ref{TEWCS_vacuum}) and~(\ref{SRTT_thermal}) with~(\ref{TEWCS_thermal}), and using the Brown-Henneaux relation $c = \frac{3}{2G}$, we find the following correspondence at leading order in $G$:
\bea
S_R(T_1:T_2) = 2 E_W(T_1:T_2).
\eea
This provides supporting evidence for the generalization of the holographic correspondence of reflected entropy to the timelike case.

\subsubsection{Comment on timelike reflected entropy via replica}\label{section_comment}

In the previous section, we find that the analytical continuation of $S_R(A:B)$ to two timelike intervals may provide a candidate CFT dual of the timelike EWCS. However, we have not yet established how to derive the timelike reflected entropy via the real-time replica method. There are several subtleties and challenges associated with this problem.

A key step in the spacelike construction of reflected entropy is the canonical purification of the reduced density matrix $\rho_{AB}$. Naturally, a similar purification procedure might be expected for the transition operator $T_{A_1A_2B_1B_2}$. As discussed in the previous section, this transition operator can be expanded in the Hilbert space $H_{A_1} \otimes H_{A_2} \otimes H_{B_1} \otimes H_{B_2}$ with bases $\ket{e^{(A_1)}_{i_1}}, \ket{e^{(A_2)}_{i_2}}, \ket{e^{(B_1)}_{j_1}}, \ket{e^{(B_2)}_{j_2}}$. To define a purification, we may introduce an auxiliary Hilbert space $H^\star_{A_1} \otimes H^\star_{A_2} \otimes H^\star_{B_1} \otimes H^\star_{B_2}$ with the corresponding basis states $\ket{e^{(A_1^\star)}_{i_1}}, \ket{e^{(A_2^\star)}_{i_2}}, \ket{e^{(B_1^\star)}_{j_1}}, \ket{e^{(B_2^\star)}_{j_2}}$.

For a positive operator such as $\rho_{AB}$, one can unambiguously define its square root $\sqrt{\rho_{AB}}$, and thus obtain the canonical purification via the procedure in~(\ref{purification_rho_AB}). However, the transition operator $T_{A_1A_2B_1B_2}$ is generically non-Hermitian. For a non-Hermitian matrix $M$, a square root may not always exist, and even if it does, it is generally non-unique. For example, if $M$ is diagonalizable, i.e., $M = S D S^{-1}$ with $D$ diagonal and $S$ invertible, then one may define a square root as $B = \pm S \sqrt{D} S^{-1}$, but this construction is highly non-unique and can be degenerate.

This ambiguity makes the definition of a ``canonical purification'' for non-Hermitian operators far more subtle than in the Hermitian case. A possible strategy is to consider evaluating $T_{A_1A_2B_1B_2}^{m/2}$ for even integers $m$, followed by analytic continuation to $m \to 1$. This approach could potentially be used to prepare a purified state for $T_{A_1A_2B_1B_2}$ using the Schwinger-Keldysh formalism. However, even if we define a purified state $\ket{\sqrt{T_{A_1A_2B_1B_2}}}$, it remains unclear whether we should define the reflected entropy $S_R(T_1:T_2)$ using the density matrix
\bea
\rho_{\text{reflected}} = \ket{\sqrt{T_{A_1A_2B_1B_2}}} \bra{\sqrt{T_{A_1A_2B_1B_2}}},
\eea
as in the standard definition~(\ref{reflected_entropy_definition}). If this is the case, then $S_R(T_1:T_2)$ would be interpreted as the EE of a reduced density matrix, and hence would always be real and non-negative. Although we indeed find $S_R(T_1:T_2) > 0$ for the vacuum state, this may not hold for more general states due to the non-Hermiticity of the transition operator.

In summary, there are substantial conceptual and technical obstacles to generalizing the canonical purification and the definition of reflected entropy to the timelike case, primarily due to the non-Hermitian nature of the transition operator. In this paper, we merely propose~(\ref{reflected_entropy_definition_analytical}) as a candidate definition of timelike reflected entropy via analytical continuation of twist operator correlators. The fundamental formulation and interpretation of timelike reflected entropy remain open questions that require further investigation.

\section{Summary and discussion}\label{section_summary}

In this paper, we investigate the concept of entanglement in time in both QFTs and their holographic duals. Previous studies have primarily focused on quantum correlations between spacelike-separated subregions, leading to fruitful results that have significantly advanced our understanding of entanglement structures in QFTs and the holographic properties of gravity, particularly through the duality between entanglement measures and bulk geometry.

The idea of defining a ``spacetime density matrix'' provides a natural generalization of entanglement to subregions that are timelike separated \cite{Milekhin:2025ycm}. This framework enables the definition of a transition operator $T_{AB}$, which is generally non-Hermitian in the timelike case and reduces to the standard reduced density matrix in the spacelike limit. The construction is fundamental and applicable to arbitrary quantum systems.

Our main focus is to further develop the notion of entanglement in time within QFTs. The transition operator $T_{AB}$ can be prepared using the Schwinger–Keldysh formalism. This allows us to define entropy-related quantities, such as the pseudo Rényi entropy and the von Neumann-like EE associated with $T_{AB}$. These quantities are computable using the real-time replica method, as demonstrated in Section~\ref{Section_entanglement_in_time}, thereby providing a concrete and well-defined notion of timelike EE.

In QFTs, evaluating these entropies is equivalent to computing correlation functions of twist operators with timelike separations. In this paper, we have considered several simple examples to illustrate this framework. However, the same setup is flexible enough to accommodate more general states. For instance, one could prepare the initial state as an excited state (see, e.g. \cite{Calabrese:2006rx,Calabrese:2007mtj,He:2014mwa}), and study the corresponding timelike EE during its dynamical evolution. It would be particularly interesting to explore the novel features that arise in such scenarios.

In this paper, we have primarily focused on the case where the subsystems $A$ and $B$ are semi-infinite, motivated by the goal of relating the EE of the transition operator $T_{AB}$ to the timelike EE defined for time intervals in \cite{Doi:2022iyj}. However, more general choices of subregions $A$ and $B$ are possible. The EE for $T_{AB}$ in such cases can be expressed in terms of more general correlation functions of twist operators. We anticipate that new features and phenomena will emerge in these extended settings. As a simple extension of Example I in Section~\ref{section_example_I}, one may consider both $A$ and $B$ to be the semi-infinite interval $[0,+\infty]$, located on different Cauchy surfaces. Analyzing such general configurations would require a more detailed understanding of correlation functions in Lorentzian spacetime. 

In Section~\ref{section_RT_surface}, we turn to the holographic dual description of timelike EE. At present, the holographic duality of timelike EE is not yet fully understood. However, several existing studies~\cite{Doi:2022iyj,Li:2022tsv,Doi:2023zaf,Heller:2024whi,Guo:2025pru} offer valuable hints. In particular, since timelike EE can, in some cases, be computed via analytical continuation of correlation functions of twist operators, it is natural to expect that a corresponding holographic dual exists for the associated RT surface.

In this paper, we adopt this perspective and construct the corresponding RT surfaces in both AdS$_3$ and higher-dimensional examples. There is some recent progress in exploring holographic timelike EE in more general backgrounds; see, e.g.,~\cite{Afrasiar:2024lsi,Giataganas:2025div,Nunez:2025gxq}. We compare analytical results with numerical computations and find excellent agreement in the cases considered. Furthermore, we observe novel features of timelike EE in AdS black hole backgrounds, especially in dimensions $d>3$.
For instance, in Section~\ref{section_higher_d_black_hole}, we investigate the case of an AdS$_5$ black hole and find that the timelike EE is real-valued. However, it differs qualitatively from the spacelike EE. While spacelike EE captures thermal entropy and increases monotonically with the size of the subregion (e.g., a strip), the perturbative calculation of timelike EE may exhibit non-monotonic behavior at finite order. Nonetheless, the full numerical results indicate that the true behavior is smooth and monotonic. In other dimensions, where the timelike EE is generally complex-valued, the real or imaginary parts similarly reflect smooth dependence on the interval size. Understanding these features may provide valuable insight into the physical meaning of timelike entanglement.

On the quantum field theory side, the quantity $\Tr(T_{AB}^n)$ can be formulated via the Schwinger–Keldysh path integral. According to the real-time AdS/CFT correspondence, such a path integral should admit a bulk dual in terms of an appropriate gravitational configuration. Inspired by the Lewkowycz–Maldacena (LM) procedure to derive the RT formula in the Euclidean signature \cite{Lewkowycz:2013nqa}, and by the real-time replica method to prove the HRT formula in Lorentzian signature \cite{Dong:2016hjy}, it is plausible that there is a similar derivation for the holographic dual of timelike EE. This would not only support the validity of our proposal but also deepen our understanding of real-time holography in the AdS/CFT correspondence.

The entanglement wedge and its cross section are important concepts for characterizing entanglement between spacelike subregions. In this paper, we explore how these concepts can be extended to the timelike case. Since the RT surface for timelike-separated regions differs significantly from that in the spacelike case, the definition of the timelike entanglement wedge must be appropriately generalized. The EWCS, which may be dual to certain boundary entanglement measures, also requires reinterpretation in this context. We define the cross section as a surface that is the stationary point of the area function of $\Sigma_{AB}$ within the generalized entanglement wedge, see Section~\ref{section_TEW}. Similarities and differences between the timelike and spacelike cases are discussed. In Section~\ref{Section_EWCS}, we present explicit examples in AdS$_3$, where we find that the timelike EWCS remains positive.

\begin{figure}[htbp]
  \centering
  \includegraphics[width=0.8\textwidth]{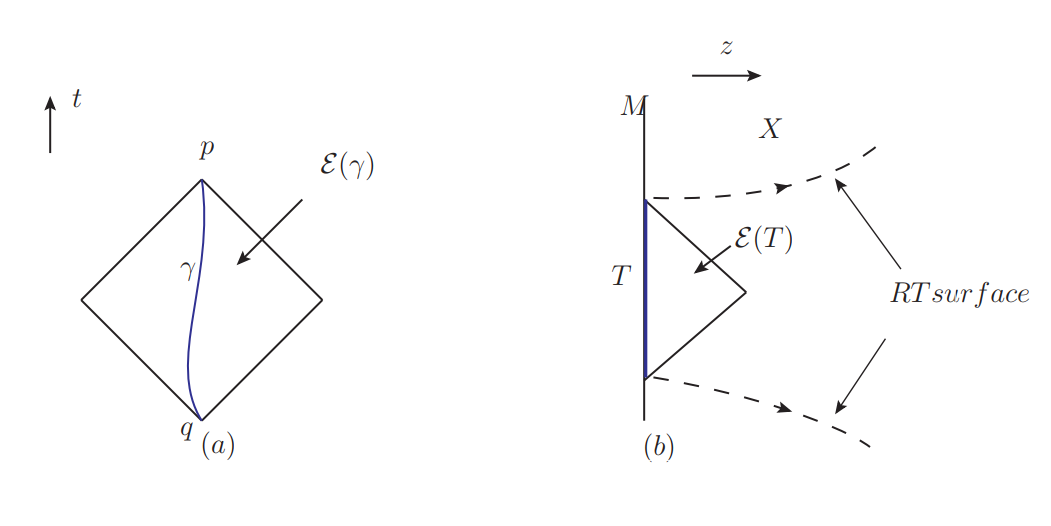}
  \caption{Figure illustrating the concept of a timelike tube. The coordinates $t$ and $z$ represent the time and holographic directions, respectively. (a) The timelike tube is constructed by deforming the timelike curve $\gamma$ while keeping the endpoints $q$ and $p$ fixed. (b) The timelike tube associated with a boundary time interval $T \subset M$ is shown. Here, $X$ denotes the bulk spacetime, and $\mathcal{E}(T)$ represents the timelike tube corresponding to $T$ in $X$. The RT surface associated with the interval $T$ is indicated by the dashed line.}
  \label{timelike_tube_fig}
\end{figure}

It is of interest to investigate whether the timelike entanglement wedge plays a similar role to its spacelike counterpart. We expect the answer to be negative. The reconstruction theorem established in \cite{Dong:2016eik} relies on the equivalence between bulk and boundary relative entropy \cite{Jafferis:2015del}. However, for the transition operator, which is generally non-Hermitian, there is no well-defined notion of relative entropy. This fundamental distinction suggests that the timelike entanglement wedge differs substantially from the spacelike case.Nonetheless, certain expected structural properties can still be observed from the perspective of operator algebras. In \cite{Strohmaier:2023hhy,Strohmaier:2023opz,Witten:2023qsv}, the author discusses the timelike tube theorem in curved spacetime. A timelike tube $\mathcal{E}(\gamma)$ associated with a timelike curve $\gamma$ consists of all points that can be reached by deforming timelike curves with fixed endpoints, as illustrated in Fig.~\ref{timelike_tube_fig}(a). In Section 3.4 of \cite{Witten:2023qsv}, a connection is drawn between the timelike tube theorem and the AdS/CFT reconstruction theorem.

Specifically, for a time interval $T$ on the boundary CFT, one can identify a corresponding timelike tube $\mathcal{E}(T)$ in the bulk, as shown in Fig.~\ref{timelike_tube_fig}(b). The timelike tube theorem then asserts
\bea\label{timelike_tube}
\mathcal{R}(T) = \mathcal{R}(\mathcal{E}(T)),
\eea
where $\mathcal{R}(\cdot)$ denotes the operator algebra associated with the specified region. Notably, the timelike tube $\mathcal{E}(T)$ in AdS coincides with the causal wedge of $T$. Hence, the timelike tube theorem~(\ref{timelike_tube}) is equivalent to the causal wedge reconstruction theorem in the context of AdS/CFT \cite{Hamilton:2006az,Hamilton:2006fh,Kabat:2011rz}.

In our study of timelike EE, we focus on a time interval $T$ on the boundary CFT. From the bulk perspective, the corresponding RT surface does not lie within the causal wedge but instead extends into the deep infrared (IR) region of the bulk. This suggests that the transition operator associated with the interval $T$ captures information beyond the causal wedge or the timelike tube defined by $T$. This observation also supports the viewpoint that timelike EE is more naturally defined between two subregions $A$ and $B$, as discussed in Section~\ref{section_example_I}, rather than solely in terms of a single time interval $T$. However, in our construction, the timelike entanglement wedge  is bounded by the interval $T$ and its associated bulk RT surface. From this perspective, the timelike entanglement wedge does not appear to serve the same role as its spacelike counterpart in the reconstruction theorem. It would be interesting to further investigate the new potential role of timelike entanglement wedges in holography.

Finally, in Section~\ref{section_EM}, we briefly discuss other entanglement measures in time. We mainly focus on two important quantities: logarithmic negativity and reflected entropy. For the negativity, we outline how to perform the partial transpose and implement the replica method to define the timelike version of logarithmic negativity. A similar procedure can be applied to other related quantities, such as the odd entropy \cite{Tamaoka:2018ned}. Unfortunately, we have not carried out explicit computations, as we are primarily interested in holographic CFTs, for which the evaluation of logarithmic negativity remains challenging, especially in the large-$c$ limit. Previous studies of the spacelike case can be found in \cite{Kulaxizi:2014nma,Kudler-Flam:2018qjo,Kusuki:2019zsp}. However, a recent work \cite{Dong:2024gud} points out that the computations in \cite{Kudler-Flam:2018qjo,Kusuki:2019zsp} may fail to capture the dominant contributions to the four-point function of twist operators. A more careful analysis of these four-point functions, particularly in the context of holographic CFTs, is an important direction for future investigation. Timelike logarithmic negativity can also be computed in spin chain models using numerical methods. It would be interesting to explore these computations in future work and investigate whether new features of timelike entanglement emerge in such models.

For reflected entropy, we define its timelike counterpart by analytically continuing the correlation functions of twist operators from the Euclidean to Lorentzian signature. We find that the result is closely related to the timelike EWCS defined in this paper. This provides supporting evidence for the existence of a holographic dual of reflected entropy in the timelike setting and offers a consistent, well-motivated justification for our definition of timelike entanglement wedge and its cross section. However, it remains unclear how to generalize the definition of reflected entropy based on purification to the timelike case. The main difficulty lies in purifying a non-Hermitian transition operator $T_{A_1A_2B_1B_2}$. We make several comments on this issue in Section~\ref{section_comment}. 

In summary, the notion of entanglement in time broadens our understanding of the entanglement structure and the holographic aspects of gravity. Although many questions remain unresolved or not yet fully understood, there are strong indications that several key results established for the spacelike case can be extended to the timelike setting. This extension may offer new insights into the role of time in holography.
~\\
~\\

{\bf Acknowledgements}
We would like to thank Bin Chen, Yun-Gui Gong, Song He, Jian-Xin Lu, Hong Lu,  Rong-Xin Miao, Carlos Nunez, Yi Pang, Dibakar Roychowdhury, Fu-Wen Shu, Jie-qiang Wu, Run-qiu Yang, Hong-bao Zhang, Jiaju Zhang, Yang Zhou and  Yu-Xuan Zhang for useful discussions.
WZG is supposed by the National Natural Science Foundation of China under Grant No.12005070 and the Hubei Provincial Natural Science Foundation of China under Grant No.2025AFB557.


\appendix

\section{Mutual information}\label{Mutualinformation_section}
In this section we would like to introduce the details on the Example II discussed in Section~\ref{section_example_II}. We need to evaluate the four-point functions of twist operators in Lorentzian spacetime~(\ref{ExampleII_trTn}), which can be obtained by analytical continuation of the Euclidean correlators \cite{Hartman:2013mia}. In $2D$ Euclidean CFTs it is more convenient to use the coordinates $w=x+ i\tau$ and $\bar w=x-i\tau$. In this section the twist operator will be denoted as $\sigma(\bar w,\bar w)$.

Specifically, we are interested in the four-point function:
\bea\label{trace_AB}
&&\langle \sigma_n(w_1,\bar w_1)
\tilde{\sigma}_n(w_2,\bar w_2)\sigma_n(w_3,\bar w_3)\tilde{\sigma}_n(w_4,\bar w_4)\rangle.\nn 
\eea
By conformal transformation we have
\bea\label{EE_conformal_block}
&&\langle \sigma_n(w_1,\bar w_1)
\tilde{\sigma}_n(w_2,\bar w_2)\sigma_n(w_3,\bar w_3)\tilde{\sigma}_n(w_4,\bar w_4)\rangle\nn \\
&&=\frac{1}{|w_1-w_3|^{4h_{\sigma_n}}|w_2-w_4|^{4h_{\sigma_n}}} \langle \sigma_n(\infty)
\tilde{\sigma}_n(1,1)\sigma_n(\eta,\bar \eta)\tilde{\sigma}_n(0,0)\rangle,
\eea
where cross ratios $\eta$ and $\bar \eta$ are given by 
\bea\label{mutual_four}
\eta=\frac{(w_1-w_2)(w_3-w_4)}{(w_1-w_3)(w_2-w_4)},\quad \bar \eta=\frac{(\bar w_1-\bar w_2)(\bar w_3-\bar w_4)}{(\bar w_1-\bar w_3)(\bar w_2-\bar w_4)}.
\eea
The four-point functions in~(\ref{EE_conformal_block}) can be decomposed into conformal blocks:
\bea\label{conformal_block_expansion}
&&\langle \sigma_n(\infty)
\tilde{\sigma}_n(1,1)\sigma_n(\eta,\bar \eta)\tilde{\sigma}_n(0,0)\rangle\nn \\
&&=\sum_p a_p \mathcal{F}(c,h_{\sigma_n},h_p,\eta)\bar{\mathcal{F}}(c,\bar h_{\sigma_n},\bar h_p,\eta),
\eea
where the summation is over the primary operator with conformal dimension $h_p$ and $\bar h_p$, $a_p$ are the OPE coefficients.
We can also define the timelike mutual information for subsystems $A_1B_1$ and $A_2B_2$:
\bea
I(A_1B_1;A_2B_2)=S(T_{A_1B_1})+S(T_{A_2B_2})-S(T_{A_1B_1A_2B_2}).\nn
\eea
In the vacuum state it is known that the mutual information in conformal invariant, that is it depends on only the cross ratios $\eta$ and $\bar \eta$. One could directly check this by using the definition of mutual information and the results~(\ref{mutual_four}). 


In this paper, we mainly focus on the CFTs with large central charge, which are expected to have bulk dual. The domain contribution of the expansion~(\ref{conformal_block_expansion}) comes from the primary operators with the lowest conformal dimensions. 

Let us firslty focus on the EE. To evaluate EE we only need the correlator~(\ref{conformal_block_expansion}) near $n\sim 1$. At large $c$ limit, for $\eta\sim 0$, the domaint contributions of the expansion of~(\ref{conformal_block_expansion}) in the s-channel comes from the vacuum family $h_p=0$, that is
\bea\label{mutual_approximation}
\langle \sigma_n(\infty)
\tilde{\sigma}_n(1,1)\sigma_n(\eta,\bar \eta)\tilde{\sigma}_n(0,0)\rangle\simeq e^{-f(h_{\sigma_n},\eta)-\bar f(\bar h_{\sigma_n},\bar \eta)},\nn
\eea
where $f$ and $\bar f$ are functions of order $O(c)$. Almost $n\sim 1$ one could analytically obtain $f$ and $\bar f$ by the monodromy method \cite{Hartman:2013mia}. Keeping the leading order of $(n-1)$ we have
\bea\label{schannel}
&&f(h_{\sigma_n},\eta)=-\frac{c(n-1)}{6}\log \eta+O((n-1)^2),\nn \\
&& \bar f(\bar h_{\sigma_n},\bar \eta)=-\frac{c(n-1)}{6}\log \bar \eta+O((n-1)^2).
\eea
While for $\eta\sim 1$ we can expand the conformal block by t-channel. In this case the function $f$ and $\bar f$ near $n\sim 1$ can also be obtained analytically, the result is
\bea\label{tchannel}
&&f(h_{\sigma_n},\eta)=-\frac{c(n-1)}{6}\log (1-\eta)+O((n-1)^2),\nn \\
&& \bar f(\bar h_{\sigma_n},\bar \eta)=-\frac{c(n-1)}{6}\log(1- \bar \eta)+O((n-1)^2).
\eea
Taking above result into~(\ref{EE_conformal_block}) one could obtain the four-point correlation functions. For the spacelike case, it is straightforward to obtain the EE which is consistent with the holographic results. 

Now we would like to discuss the case in the view of correlators of twist operators. Consider the EE for $T_{A_1A_2B_1B_2}$ for the intervals on the time coordinates as shown in Fig.~\ref{two_interval_write}. The timelike EE can be obtained by the following analytical continuation:
\bea\label{timelimeEE_analytical}
&&S(T_{A_1A_2B_1B_2})\nn \\
&&=-\partial_n\langle \sigma_n(w_1,\bar w_1)
\tilde{\sigma}_n(w_2,\bar w_2)\sigma_n(w_3,\bar w_3)\tilde{\sigma}_n(w_4,\bar w_4)\rangle|_{\tau_i\to i t_i +\epsilon_i,x_i\to 0},\nn
\eea
with $\epsilon_1>\epsilon_2>\epsilon_3>\epsilon_4>0$. With the analytical continuation we find the cross ratios become
\bea\label{cross_ratio_timelike}
&& \eta_t:=\frac{(t_1-t_2)(t_3-t_4)}{(t_1-t_3)(t_2-t_4)}+i \tilde{\epsilon}\nn \\
&& \bar \eta_t:=\frac{(t_1-t_2)(t_3-t_4)}{(t_1-t_3)(t_2-t_4)}-i \tilde{\epsilon},
\eea
where $\tilde{\epsilon}$ are constant depending on $\epsilon_i$ and the coordinates $t_i$ which vanishing at the limit $\epsilon_i \to 0$. We should note that both $\eta_t$,$\bar \eta_t$ and $1-\eta_t$, $1-\bar \eta_t$ have a positive real part.  

For the large $c$ theory, we can also evaluate the four-point functions by the approximation~(\ref{mutual_approximation}). For $\eta_t \sim 0$, we expect that the s-channel will give the dominant contributions to the correlator $\langle \sigma_n(\infty)
\tilde{\sigma}_n(1,1)\sigma_n(\eta_t,\bar \eta_t)\tilde{\sigma}_n(0,0)\rangle$. Thus, according to~(\ref{schannel}) the correlator $\langle \sigma_n(\infty)
\tilde{\sigma}_n(1,1)\sigma_n(\eta,\bar \eta)\tilde{\sigma}_n(0,0)\rangle$ would be real. For $\eta_t\sim 1$, we expect that the t-channel would domain the contribution. According to~(\ref{tchannel}) this correlator will also be real. Thus in both phases the imaginary part of the timelike EE of two intervals comes from the factor $$\frac{1}{|w_1-w_3|^{4h_{\sigma_n}}|w_2-w_4|^{4h_{\sigma_n}}}$$ in~(\ref{EE_conformal_block}). 

With some calculations, for $\eta_t \sim 0$ by using~(\ref{EE_conformal_block}) and~(\ref{timelimeEE_analytical}) we find
\bea \label{timlikeEE_two_interval_phase1}
S(T_{A_1A_2B_1B_2})=\frac{c}{3}log\frac{t_2-t_1}{\delta}+\frac{c}{3}\log \frac{t_4-t_3}{\delta}+\frac{i\pi c}{3}.
\eea
For $\eta_t \sim 1$ one could obtain
\bea\label{timlikeEE_two_interval_phase2}
S(T_{A_1A_2B_1B_2})=\frac{c}{3}log\frac{t_3-t_2}{\delta}+\frac{c}{3}\log \frac{t_4-t_1}{\delta}+\frac{i\pi c}{3}.
\eea
Note that the imaginary part of both phases is the constant $\frac{i\pi c}{3}$, which is double the case of the single interval. The result is also consistent with the holographic explanation of timelike EE in \cite{Doi:2023zaf}. The mutual information can be evaluated using the results of $S(T_{A_1B_1})$ and $S(T_{A_2B_2})$. It is obvious that the mutual information is real.

\section{Replica trick to calculate negativity}\label{negativity-replica-trick-section}

In this appendix, we detail the replica trick used to calculate logarithmic negativity via twist operator correlation functions. The method involves computing the traces of integer powers of the partially transposed density matrix, $\rho_{AB}^{T_B}$. These traces take different forms depending on whether the power $n$ is an even integer ($n_e$) or an odd integer ($n_o$):
\bea
\Tr (\rho_{AB}^{T_B})^{n_e} = \sum_{\lambda_i > 0} |\lambda_i|^{n_e} + \sum_{\lambda_i < 0} |\lambda_i|^{n_e}, \label{even-power} \\
\Tr (\rho_{AB}^{T_B})^{n_o} = \sum_{\lambda_i > 0} |\lambda_i|^{n_o} - \sum_{\lambda_i < 0} |\lambda_i|^{n_o}. \label{odd-power}
\eea
Here, $\lambda_i$ are the eigenvalues of $\rho_{AB}^{T_B}$. The trace norm required for negativity is obtained by analytically continuing the expression for even powers, $\Tr(\rho_{AB}^{T_B})^{n_e}$, to the limit $n_e \to 1$. This leads to the replica limit formula
\bea
\mathcal{E} = \lim_{n_e \to 1} \log \Tr(\rho_{AB}^{T_B})^{n_e}.
\eea


\begin{figure}[htbp]
  \centering
  \includegraphics[width=0.5\textwidth]{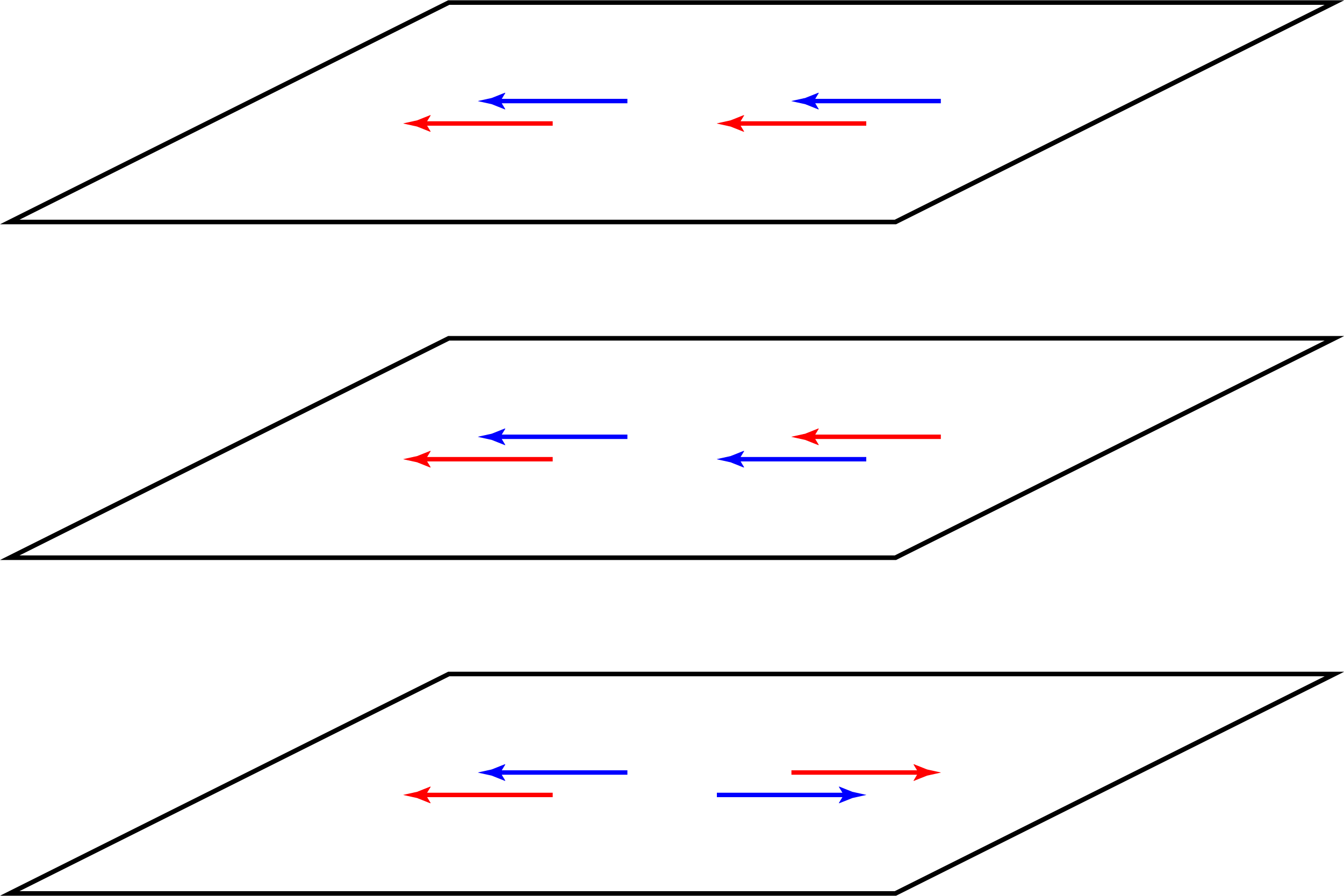}
  \caption{Illustration of the replica manifold connections. Top: The sheet structure for $\Tr(\rho_{AB}^n)$. Middle: For $\Tr(\rho_{AB}^{T_B})^n$, the connections along the cut for subsystem $B$ are crossed. Bottom: Applying the $C$ operation flips the orientation of the entire $B$ region, resulting in a configuration for $\Tr(\rho_{AB}^{C_B})^n$ that is often more symmetric.}
  \label{TC-show}
\end{figure}




In a quantum field theory path integral, $\Tr(\rho_{AB}^{T_B})^n$ is computed on an n-sheeted Riemann surface. As shown in \cite{Calabrese:2012nk}, the calculation is often simplified by introducing an operator $C$ that reverses the orientation of the cut along the region $B$ (see Fig.~\ref{TC-show}). This transformation modifies the matrix to $\rho_{AB}^{C_B} = C \rho_{AB}^{T_B} C$ but leaves the trace invariant: $\Tr(\rho_{AB}^{T_B})^n = \Tr(\rho_{AB}^{C_B})^n$. The utility of this transformation is that it can lead to a more symmetric arrangement of the twist operators that implement the replicated geometry.

For the case of two disjoint spacelike intervals, $A=[x_1, x_2]$ and $B=[x_3, x_4]$, this procedure yields a four-point function of twist ($\sigma_n$) and anti-twist ($\tilde{\sigma}_n$) operators:
\bea
\Tr (\rho_{AB}^{T_B})^n = \braket{\sigma_n(x_1) \tilde{\sigma}_n(x_2) \tilde{\sigma}_n(x_3) \sigma_n(x_4)}.
\eea


We apply a similar logic to the timelike configuration depicted in Fig.~\ref{timelike-negativity-replica-trick}. The intervals $A_2$ and $B_2$ are treated as finite timelike regions. The operation $C$, applied to region $B_2$, swaps the twist operators at its endpoints. Subsequently, we take a semi-infinite limit, pushing one endpoint of each interval to the far future. This procedure results in the final configuration used in our calculations, where the operator at the finite-time endpoint of $B_2$ is a twist operator $\sigma_n$, and the operator at the finite-time endpoint of $A_2$ is an anti-twist operator $\tilde{\sigma}_n$.

\section{RT surface of timelike interval in complexified metric}
Before calculating the timelike EWCS, we should get the form of the timelike RT surface. We use the complex geometric interpretation to calculate it.

\subsection{Global metric}\label{Appendix_global}
We begin by considering an interval on the boundary of global AdS$_3$. The endpoints are located at $(t_1, r_\infty, \theta_1)$ and $(t_2, r_\infty, \theta_2)$, with $t_2 > t_1$. The Lorentzian metric for global AdS$_3$ is given by
\bea
ds^2 =  -(1+r^2)dt^2 + \frac{dr^2}{1+r^2} + r^2 d\theta^2.
\eea
By performing a Wick rotation to Euclidean time, $t \to -i\tau$, the metric becomes
\bea
ds^2 =  (1+r^2)d\tau^2 + \frac{dr^2}{1+r^2} + r^2 d\theta^2.
\eea

To find the RT surface anchored to this interval, we parameterize it using the radial coordinate $r$, such that $\tau = \tau(r)$ and $\theta = \theta(r)$. The Euler-Lagrangian equations yield two conserved quantities, $p_\tau$ and $p_\theta$:
\bea 
p_\tau &= \frac{(1+r^2)^{3/2}\tau'}{\sqrt{1 + (1+r^2)^2\tau'^2 + (1+r^2)r^2\theta'^2}},\nn \\
p_\theta &= \frac{\sqrt{1+r^2}r^2 \theta'}{\sqrt{1 + (1+r^2)^2\tau'^2 + (1+r^2)r^2\theta'^2}}.
\eea 

In the complex geometry interpretation, we assume that the surface has a turning point at a complex radial coordinate $r=r_\tau$, where the derivatives $\tau'(r)$ and $\theta'(r)$ diverge. This leads to the following set of conditions:
\bea 
& p_\theta^2 - r_\tau^2 + p_\tau^2r_\tau^2 + p_\theta^2r_\tau^2 - r_\tau^4 = 0,\nn \\
& \int_{r_\infty}^{r_\tau} \tau' dr = \frac{\tau_1 - \tau_2}{2}, \nn \\
& \int_{r_\infty}^{r_\tau} \theta' dr = \frac{\theta_2 + \theta_1}{2}.
\eea 

Solving this system of equations yields
\bea 
&& p_\tau = \frac{\sinh(\tau_2 - \tau_1)}{\cos(\theta_2 + \theta_1) - \cosh(\tau_2 - \tau_1)},\nn \\
&& p_\theta = \frac{\sin(\theta_2 + \theta_1)}{\cos(\theta_2 + \theta_1) - \cosh(\tau_2 - \tau_1)},\nn \\
&& r_\tau = \sqrt{2} \frac{i \cos(\frac{\theta_2 + \theta_1}{2})}{\sqrt{\cos(\theta_2 + \theta_1) - \cosh(\tau_2 - \tau_1)}}.
\eea 

For a purely timelike interval, we can, without loss of generality, set the angular positions be equal $\theta_1=\theta_2=\theta_0=0$ and choose a symmetric time interval $-t_1=t_2=\Delta t$. After rotating back to the Lorentzian signature, the profile of the RT surface is described by the relation
\bea
r(t) = \frac{1}{|\cos(\Delta t)|\sqrt{\tan^2(t)-\tan^2(\Delta t)}}.
\eea

\subsection{RT Surface in the BTZ Black Hole}\label{section_RT_BTZ}

We now apply the same procedure to a timelike interval in the BTZ black hole background. The Lorentzian BTZ metric is given by
\begin{align}
ds^2 &= \frac{1}{z^2}\left( -f(z)\, dt^2 + \frac{dz^2}{f(z)} + dx^2 \right),
\end{align}
where $f(z) = 1 - \frac{z^2}{z_H^2}$ and $z_H$ are the positions of the event horizon. Wick rotating $t \to -i\tau$ gives the Euclidean metric
\begin{align}
ds^2 = \frac{1}{z^2}\left( f(z)\, d\tau^2 + \frac{dz^2}{f(z)} + dx^2 \right).
\end{align}

We consider a timelike interval on the boundary with endpoints at $(\tau_1, x_1)$ and $(\tau_2, x_2)$. To determine the RT surface anchored to these points, we parameterize it by the holographic coordinate $z$, such that $\tau = \tau(z)$ and $x = x(z)$. The Euler–Lagrange equations again yield two conserved quantities, $p_\tau$ and $p_x$:
\begin{align}
p_\tau &= \frac{f(z)\, \tau'}{z^2 \sqrt{f(z)\tau'^2 + x'^2 + \frac{1}{f(z)}}},\nn\\
p_x &= \frac{x'}{z^2 \sqrt{f(z)\tau'^2 + x'^2 + \frac{1}{f(z)}}}.
\end{align}

Assuming the surface has a unique turning point at $z = z_\tau$, where both $\tau'$ and $x'$ diverge, we impose the following conditions:
\bea
& p_x^2 z_\tau^4 + z_H^2 - z_\tau^2 \left[1 + z_H^2(p_\tau^2 + p_x^2)\right] = 0, \label{btz_constraint} \nn\\
& \int_0^{z_\tau} \tau'(z)\, dz = \frac{\tau_2 - \tau_1}{2},\nn\\
&\int_0^{z_\tau} x'(z)\, dz = \frac{x_2 - x_1}{2}.
\eea

Solving this system provides explicit expressions for the conserved quantities and the turning point:
\begin{align}
p_\tau &= \frac{\sin\left( \frac{\tau_2 - \tau_1}{z_H} \right)}{z_H\left[\cos\left( \frac{\tau_2 - \tau_1}{z_H} \right) - \cosh\left( \frac{x_2 - x_1}{z_H} \right)\right]}, \\
p_x &= \frac{\sinh\left( \frac{x_2 - x_1}{z_H} \right)}{z_H\left[\cosh\left( \frac{x_2 - x_1}{z_H} \right) - \cos\left( \frac{\tau_2 - \tau_1}{z_H} \right)\right]}, \\
z_\tau &= \sqrt{\frac{z_H \left[ \cosh\left( \frac{x_2 - x_1}{z_H} \right) - \cos\left( \frac{\tau_2 - \tau_1}{z_H} \right) \right]}{1 + \cosh\left( \frac{x_2 - x_1}{z_H} \right)}}.
\end{align}

To describe the RT surface in the original Lorentzian spacetime, we Wick rotate back via $\tau \to it$. For a symmetric, purely timelike interval defined by $-t_1 = t_2 = \Delta t$ and $x_1 = x_2 = 0$, the profile of the RT surface is
\begin{align}
z(t) = z_H \sqrt{\frac{\cosh\left( \frac{2t}{z_H} \right) - \cosh\left( \frac{2\Delta t}{z_H} \right)}{1 + \cosh\left( \frac{2t}{z_H} \right)}}.
\end{align}

This RT surface has a complex-valued turning point, consistent with the analytic continuation framework discussed in the main text.  Its shape is analogous to that of extremal surfaces in pure AdS, and its near-boundary divergence can be regularized by introducing a cutoff at $z = \delta$.

\section{Calculation of timelike EWCS}
In this appendix, we provide the detailed derivation of the timelike EWCS and discuss the procedure for determining the relevant saddle point.

\subsection{Poincaré coordinate}\label{Poincare_crosssection}

We consider two timelike symmetric intervals $A$ and $B$ with endpoints specified by $-t_1=t_4=\Delta t_1$ and $-t_2=t_3=\Delta t_2$ in the Poincaré patch of AdS$_3$. We denote the corresponding RT surfaces by $\Gamma_r$ for interval $A$ and $\Gamma_b$ for interval $B$. The timelike EWCS, denoted as $\Sigma_{AB}$, is defined as the geodesic connecting the point $(t_r,z_r)=(t_r,\sqrt{t_r^2-\Delta t_1^2})$ on $\Gamma_r$ and $(t_b,z_b)=(t_b,\sqrt{t_b^2-\Delta t_2^2})$ on $\Gamma_b$.

The length of the $\Sigma_{AB}$ is given by
\begin{align}
\text{Area}(\Sigma_{AB}) = \mathcal{A}(t_r,t_b)&= \int_{z_b}^{z_r} \frac{1}{z} \sqrt{1-(\frac{dt}{dz})^2} dz \nn \\
 &= \log(\frac{z}{1+ \sqrt{p^2 z^2 +1}})|_{z_b}^{z_r},\nn 
\end{align}
where the derivative of the geodesic satisfies $t'(z) = \frac{dt}{dz} = \frac{pz}{\sqrt{1+p^2z^2}}$ which follows from the relation $\frac{t'(z)^2}{z^2(1-t'(z)^2)}=p^2$. Parameter $p^2$ is determined by the boundary conditions and is explicitly given by
\bea
p^2 = \frac{4 \left(t_r-t_b\right)^2}{-2 \Delta t_1^2 \left(2 t_b \left(t_r-t_b\right)+\Delta t_2^2\right)+4 \Delta t_2^2 t_r \left(t_r-t_b\right)+\Delta t_1^4+\Delta t_2^4}\nn.
\eea

Substituting $z_r$, $z_b $ and $p^2$ into the expression for the area, we obtain
\bea
\mathcal{A}(t_r,t_b) = \log\left(\frac{\sqrt{t_r^2 - \Delta t_1^2} \left(1 + \sqrt{1 + p^2(t_b^2 - \Delta t_2^2)}\right)}{\sqrt{t_b^2 - \Delta t_2^2} \left(1 + \sqrt{1 + p^2(t_r^2 - \Delta t_1^2)}\right)}\right).
\eea

The saddle points are determined by solving the extremality conditions
\bea
\partial_{t_r} \mathcal{A}(t_r,t_b) = 0, \quad \partial_{t_b} \mathcal{A}(t_r,t_b) = 0.\nn
\eea
The solutions include
\bea
(t_r, t_b) = (0,0), (\pm \Delta t_1, \pm \Delta t_2), (\pm \Delta t_1, \mp \Delta t_2).
 \eea
 
In holography, the EWCS corresponds to the minimal-area surface connecting the two RT surfaces, capturing the correlations between the subsystems. Among the saddle points, $(t_r, t_b)=(0,0)$ yields the minimal geodesic length,and thus dominates in the semiclassical approximation. Therefore, the area of the timelike EWCS is
\bea
\mathcal{A}(0,0) = \log \frac{\Delta t_1}{\Delta t_2}.
\eea

\subsection{Global metric}\label{Appendix_global_EWCS}
To compute the length of a geodesic between two points in global AdS$_3$, it is often convenient to embed AdS$_3$ as a hyperboloid in a flat (2+2)-dimensional spacetime equipped with the metric
\bea
ds^2 = -dX_0^2 + dX_1^2 + dX_2^2 - dX_3^2,
\eea
where the signature is $(-++-)$. For two points $(t_r,r_r,\theta_0)$ and $(t_b,r_b,\theta_0)$ on RT surfaces, their positions in the embedding space are given by
\bea
&& X_0 = \sqrt{r^2+1} \cos(t), \nn \\
&& X_1 = r \cos(\theta), \nn \\
&& X_2 = r \sin(\theta), \nn \\
&& X_3 = \sqrt{r^2+1} \sin(t).
\eea

Let the embedding coordinates of these two points be denoted by $\mathrm{X}^{(r)}$ and $\mathrm{X}^{(b)}$. The geodesic distance $d$ between them in AdS$_3$ is then
\bea
d = \arccosh( -\mathrm{X}^{(r)} \cdot \mathrm{X}^{(b)} ).
\eea

We now consider two timelike symmetric intervals $A$ and $B$ with endpoints $-t_1=t_2=\Delta t_1$ and $-t_2=t_3=\Delta t_2$. The radial coordinates of the corresponding RT surfaces are obtained as
\bea
&& r_r = \frac{1}{|\cos(\Delta t_1)|\sqrt{\tan^2(t_r)-\tan^2(\Delta t_1)}},\nn \\
&& r_b = \frac{1}{|\cos(\Delta t_2)|\sqrt{\tan^2(t_b)-\tan^2(\Delta t_2)}}.
\eea

The length of the $\Sigma_{AB}$ is given by
\begin{align}
\text{Area}(\Sigma_{AB})
&= \mathcal{A}(t_r, t_b) \nn \\
&= \arccosh\Bigg( 2\cos(t_b - t_r) \sqrt{\frac{\cos^2(\Delta t_1)}{\cos(2 \Delta t_1) - \cos(2t_r)}} \nn \\
&\qquad\qquad \times \sqrt{\frac{\cos^2(\Delta t_2)}{\cos(2 \Delta t_2) - \cos(2t_b)}} \nn \\
&\qquad\qquad - \frac{1}{\sqrt{\tan^2(t_r)-\tan^2(\Delta t_1)}\sqrt{\tan^2(t_b)-\tan^2(\Delta t_2)}} \nn\\
&\qquad\qquad \times \frac{1}{\sqrt{\cos^2(\Delta t_1)\cos^2(\Delta t_2)}} \Bigg).
\end{align}

The saddle point is again determined by the extremality conditions $\partial_{t_r} \mathcal{A}(t_r, t_b) = 0$ and $\partial_{t_b} \mathcal{A}(t_r, t_b) = 0$, which yields $(t_r,t_b)=(0,0)$ as the dominant solution. Evaluating at the saddle point gives the area of the timelike EWCS
\begin{align}
\mathcal{A}(0,0) =& \arccosh \left( -\cot(\Delta t_1)\cot(\Delta t_2)+ \frac{1}{\sin(\Delta t_1)\sin(\Delta t_2)} \right) \nn\\
=& \log \left( \frac{\tan{\frac{\Delta t_1}{2}}}{\tan \frac{\Delta t_2}{2}} \right).
\end{align}

\subsection{BTZ black hole}\label{Appendix_BTZ_EWCS}

The thermal state in a $(1+1)$-dimensional CFT is dual to the planar BTZ black hole, whose metric is given by
\bea
ds^2 = \frac{-f(z) dt^2 + \frac{dz^2}{f(z)} + dx^2}{z^2},
\eea
where $f(z)=1-\frac{z^2}{z_H^2}$, and $z=z_H$ denotes the location of the black hole horizon.

We again consider two symmetric timelike intervals $A$ and $B$ with endpoints $-t_1=t_4=\Delta t_1$ and $-t_2=t_3=\Delta t_2$. The RT surfaces corresponding to these intervals, denoted by $\Gamma_r$ and $\Gamma_b$, have bulk points given by
\bea
&& z_r = z_H \sqrt{\frac{\cosh(\frac{2 t_r}{z_H})-\cosh(\frac{2 \Delta t_1}{z_H})}{1+\cosh(\frac{2 t_r}{z_H})}},\nn \\
&& z_b = z_H \sqrt{\frac{\cosh(\frac{2 t_b}{z_H})-\cosh(\frac{2 \Delta t_2}{z_H})}{1+\cosh(\frac{2 t_b}{z_H})}}.
\eea

To compute the length of the timelike  EWCS connecting these points, we embed the BTZ black hole into $\mathbb{R}^{2,2}$ with the coordinates
    \begin{align}
    X_0 =& \frac{z_H}{z} \sqrt{f(z)} \sinh(\frac{t}{z_H}), \nn \\
    X_1 =& \frac{z_H}{z} \cosh(\frac{x}{z_H}), \nn \\
    X_2 =& \frac{z_H}{z} \sinh(\frac{x}{z_H}), \nn \\
    X_3 =& \frac{z_H}{z} \sqrt{f(z)} \cosh(\frac{t}{z_H}).
    \end{align}

The embedding-space inner product between points $X^{(r)}$ and $X^{(b)}$ is
\begin{equation}
    X^{(r)} \cdot X^{(b)} = -X_0^{(r)}X_0^{(b)} + X_1^{(r)}X_1^{(b)} + X_2^{(r)}X_2^{(b)} -X_3^{(r)}X_3^{(b)}.
\end{equation}
The geodesic distance is therefore
\begin{equation}
    d = \arccosh(- X^{(r)} \cdot X^{(b)} ).
\end{equation}

Thus, the length of the $\Sigma_{AB}$ is
\begin{align}
\text{Area}(\Sigma_{AB})
&= \mathcal{A}(t_r, t_b) \nonumber \\
&= \operatorname{arccosh}\left(
    \frac{
        \begin{gathered}
            2\cosh\left(\frac{t_r+t_b}{z_H}\right) \cosh\left(\frac{\Delta t_1}{z_H}\right) \cosh\left(\frac{\Delta t_2}{z_H}\right) \\
            - 2\cosh\left(\frac{t_r}{z_H}\right) \cosh\left(\frac{t_b}{z_H}\right)
        \end{gathered}
        }{
        \begin{gathered}
            \sqrt{\cosh\left(\frac{2t_r}{z_H}\right) - \cosh\left(\frac{2\Delta t_1}{z_H}\right)} \\
            \times \sqrt{\cosh\left(\frac{2t_b}{z_H}\right) - \cosh\left(\frac{2\Delta t_2}{z_H}\right)}
        \end{gathered}
        }
\right).
\end{align}

At the saddle point $(t_r,t_b) = (0,0)$, we obtain the area of the timelike EWCS
    \begin{align}
    \mathcal{A}(0,0) &= \arccosh\left(\coth(\frac{\Delta t_1}{z_H})\coth(\frac{\Delta t_2}{z_H})- \frac{1}{\sinh(\frac{\Delta t_1}{z_H})\sinh(\frac{\Delta t_2}{z_H})} \right) \nn \\
    &= \log\left( \frac{\tanh(\frac{\Delta t_1}{2z_H})}{\tanh(\frac{\Delta t_2}{2z_H})} \right).
    \end{align}


\end{document}